\documentclass[aps,prb,groupedaddress,twocolumn,showpacs,longbibliography,10pt]{revtex4-2}
\usepackage{amsmath}
\usepackage{amsthm, amssymb,bbold}    
\usepackage{braket}              
\usepackage{graphicx}
\usepackage{color}
\usepackage[sf]{subfigure}
\usepackage{times, verbatim}      
\usepackage{mathtools,tikz}
\usetikzlibrary{matrix}
\usepackage{natbib}
\usepackage{multirow}
\usepackage{hyperref}
\usepackage{blindtext}
\hypersetup{
       colorlinks=true,
}

\begin{document}

\title{
Approximate two-body generating Hamiltonian for the PH-Pfaffian wavefunction
}

\author{{Kiryl Pakrouski$^{1,2}$}         }
\affiliation{$^{1}$Institute for Theoretical Physics, ETH Zurich, 8093 Zurich, Switzerland}
\affiliation{$^{2}$Department of Physics, Princeton University, Princeton, NJ 08544}

\begin{abstract}

We present two 2-body Hamiltonians that approximate the exact PH-Pfaffian wavefunction with their ground states for all the system sizes where this wavefunction has been numerically constructed to date. The approximate wavefunctions have high overlap with the original and reproduce well the low-lying entanglement spectrum and structure factor. The approximate generating Hamiltonians are obtained by an optimisation procedure where three to four pseudopotentials are varied in the neighbourhood of second Landau level Coulomb interaction or of a non-interacting model. They belong to a finite region in the variational space of Hamiltonians where each point approximately generates the PH-Pfaffian. We diagonalize the identified Hamiltonians for up to 20 electrons and find that for them the PH-Pfaffian shift appears energetically more favorable. Possibility to interpret the data in terms of composite fermions is discussed.

 \end{abstract}

\date{\today}


\maketitle

\section{Introduction}

Fractional quantum Hall effect at the filling factor $\nu=5/2$ \cite{Willett1987} is an interesting topological \cite{Wen1990,*Wen1992} state of matter that may potentially be used as a building block of a topological quantum computer \cite{RevModPhys-Nayak-Simon-2008,RevModPhys-Hansson-Simon-2017}. To large extent our understanding of the phenomenon rests on the model wavefunctions proposed to describe the state and their properties tested in experiment or numerically in small systems.
Until recently the two leading candidate wavefunctions were the Moore-Read Pfaffian \cite{MR1991nonabelions} and its particle-hole conjugate: anti-Pfaffian \cite{Levin-Apf,Nayak-Apf}.  The finite magnetic field used in experiment means finite energy difference between Landau levels \footnote{that may be on the order of the intra-Landau-level Coulomb energy scale} and leads to "Landau level mixing" that generates the particle-hole (PH) symmetry breaking terms in the effective models that could favour one of the two descriptions.

Although the two wavefunctions are topologically distinct, in the small systems the exact numerics is limited to, the competition between them is quite close. For example, including the first five PH-symmetry breaking terms one finds that Pfaffian is favoured \cite{Pakrouski-etal-2016} (see also \cite{Peterson-Nayak-2013,Jain-Toke-Wojs-2010}) while a more precise model including first 6 \cite{Rezayi-2017} terms determines anti-Pfaffian (see also \cite{Simon-Rezayi-2013,*Rezayi-simon-2011,Zaletel-etal-2015}).

Particle-Hole symmetric Pfaffian (PH Pfaffian) \cite{SonPRX.5.031027} (also see a related earlier work \cite{Jolicoeur2007}) is the third very recent candidate that has received support from some experimental observations \cite{Banerjee-etal-2018,dutta2021NewPHPfThermal} but not from numerics \cite{Mishmash-etal-2018,Balram-etal-2018,Mross2020QMC56PHPf} (in contrast to Pfaffian and anti-Pfaffian: \cite{Morf1998-spin,Rezayi-Haldane2000,feiguin-DMRG-2008,Peterson-Jolicoeur-2008,XinWan-etal-disk-2008,Sheng-Haldane-2011}). Under the assumption that the signatures of the PH-Pfaffian never show up in the numerics and it thus can not be stabilised by any Hamiltonian relevant for $\nu=5/2$ a number of alternative scenarios explaining the quantized thermal Hall conductance $\kappa_{xy}$ measurements \cite{Banerjee-etal-2018} have been explored recently \cite{Feldman2016,Simon-aPfGS-2018,Wang-etal-2018,Mross-etal-2018,Lian-Wang-2018,Simon-etal-pf-apf-2020,PhysRevB.98.167401,PhysRevB.98.167402,PhysRevB.99.085309,PhysRevLett.125.157702,PhysRevLett.125.236802,PhysRevLett.124.126801,PhysRevB.102.205104,PhysRevLett.125.016801}.

In this work, we report on a 2-body Hamiltonian that is a deformation of the second Landau level (SLL) Coulomb interaction and whose ground state approximates well the PH-Pfaffian wavefunction as written down in Ref. \cite{Feldman2016} and in all the finite-size systems where this model wavefunction has been numerically constructed until now \cite{rezayi2021energetics}. No such Hamiltonian has been reported to date \footnote{although it has been argued \cite{Sun2020PRB} that in principle the PH-Pfaffian topological order can exist in a translationally and rotationally invariant system} and we hope that it will be instrumental for further studies.

In particular, we are using the wavefunction translated \cite{Fano-etal-1986} into the spinor coordinates \cite{Haldane-1983} in the spherical geometry where it reads

\begin{gather}
\lvert\Psi_{\text{PH-Pf}}(\{{\bm r_k}\})\rangle= \notag \\
P_{LLL} \text{Pf}_{k,l}\left\{ \frac{1}{\bar{u}_k\bar{v}_l-\bar{u}_l\bar{v}_k}\right\}
\prod_{k>l}(u_k v_l-u_l v_k)^2,
\end{gather}
where $P_{LLL}$ stands for projection on the lowest Landau level.

Computing the actual weights of the wavefunction in the fermion occupation basis is described in detail in Ref. \cite{rezayi2021energetics}. It is analogous to the calculation of Coulomb matrix elements on the sphere (given in the appendix of Ref. \cite{Fano-etal-1986}) upon substitution of the Coulomb potential $1/r$ with $1/r^2$.

Particle-hole symmetry requires that exactly half of the available single-particle states are filled with electrons such that $N_\Phi + 1 = 2N_e$, where $N_\Phi$ is the number of flux quanta through the spherical surface and $N_e$ - number of (spin-polarized) electrons. The shift $X$ \cite{Wen1990shift} is a quantum number that distinguishes different topological phases on the sphere, for $\nu=5/2$ states it is defined by the equation $N_\Phi = 2N_e - X$. We observe that $X = 1$ for the PH-Pfaffian while for Pfaffian and anti-Pfaffian wavefunctions we have $X_{Pf} = 3$ and $X_{aPf} = -1$ respectively. A direct consequence of this is that the finite-size calculations at PH-Pfaffian and anti-Pfaffian shifts are performed in different Hilbert spaces which one should keep in mind when comparing them.

\section{Optimisation approach}

An optimisation approach~\cite{pakrouski2019automatic} is used to determine the approximate 2-body generating Hamiltonian. It varies the pseudopotentials and searches the vicinity of a reference interaction for the points that are best according to a certain criteria or target function that is a weighted sum of the relevant properties. The highest contributions are given to the high overlap with the PH-Pfaffian wavefunction and small energy variance of the reference state $(\sigma^{rel}_E)^2 = \frac{ \braket{\psi_{r}|H_O^2|\psi_{r}} - \braket{\psi_{r}|H_O|\psi_{r}}^2}{ \braket{\psi_{r}|H_O|\psi_{r}}^2},$ which quantifies how close it is to being an eigenstate of the variational Hamiltonian. Further properties accounted in the search are: the total angular momentum of the ground state, gap, deviation from the reference interaction and ground state energy. All the system sizes where the model wavefunction is available (plus $N_e$=16) are used and contribute to the target function being optimised proportionaly to the size of their Hilbert space. As we work on the sphere at fixed $L_z=0$ we need to remember that the subspace, relevant for describing FQHE ground state, is given by the condition $L=0$. There are 3;7;24;127 $L=0$ states for systems with 6;8;10;12 electrons respectively. This ensures that the problem of "fitting" four variational parameters (pseudopotentials) is not trivial or over-parametrised as long as systems with at least 12 electrons are used (we use up to 16 electrons).  

The Hamiltonian is parametrized by the 2-body pseudopotentials in spherical geometry \cite{Haldane-1983}. The search is performed following the non-linear conjugate gradient descent algorithm with the Hestenes update rule (see \cite{pakrouski2019automatic} for details).

The method \cite{pakrouski2019automatic} can be viewed as mapping the variational parameters (pseudopotentials) into the feature space (Hilbert space) through diagonalizing the corresponding Hamiltonian and taking its ground state wavefunction. The overlap then defines a kernel in the feature space and other kernel methods of machine learning could potentially be used on top. Thus the method \cite{pakrouski2019automatic} may be considered a very simple example of the kernel-based Machine Learning \cite{Killoran2019QMLInFeatureHilbSp}.

We also have attempted to find the exact 2-body generating Hamiltonian following the "covariance matrix" methods \cite{Chertkov2018InvMethod,Thomale2018HMethod,Qi2019HMethod} without success as zero eigenvalues required by these methods were absent for the covariance matrices constructed for the problem in question. This may be an indication that such an exact 2-body Hamiltonian doesn't exist and an approximate Hamiltonian such as the one presented in this work and found using \cite{pakrouski2019automatic} is the best one can do if restricted to 2-body terms only.

\section{The approximate generating Hamiltonian}

\begin{table}
\caption{\label{tab:ppsOfCV7ANDMV3} The lowest pseudopotentials of the approximate PH Pfaffian generating Hamiltonians and the reference SLL Coulomb values. All higher pseudopotentials of CV7 are identical to the SLL Coulomb (values are given in Table \ref{tab:ppsUsed}) and are 0 for MV3.}
\begin{center}
\begin{tabular}{|c|c|c|c|}
\hline
		& CV7					& 	MV3 					&	 SLL Coulomb\\
\hline
$V_1$	&	1					&	1					&	1		\\
\hline
$V_3$	&	0.694456627311176		&	0.433617799341989		&	0.773278825612202		\\
\hline
$V_5$	&	0.665960300533016		&	0.370884676389928		&	0.576859542105588		\\
\hline
$V_7$	&	0.448785272954577		&	0					&	0.487302680505104		\\
\hline
$V_9$	&	0.52955224410569		&	0.164743667925535		&	0.433005097996708		\\
\hline
\end{tabular}
\end{center}
\end{table}

The approximate 2-body generating Hamiltonian is determined in the vicinity of the two reference interactions: SLL Coulomb given by the pseudopotentials computed for 20 electrons and the non-interacting system $H=0$. The explicit formula for Coulomb pseudopotentials is given by the Eq. (6) of Ref. \cite{TokeJain2006NLLCoulPPs}. It allows one to follow the slight dependence of the pseudopotentials on the system size which is not crucial for our purposes of finding an approximate Hamiltonian in the neighbourhood of these reference values.

In case of Coulomb we vary four pseudopotentials $V_3$ through $V_9$ while "freezing" all others to their reference values. In case of the "minimal model" we only vary three pseudopotentials $V_3$, $V_5$, $V_9$, while freezing the rest. Although one could achieve better fits varying more pseudopotentials it is commonly believed that usually only the lowest $V_m$ with $m\le 9$ have physical significance. Another reason for limiting the number of variational parameters is the desired simplicity of the resulting model.

The optimisation results depend on the significance weights that we assign to various criteria contributing to the target function. Furthermore, because the optimisation problem is non-convex different results might in general be obtained for different initial conditions. Combined together this leads to a certain freedom as to what results should be identified as the best. In Table \ref{tab:ppsOfCV7ANDMV3} we give two of the possible solutions with ids "CV7" (near SLL Coulomb) and "MV3" (minimal model).

\begin{figure}[t]
     \centering
     \includegraphics[width=\columnwidth]{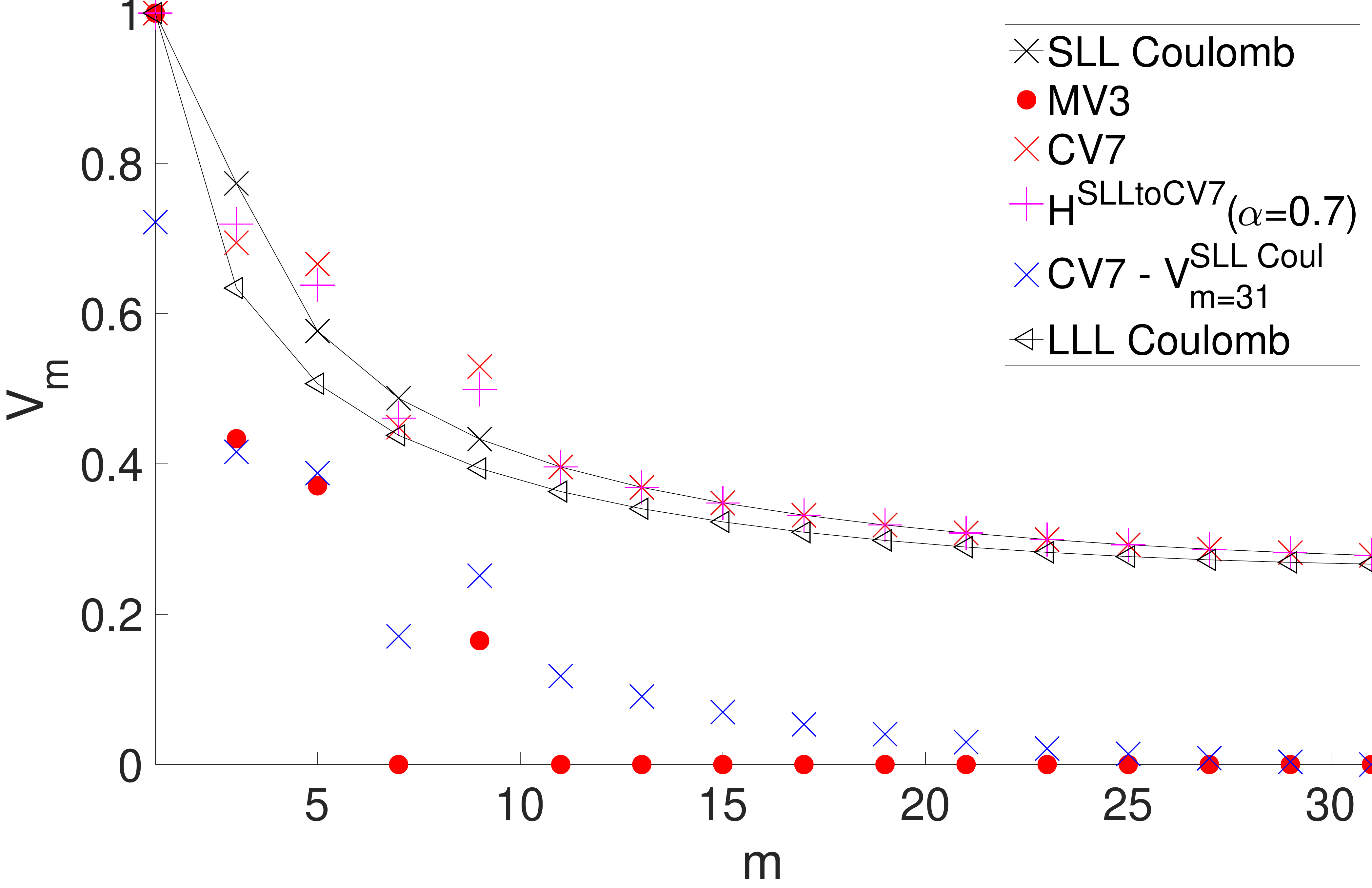}
    \caption{\label{fig:learnedPPsPlot}
     Coulomb interaction in first and second Landau levels is shown with black triangles and crosses respectively. Red circles and crosses are the learned Hamiltonians MV3 and CV7. Magenta "+" data indicates the interaction closest to SLL Coulomb that is still a reasonable approximation of PH-Pfaffian: $0.3*H^{SLL Coul} + 0.7*H^{CV7}$. Blue crosses correspond to the CV7 interaction shifted down by a constant -$V^{SLL Coul}_{31}$.
}
\end{figure}

Fig.~\ref{fig:learnedPPsPlot} shows the learned pseudopotentials plotted together with the reference SLL and LLL Coulomb interactions. Compared to the SLL Coulomb $V_5$ and $V_9$ are the pseudopotentials that differ the most and this deformation is in the direction opposite to the LLL Coulomb (the named pseudopotentials are increased but would need to be decreased to obtain the LLL Coulomb). We also notice that in all the solutions we have obtained $V_3$ and $V_7$ are decreased and $V_5$ and $V_9$ are increased relative to the SLL Coulomb.

\begin{table}
\caption{\label{tab:learnedHperformance} Overlaps and energy variances for the two approximate generating Hamiltonians and the lowest Landau level Coulomb interaction}
\begin{center}
\begin{tabular}{|c|c|c|c|c|c|}
\hline
	 					& 	MV3 				& CV7				& LLL Coulomb \\
\hline
$\braket{\psi_{o}|\psi_{r}}(6)$	&	0.99142087		&	0.99210070			&	0.98628980\\
\hline
$\braket{\psi_{o}|\psi_{r}}(8)$	&	0.98009459		&	0.96521979			&	0\\
\hline
$\braket{\psi_{o}|\psi_{r}}(10)$	&	0.97161429		&	0.96271627			&	0\\
\hline
$\braket{\psi_{o}|\psi_{r}}(12)$	&	0.95538874		&	0.98570050			&	0.92987018\\
\hline
$\braket{\psi_{o}|\psi_{r}}(14)$	&	0.95878369		&	0.97167128			&	0\\
\hline
$(\sigma^{rel}_E)^2(6)$		& 	2.4646E-4			&	1.4254E-5			&	1.3853E-4\\
\hline
$(\sigma^{rel}_E)^2(8)$		& 	4.0823E-4			&	5.2169E-5			&	4.6648E-5\\
\hline
$(\sigma^{rel}_E)^2(10)$		&	2.1047E-4			&	1.4792E-5			&	3.6334E-5\\
\hline
$(\sigma^{rel}_E)^2(12)$		&	9.3918E-5			&	4.5660E-6			&	3.5453E-5\\
\hline
$(\sigma^{rel}_E)^2(14)$		& 	1.6555E-4			&	8.4828E-6			&	1.7261E-5\\
\hline
\end{tabular}
\end{center}
\end{table}

Table~\ref{tab:learnedHperformance} contains the information on how close the Hamiltonians found are to being exact generating Hamiltonians of the PH-Pfaffian wavefunction. It lists the overlaps between the Hamiltonian ground state and the exact model wavefunction as well as the relative energy variance in exact PH-Pfaffian for individual system sizes. We also include the information on the LLL Coulomb that was noticed \cite{Balram-etal-2018,Mishmash-etal-2018} to provide a relatively large overlap with PH Pfaffian for 6 and 12 electrons. Relative energy variance quantifies how far a Hamiltonian is from having a particular wavefunction as an eigenstate. To put these numbers in perspective we provide the same data for the second Landau level Coulomb interaction and the anti-Pfaffian model wavefunction in Table \ref{tab:SLLPLusAPF}. We observe that the precision with which CV7 approximates PH-Pfaffian is not worse than the one for SLL Coulomb and anti-Pfaffian.

The fact that the energy variance for the LLL Coulomb interaction ground state for 8 electrons is lower than that for CV7 highlights the crucial importance of using the multi-variable target function when searching for an approximate generating Hamiltonian. Minimizing the energy variance alone would not necessarily lead to a desired solution as it, for example, carries no information about the position of the model wavefunction in the Hamiltonian spectrum and we might find a Hamiltonian that produces it as an exact but highly excited state. On the other hand, we know that energy variance is positive-definite and assumes its minimal possible value on the desired exact solution and it is therefore advantageous to include it as a part of multi-variable target function in order to reduce the effective size of the variational space and arrive at the Hamiltonian for which the model wavefunction is (almost) an eigenstate. We should also note that finite-size effects are commonplace in exact diagonalisation studies of the fractional quantum Hall effect and the relatively small system with 8 electrons may just as well be simply deviating from the common trend because of its size.

Entanglement spectrum is a way of interpreting the singular value decomposition (also known as Schmidt decomposition) of a quantum system
\begin{gather}
\ket{\psi} = \sum_i s_i \ket{\psi^i_A} \otimes \ket{\psi^i_B},
\end{gather}
where the system is thought of to be made of two parts $A$ and $B$. Working on a sphere we will consider the separation along the equator following Ref. \cite{LiHaldane2008EntSpec}.

Using $s_i = e^{-\frac{\xi_i}{2}}$ we can interpret $\xi_i$ as energy levels \cite{LiHaldane2008EntSpec} and observe that the corresponding thermodynamic entropy becomes identical to the entanglement entropy $S=\sum_i \xi_i e^{-\xi_i}$ which is a widely used measure of entanglement.

The total projection of the angular momentum in sub-system A ($L^z_A$) is a good quantum number that could be used to label each singular value. The level counting in the low-energy part of entanglement spectrum is thought \cite{LiHaldane2008EntSpec} to be a characteristic signature of the underlying topological phase.

An approximate generating Hamiltonian might not reproduce the complete model wavefunction exactly. But it should at least be reasonably reproducing its universal topological fingerprint encoded in the lowest levels of the entanglement spectrum. For example it is known that the Coulomb interaction \cite{LiHaldane2008EntSpec} and some of its deformations \cite{Pakrouski-etal-2016} do reproduce the level counting of the model Pfaffian wavefunction at the Pfaffian shift on the sphere.

\begin{figure}
	\centering
	\includegraphics[width=\columnwidth]{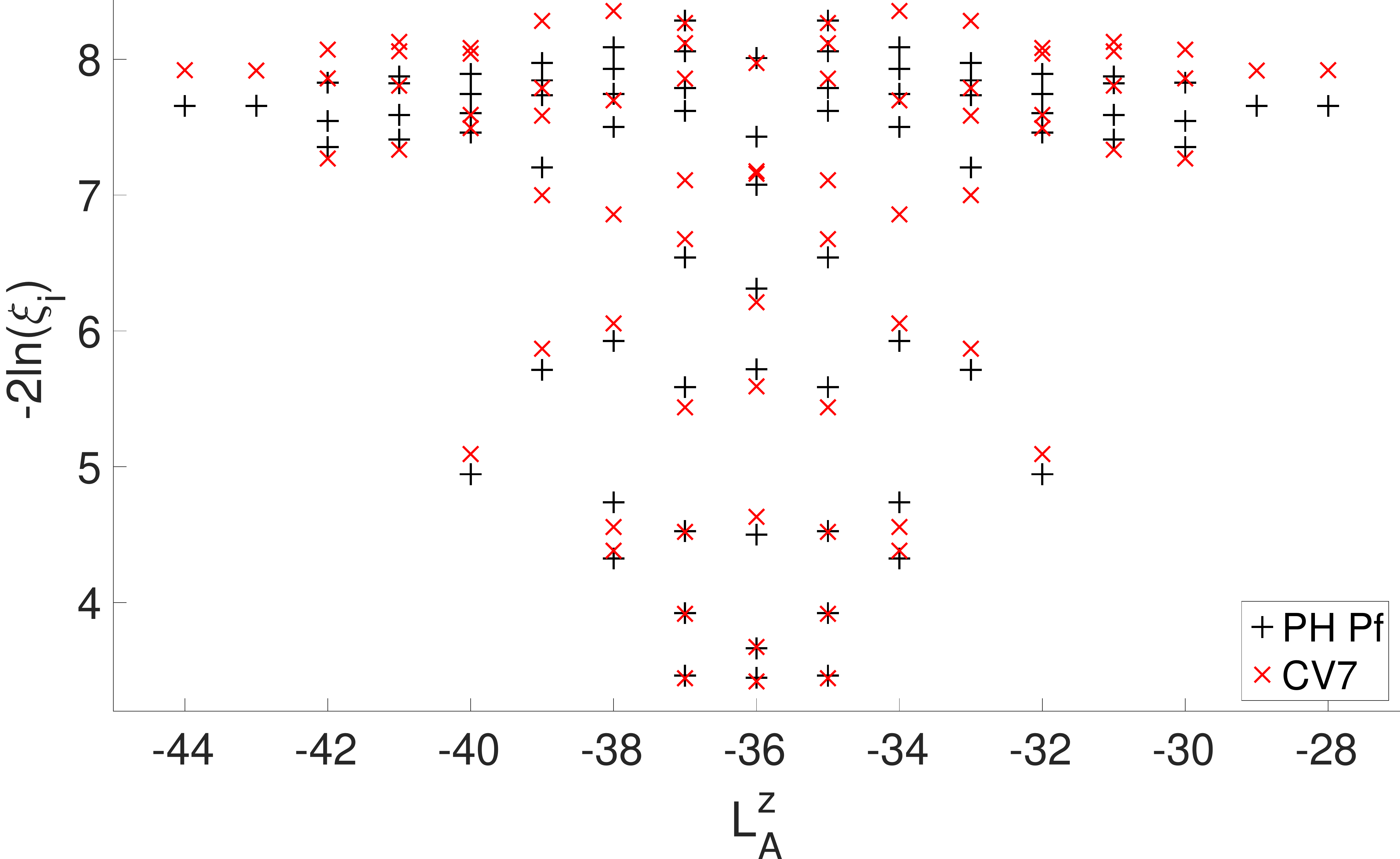}
	\includegraphics[width=\columnwidth]{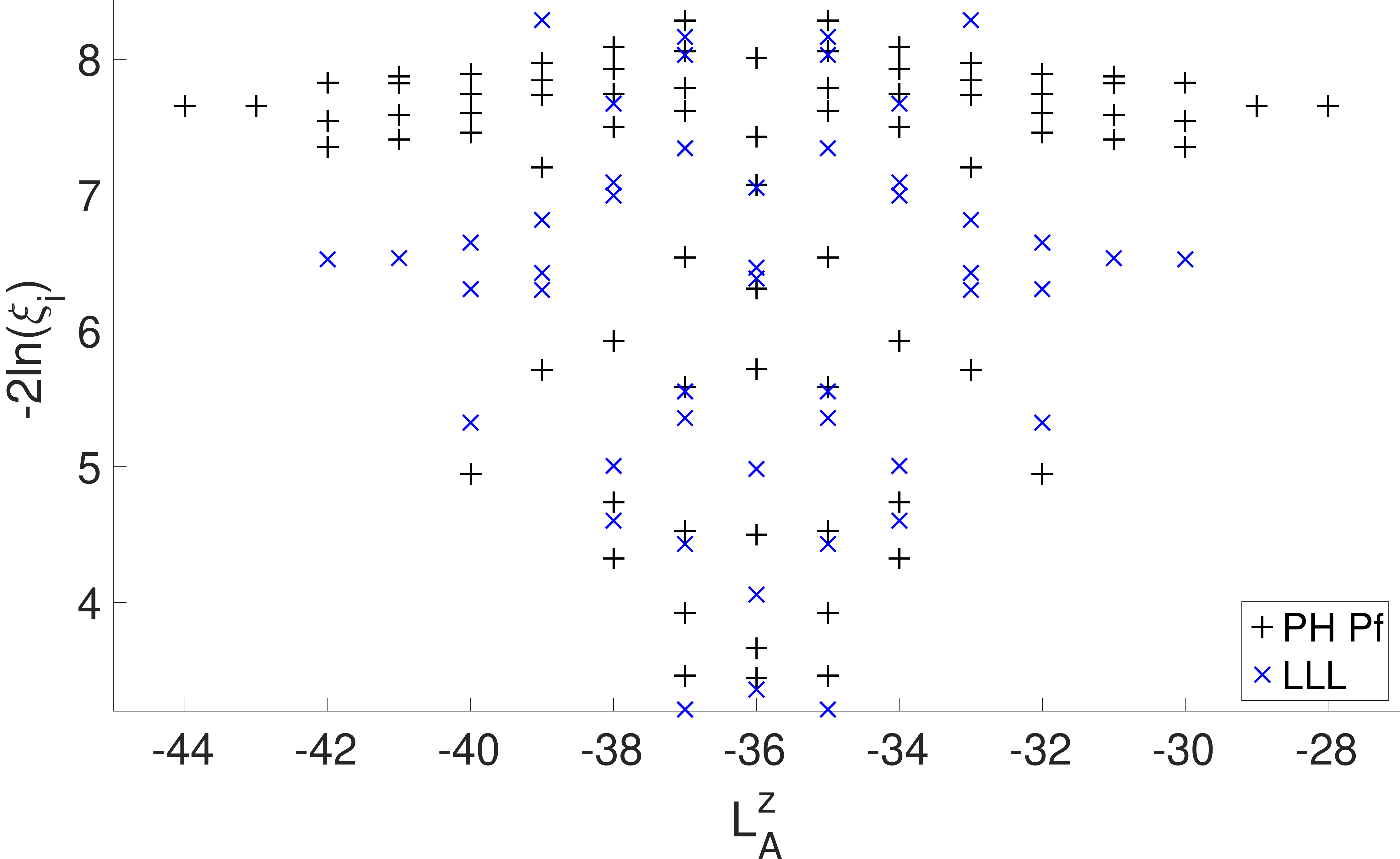}
	\includegraphics[width=\columnwidth]{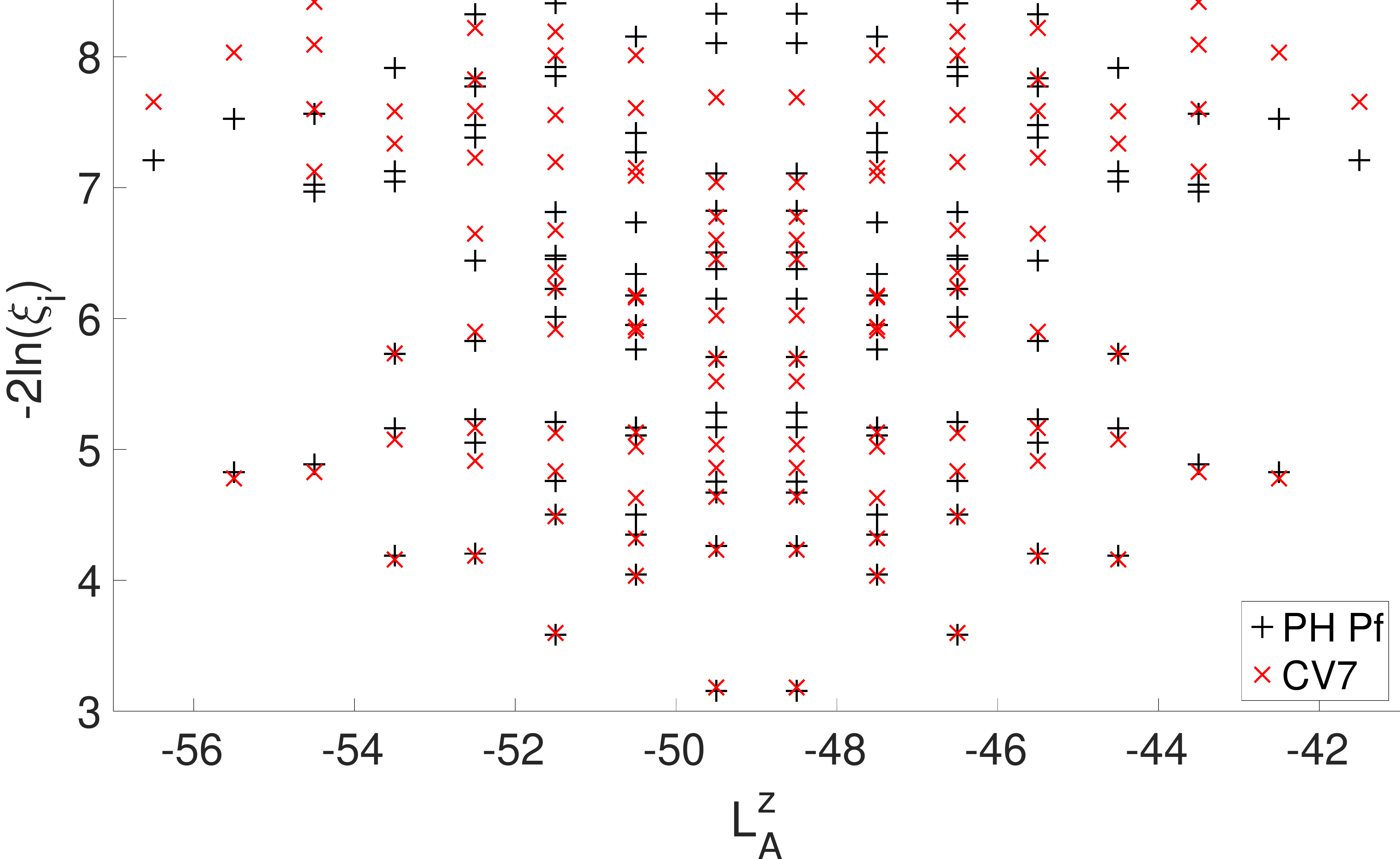}
	\caption{Entanglement spectrum for 12 (Top Panel) and 14 (Bottom Panel) electrons computed in the model wavefunction and the ground state of the learned Hamiltonian near SLL Coulomb (CV7). Middle panel is obtained for 12 electrons and LLL Coulomb ground state. The cut is made at the "equator" of the sphere.} 
		\label{fig:entSpectrum1214PHPfVsLearned}
\end{figure}

In Fig. \ref{fig:entSpectrum1214PHPfVsLearned} we show the entanglement spectra calculated for the ground state of CV7 and the exact PH-Pfaffian for 12 and 14 electrons. We observe that the structure of the low-lying levels is reproduced well. For comparison, the middle panel of Fig. \ref{fig:entSpectrum1214PHPfVsLearned} shows the data for the LLL Coulomb ground state. We observe that the learned Hamiltonians (see Appendix Fig. \ref{fig:entr12CV7vsLLL} for the MV3 data) are substantially better than the LLL Coulomb in reproducing the structure of the PH-Pfaffian entanglement for 12 electrons (for 14 electrons the ground state of LLL Coulomb has $L\ne 0$).

A central question related to the PH-Pfaffian wavefunction and universality class is whether it could be a valid description of uniform gapped FQHE state observed at $\nu=5/2$ in experiment. To make a step towards answering this question we study the static structure factor of the model wavefunction and the ground states of the learned Hamiltonians.

It has been argued \cite{GMPQtoThe4thPRL1985,GMPQtoThe4th,haldane_quantum_1990} that the projected static structure factor 

\begin{gather}
\bar{S}(q) = \frac{1}{N} \braket{\bar{\rho}^\dagger_q \bar{\rho}_q}
\end{gather}
(with $N$ - number of particles and $\bar{\rho}_q$ - the Fourier transform of the density operator projected onto the lowest Landau level) must quite universally vanish as $|q|^4$ or faster in order for the state in which it is evaluated to be a gapped FQHE state.

 The corresponding quantity on the sphere with $2S+1$ single-particle basis states (for $L\ne0$)\footnote{$S_0(0)=0$} is \cite{HaldaneUnpublished}
 \begin{gather}
S_0(L) = \frac{1}{2S+1} \braket{\bar{\rho}^\dagger_{LM} \bar{\rho}_{LM}},
\end{gather}
where $L$ is the total angular momentum and $M$ - its projection.

For $L\ne0$ it can be evaluated \cite{HaldaneUnpublished} as \footnote{while the structure factor is defined to be 0 for $L=0$ the equation evaluates to $\frac{N^2}{2Q+1}$ for $L=0$}
 \begin{gather}
S_0(L) = \frac{2L+1}{2S+1} \sum_{m,m'} \braket{ S,m | L,0; S,m } \braket{ S,m' | L,0; S,m' } *\notag \\
\braket{n(m)n(m')}.
 \label{eq:structFact}
\end{gather}
Here $\braket{ J,M | J_1,m_1; J_2,m_2 }$ is a Clebsch-Gordan coefficient that couples two particles with total angular momentum (projections) of $J_1$ and  $J_2$ ($m_1$ and $m_2$) into a state with total angular momentum $J$ and projection $M$. Operator $n(m)$ is the number operator in the "orbital" $m$.

\begin{figure}
	\centering
	\includegraphics[width=0.49\columnwidth]{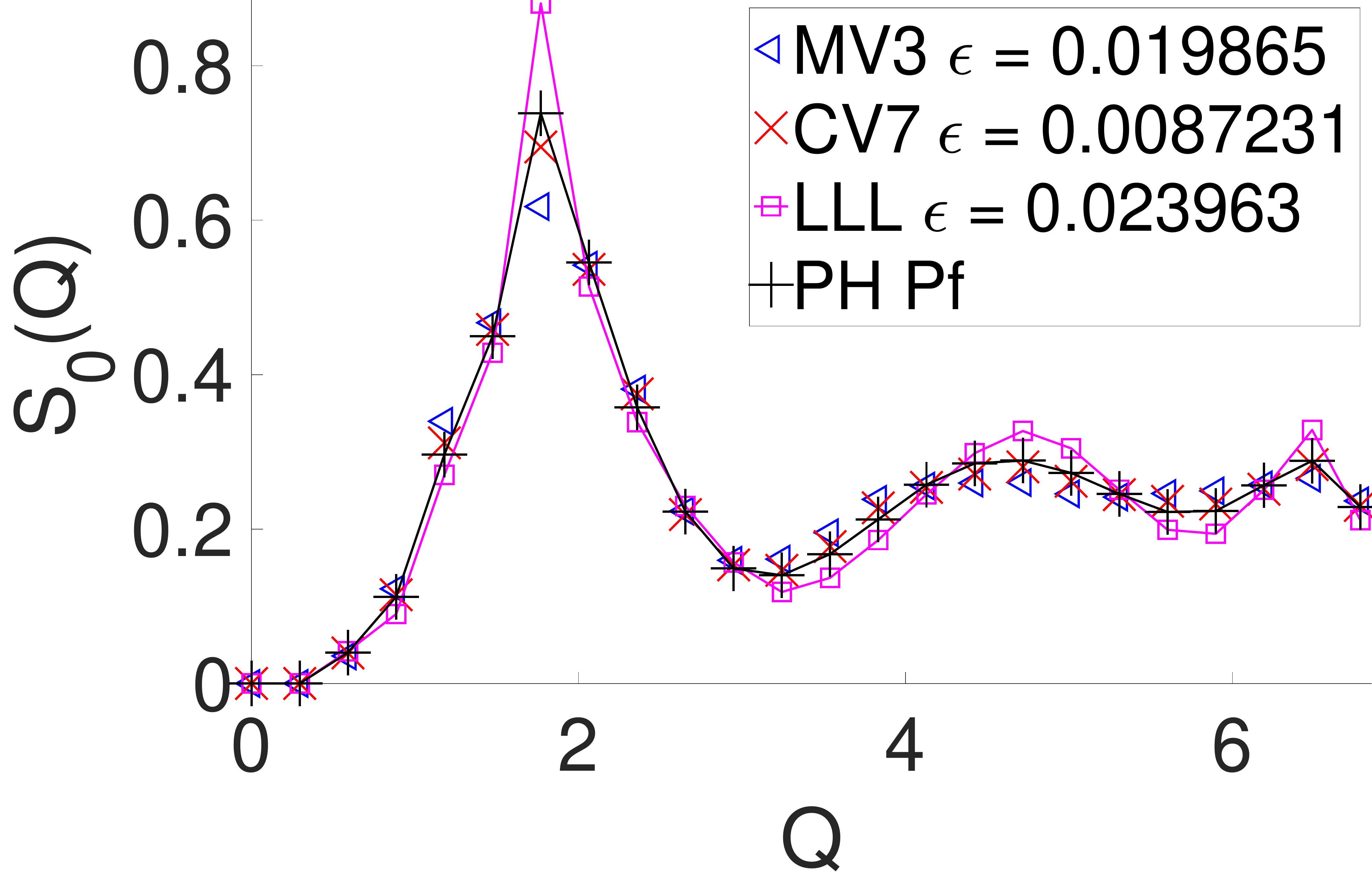}
	\includegraphics[width=0.49\columnwidth]{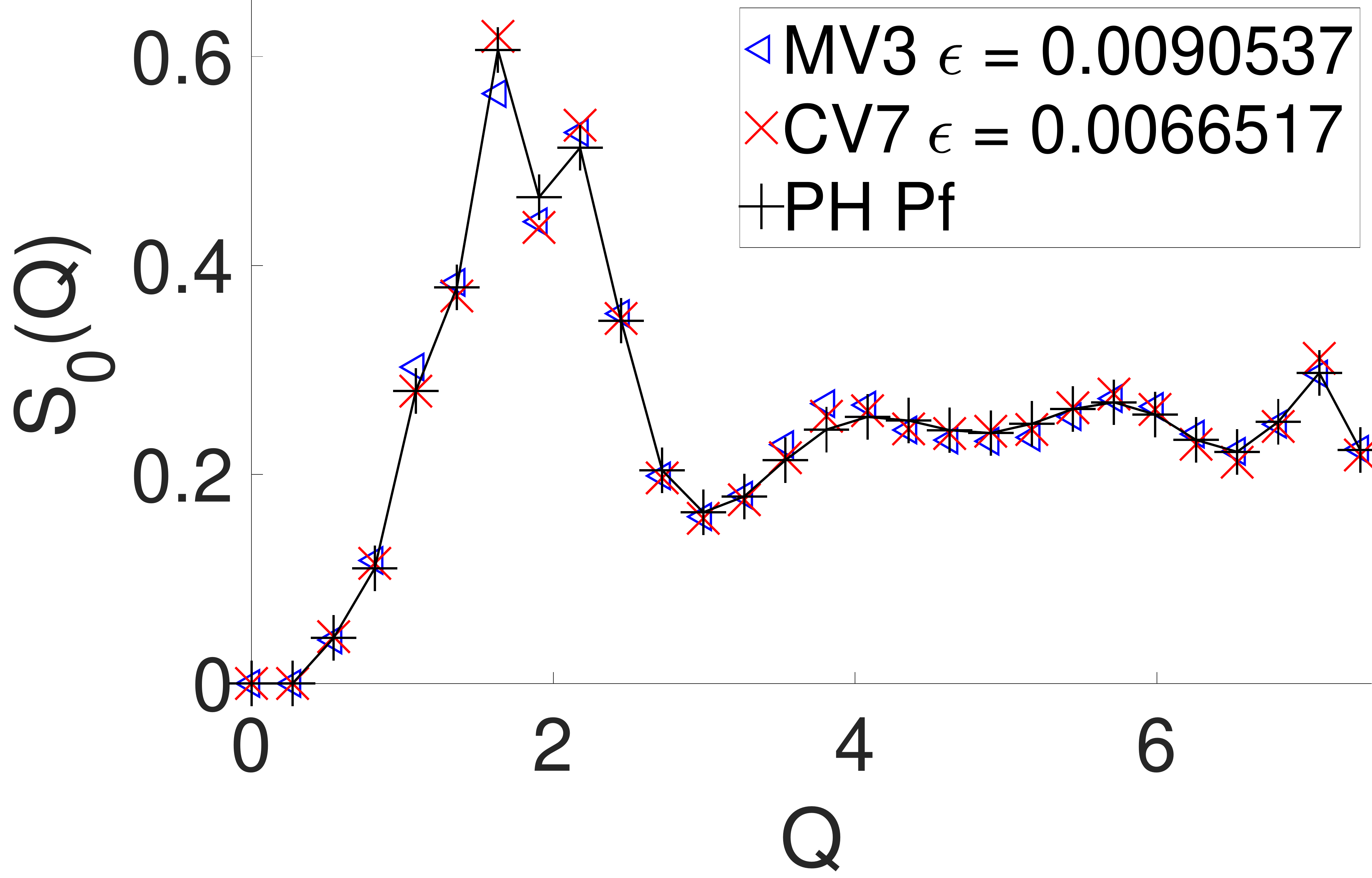}
	\caption{Structure factor $S_0(Q)$ for 12 (Left Panel) and 14 (Right Panel) electrons computed in the model PH-Pfaffian wavefunction and the ground states of the learned Hamiltonians and the LLL Coulomb interaction. The normalized deviation from the PH-Pfaffian data is $\epsilon = \sum^{L_{max}}_{L=1} (|S_0(Q) - S^{PHPf}_0(Q)|) / L_{max}$}.
		\label{fig:SofQ1214PHPfVsLearned}
\end{figure}

Fig. \ref{fig:SofQ1214PHPfVsLearned} shows the structure factor \eqref{eq:structFact} as a function of $Q=\sqrt{L^2/S}$ for 12 and 14 electrons computed in the model wavefunction and in the ground states of relevant Hamiltonians. We observe that the learned Hamiltonian CV7 is best at reproducing the original structure factor while the data obtained in the LLL Coulomb ground state shows more pronounced oscillations and also grows slower at small $Q$ than all other data.

Although the two Hamiltonians CV7 and MV3 are visually quite different (Fig. \ref{fig:learnedPPsPlot}) their ground states have relatively high overlap above 0.98 up to 20 electrons (Tab. \ref{tab:CV7vsMV3}).

If we interpolate between them $H(\alpha)=(1-\alpha) H_{MV3} + \alpha H_{CV7}$ (Fig. \ref{fig:interpolateMV3toCV7}) we do not observe any sign of gap closing and the PH-Pfaffian overlap stays above 0.955 for all system sizes at all interpolation points. This suggests that the two Hamiltonians CV7 and MV3, actually belong to a single continuously connected region in the parameter space defining the Hamiltonians that approximately generate PH-Pfaffian \footnote{Conditioned on the state at the two points MV3 and CV7 being gapped in the first place, as discussed later finite size effects prohibit a definitive answer}. This conclusion is supported by the entanglement spectrum data computed at every third interpolation step (Fig. \ref{fig:entanglementMV3toCV7}) for $N_e=12,14$. The original level counting of the PH-Pfaffian (forming a "signature" of a topological phase \cite{LiHaldane2008EntSpec}) is preserved for every interpolation step. A further evidence is the good agreement between the low-lying entanglement spectra of CV7 and MV3 for $N_e=18,20$ (Fig. \ref{fig:EntSpectrumCV7vsMV3}).

An interesting direction for the future studies would be to use the presented data for mapping out the full subspace of Hamiltonians that approximate PH-Pfaffian. Given the observed distinction between CV7 and MV3 defined in terms of pseudopotentials (Fig.~\ref{fig:learnedPPsPlot}) it is possible that this is best done in terms of other variables. It would also be interesting to understand the similarities between the two learned Hamiltonians as real-space interactions with a certain screening.

Blue crosses in Fig. \ref{fig:learnedPPsPlot} show the CV7 pseudopotentials shifted downwards by a constant equal to the largest pseudopotential that was used in the optimisation procedure $V^{SLL Coul}_{31}=V^{CV7}_{31}$. Note how close the resulting $V_3$ and $V_5$ become to the values in MV3. Together with $V_9$ increased over $V_7$ this might be the underlying general feature that is required for approximating PH-Pfaffian.

The available data indicates that the learned Hamiltonians CV7 and MV3 represent a reasonable approximation of the PH-Pfaffian wavefunction for the system sizes with up to 14 electrons. We will now attempt to extract additional information about the PH-Pfaffian state from these Hamiltonians. Note however, that this program may only be successful if the small system sizes we used for Hamiltonian learning already contained enough information representative of the PH-Pfaffian phase \cite{pakrouski2019automatic} which can not be verified before the PH-Pfaffian wavefunction for larger sizes is available.  

\section{Physics of the approximate generating Hamiltonian \label{sec:PhysOfApproxH}}
First we would like to understand how the learned Hamiltonians compare to the Coulomb interaction in the first and second Landau levels keeping in mind that the SLL Coulomb interaction ground state was shown to be adiabatically connected to the three-body interaction exactly generating the MR Pfaffian state \cite{Morf2010CoulSameUCAsMR} at the Pfaffian shift on the sphere.

\begin{figure}[t]
    \centering
\includegraphics[width=\columnwidth]{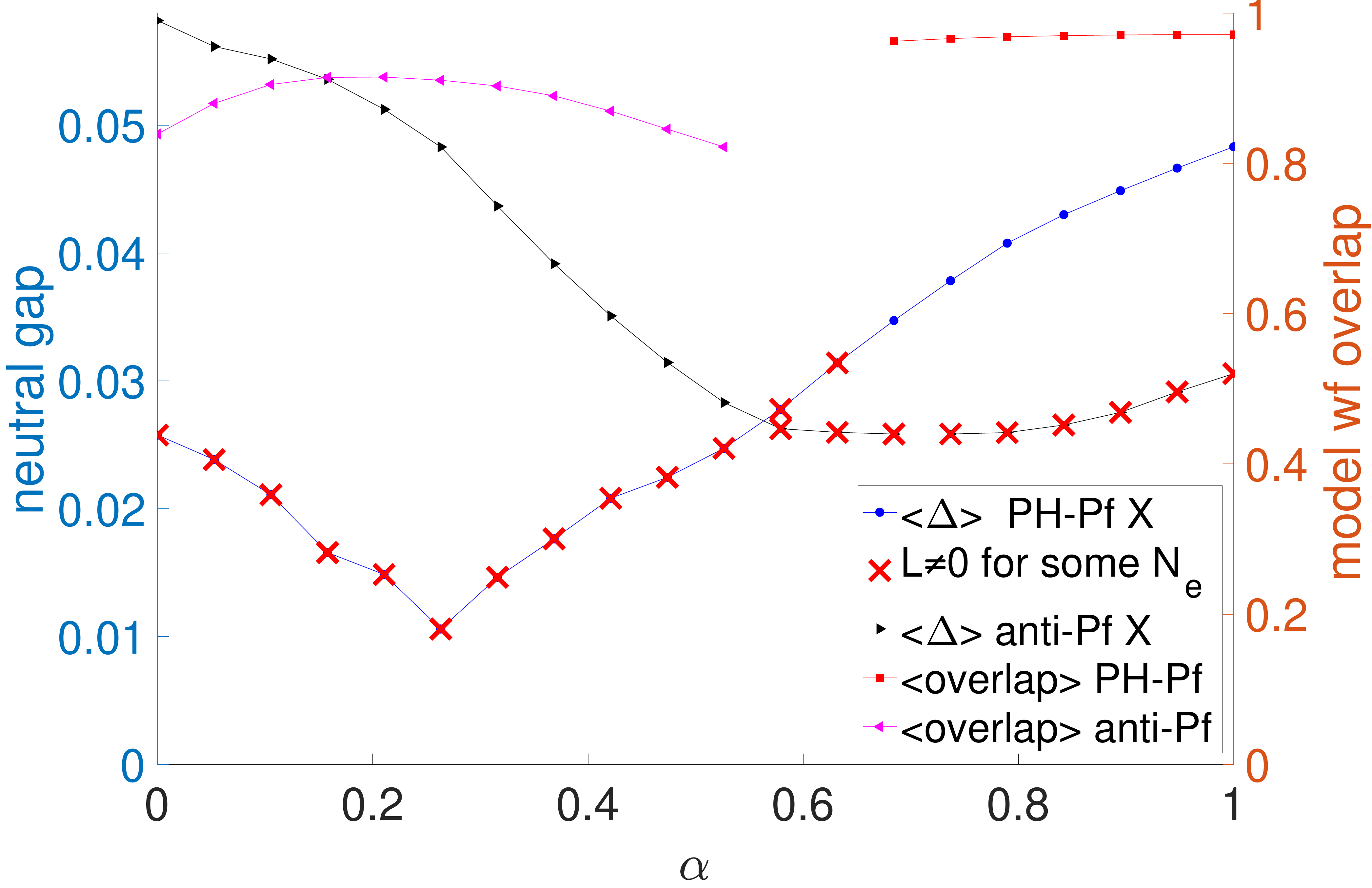}
    \caption{\label{fig:interpolateSLLToCV7}
    Interpolation between the SLL Coulomb interaction ($\alpha=0$) and the learned Hamiltonian CV7 ($\alpha=1$). $<>$ stands for averaging over the available system sizes. Overlap is averaged over systems with 8-14 particles and gap - over systems with 8-16 particles. 
}
\end{figure}

When interpolating between SLL Coulomb and CV7 ($H(\alpha)=(1-\alpha) H_{Coul} + \alpha H_{CV7}$) we perform the diagonalisation at both the PH-Pfaffian and anti-Pfaffian shifts. For each of the system sizes we keep track of the neutral gap (difference between the lowest eigenvalues) and the overlap with the relevant model wavefunction. Fig. \ref{fig:interpolateSLLToCV7} shows the cumulative averaged data for the system sizes with 8 to 16 electrons (holes). If at a given $\alpha$ we find an $L\ne 0$ ground state for any of the available system sizes (up to $N_e$=18) we mark the gap with a red cross indicating that no fractional quantum Hall state is possible for that interaction at the corresponding shift.

Both PH-Pfaffian and anti-Pfaffian appear to be stabilized in the accessible finite systems in the vicinity of the CV7 or SLL Coulomb interactions at the PH-symmetric and anti-Pfaffian shifts accordingly. As we interpolate between the two interactions the gap at both shifts appears to decrease and have a minimum in the region of $\alpha \approx 0.5-0.7$.  For most individual system sizes (Fig. \ref{fig:interpolateToCV7SM}) the gap actually closes accompanied by a sharp drop of the overlap (where available). This however is not the case for the "closed shell" \cite{rezayi2021energetics} systems such as 12 particles.

At least some of the system sizes have non-uniform ($L\ne 0$) ground states in the same region $\alpha \approx 0.5-0.7$. Taken together these observations would be consistent with a phase transition from PH-Pfaffian to anti-Pfaffian universality classes as we tune the interaction between CV7 and SLL Coulomb. Further, there might be an intermediate phase with broken spatial symmetry. The detailed characterization of the possible phase transition however goes beyond the scope of this work and due to the extremely small energy gaps would require access to significantly larger system sizes. Such a study may also be best performed in another geometry - on the sphere the two universality classes appear at different shifts which would complicate their direct comparison.

If the learned Hamiltonian CV7 is deformed in the direction of the LLL Coulomb the PH-Pfaffian state is destroyed much faster (Fig. \ref{fig:interpolateLLLtoCV7}). For $N_e=18$ (not shown) there is no single datapoint in that direction with $L=0$, meaning no valid FQHE state is possible if we move towards LLL Coulomb. The system with 12 electrons again does not seem to close the gap during the interpolation while the first excited state changes from $L=6$ at LLL Coulomb to $L=2$ for CV7.

The SLL Coulomb-to-CV7 interpolation data (Fig. \ref{fig:interpolateSLLToCV7}) allows us to conclude that the found Hamiltonian CV7 is a reasonable approximation for PH-Pfaffian in a finite region in the parameter space rather than at a single special point.

\begin{figure}
	\centering
	\includegraphics[width=\columnwidth]{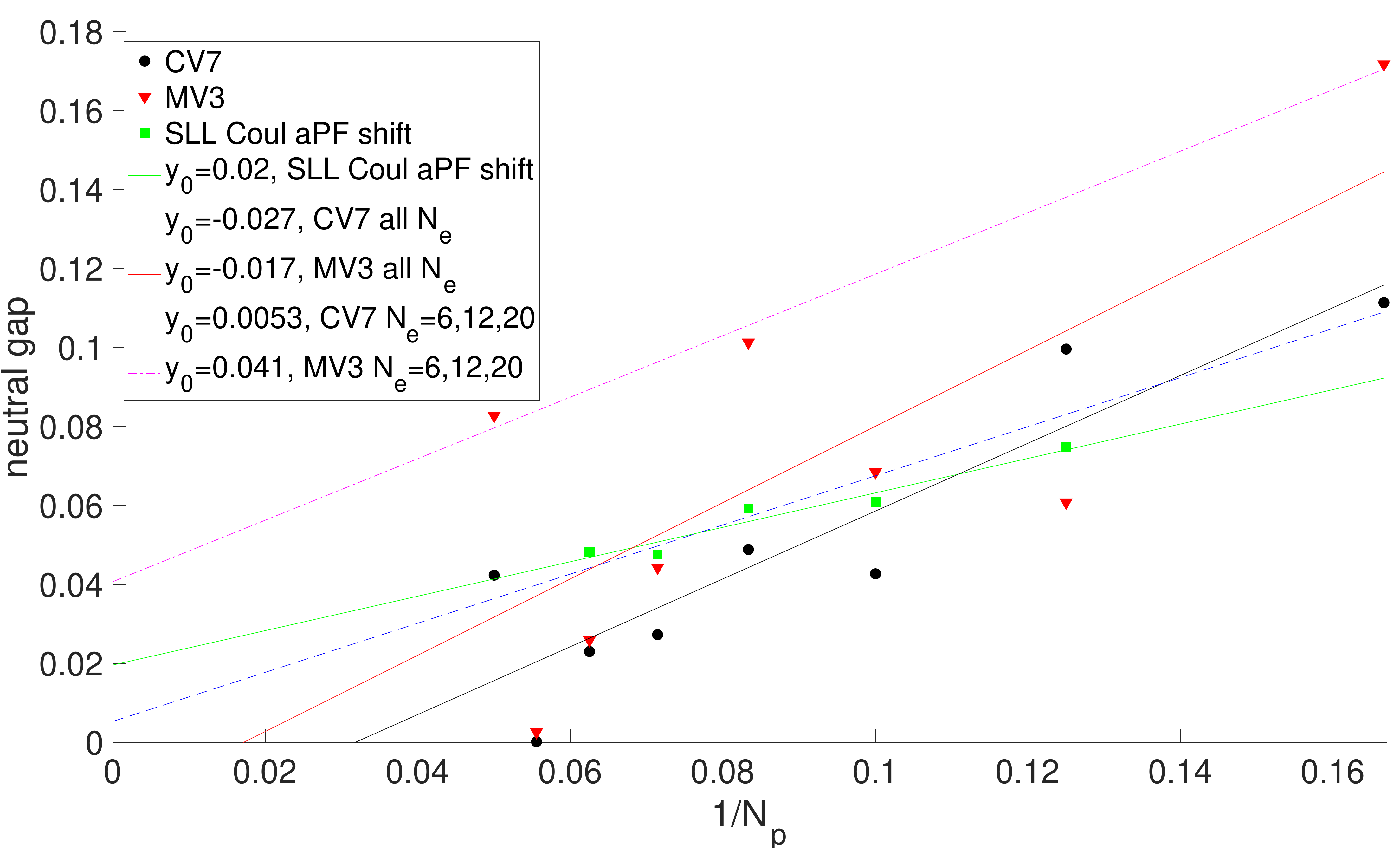}
	\includegraphics[width=\columnwidth]{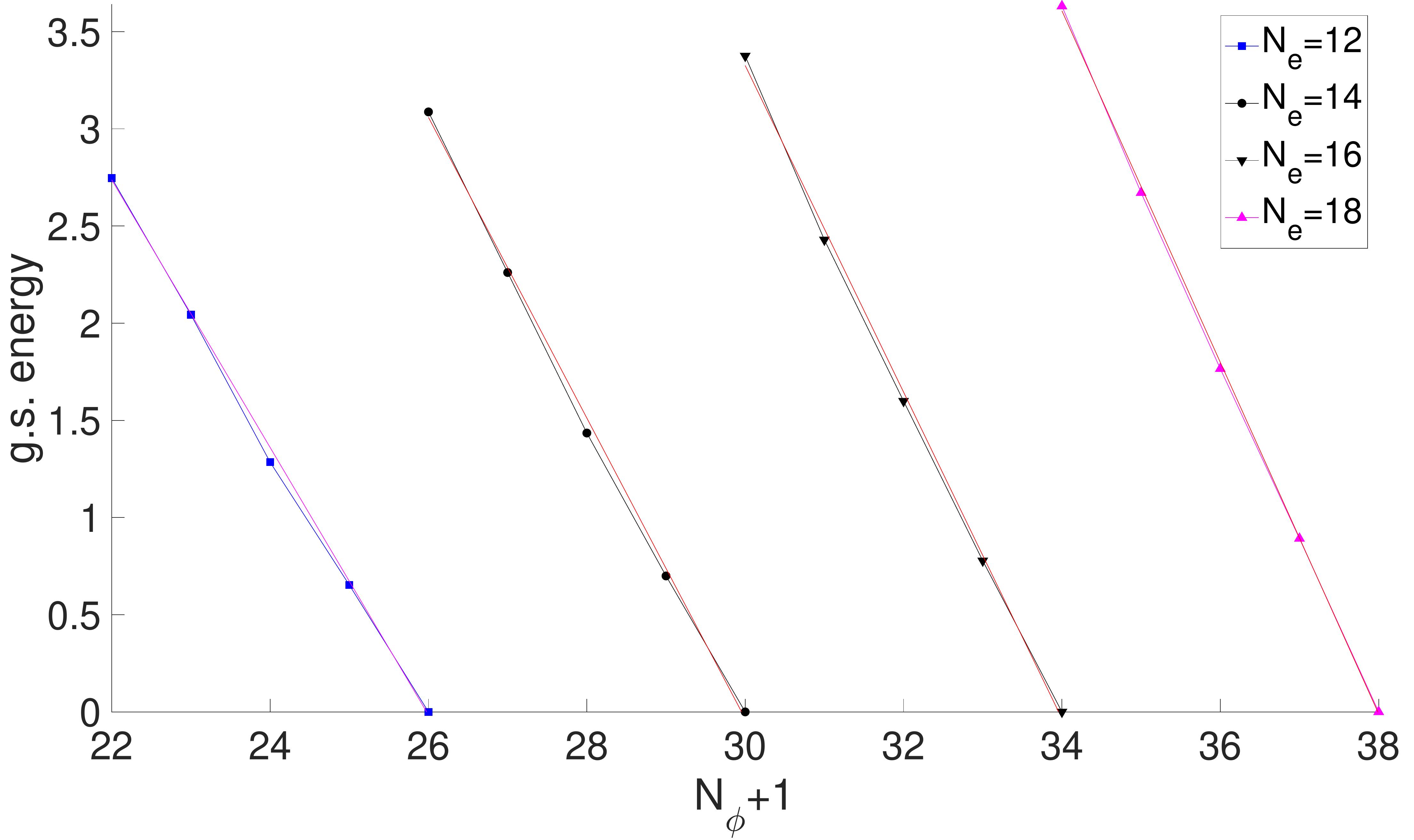}
	\caption{\label{fig:genGapsBothLearned} Top Panel: Black circles and red triangles - energy gaps for the two learned Hamiltonians CV7 and MV3 and system sizes of 6 to 20 electrons. For each data set two linear fits are displayed: using all data points (black solid line for CV7 and red solid line for MV3) and using the "closed shell" sizes of 6,12,20 electrons (blue dashed line for CV7 and magenta dot-dashed line for MV3). Green squares - energy gaps for SLL Coulomb interaction at anti-Pfaffian shift computed for 8-16 holes. The constant term of the linear fit is indicated in the legend. Bottom Panel: Lowest energy at the PH-Pfaffian and nearby shifts calculated for CV7 and $N_e=12,14$. For each $N_e$ we subtract the energy at maximum flux to make them comparable. The linear fit shown in red and magenta is over the four points excluding the flux corresponding to the PH-Pfaffian shift.
}
\end{figure}

The neutral energy gaps for the two learned Hamiltonians are shown in the top panel of Fig. \ref{fig:genGapsBothLearned}. The analysis is complicated by both the finite-size effects and the fact that the closed-shell systems with 6,12 and 20 electrons have much higher gaps as if they stemmed from a different dataset. Therefore the data doesn't seem to allow a reliable extrapolation and a conclusion if the corresponding states are gapped in the thermodynamic limit. For comparison we also show the more consistent data at anti-Pfaffian shift and SLL Coulomb interaction.

For CV7 we also compare the ground state energy at various shifts at fixed electron number (Bottom Panel of Fig. \ref{fig:genGapsBothLearned}). We observe that the energy at the PH-Pfaffian shift is lower than the average at nearby flux values. For 12 electrons energy at the PH-Pf shift is $E(24) = 36.2963$ while the average of energies at the Pfaffian and anti-Pfaffian shifts is $(E(22)+E(26))/2 = 36.3845$. For 14 electrons corresponding energies are 47.5120 and 47.6207 again favouring PH-Pfaffian shift. The energies and angular momenta at various fluxes and electron numbers for CV7 and MV3 are listed in Table \ref{tab:sweepFlux} and support the discussed trend. The fact that $L\ne 0$ at $\pm 1$ flux is consistent with the PH-Pfaffian flux being most energetically favourable and the nearby states corresponding to quasiparticles. 

The systems with $N_e$ = 6,12,20 electrons stand out by every possible measure. In Fig.~\ref{fig:genGapsBothLearned} we observe that the neutral gap for these system sizes is significantly higher and extrapolates to a positive value while a linear fit including all data extrapolates to negative values. Note that the model wavefunctions themselves are also substantially different as can be seen from the structure factor plots (Fig.~\ref{fig:strFactsLearnedH}).

The special properties of these systems are consistent with them corresponding to closed-shell configurations of composite fermions \cite{Jain-CF-1989} with the effective flux 1 where a system with $(\tilde{n}+1)(\tilde{n}+2)$ electrons completely fills all $\Lambda$-levels up to $\tilde{n}$ so that maximum filled level for $N_e=$6,12,20 is $\tilde{n}=$1,2,3. Further numerical data consistent with this assumption is assembled in Appendix \ref{sec:AppendixCFInterpret}.

To discuss whether the state described by the model wavefunction and approximated by the ground states of the learned Hamiltonian is gapped we perform the scaling analysis of the structure factor $S_0(Q)$ \eqref{eq:structFact}. For a gapped state the structure factor should grow as $Q^\alpha=L^\alpha S^{-\alpha/2}$ with $\alpha\ge 4$ \cite{GMPQtoThe4thPRL1985,GMPQtoThe4th,haldane_quantum_1990}. Consider the two smallest values $L^*=2,3$ and plot $ln(S_0(L^*))$ against $ln(S)$. For sufficiently large system sizes we should be able to read off $-\alpha/2$ from the slope of the linear fit to the data. We do obtain $\alpha\ge4$ for such an analysis performed for the anti-Pfaffian model wavefunction and {\it large enough ($N_h\ge 12$) systems} (Fig. \ref{fig:strFactPhPfVSAPF}). However, as the linear fit using smaller system sizes shows (Fig. \ref{fig:strFactPhPfVSAPF}) the answer wouldn't have been so clear even for the anti-Pfaffian had only the systems with $N_h\le 14$ been available. As the previous literature suggested \cite{Mishmash-etal-2018,Balram-etal-2018,rezayi2021energetics} the gap of the PH-Pfaffian must be much smaller should it be gapped. It is therefore to be expected that the "reliable" system size threshold is higher for the PH-Pfaffian.

\begin{figure}[t]
     \centering
     \includegraphics[width=\columnwidth]{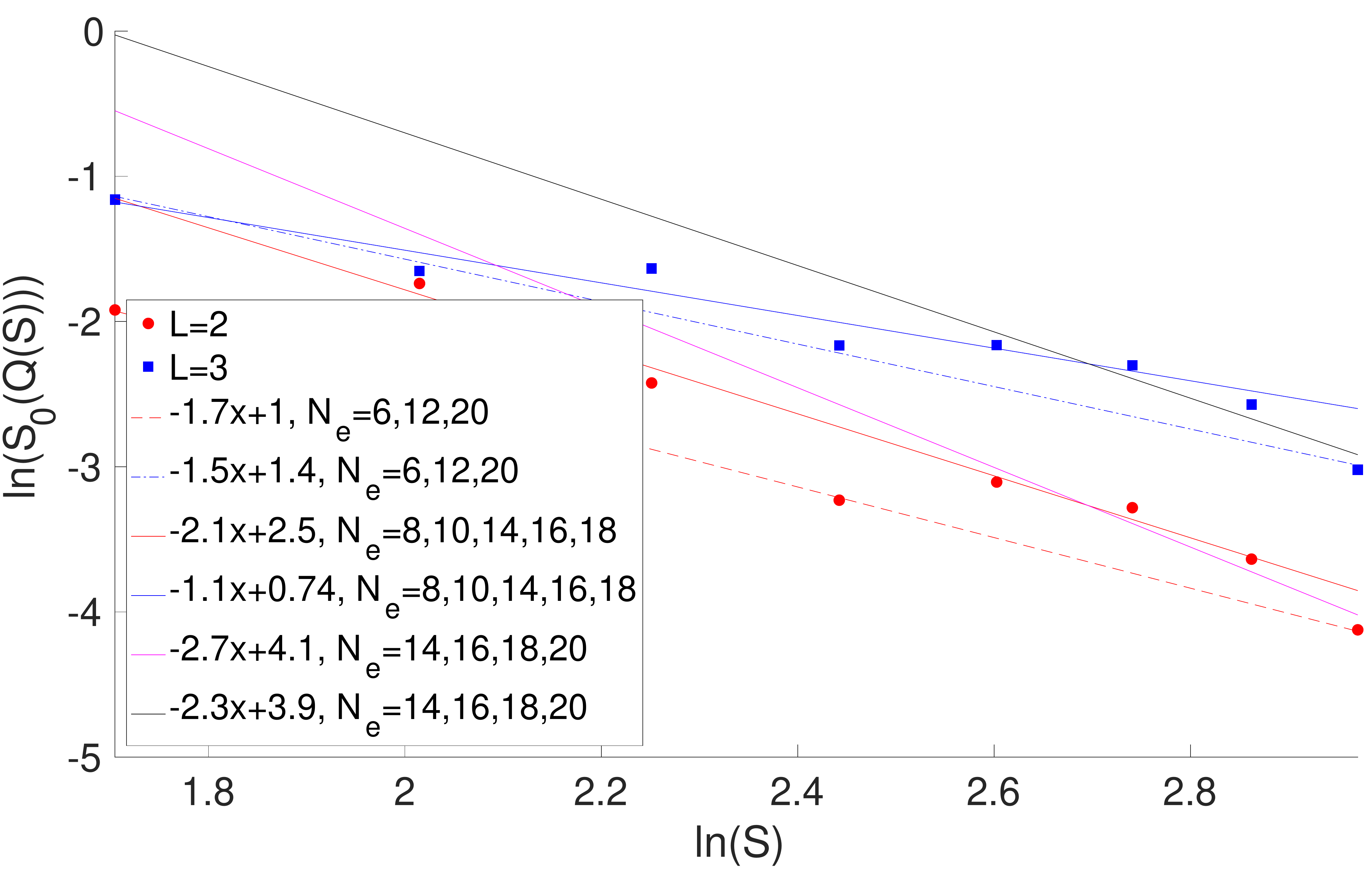}
     \includegraphics[width=\columnwidth]{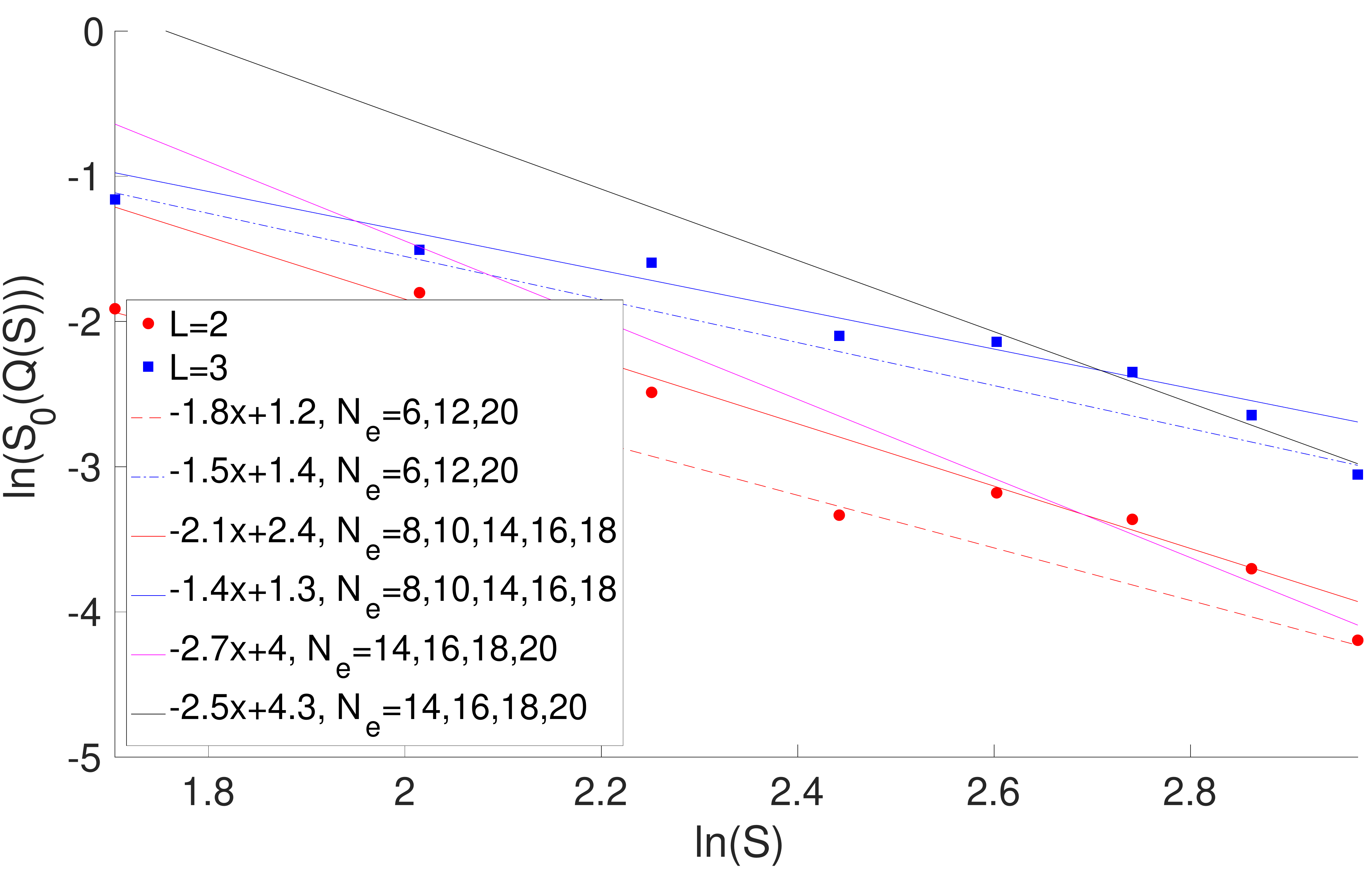}
     \includegraphics[width=\columnwidth]{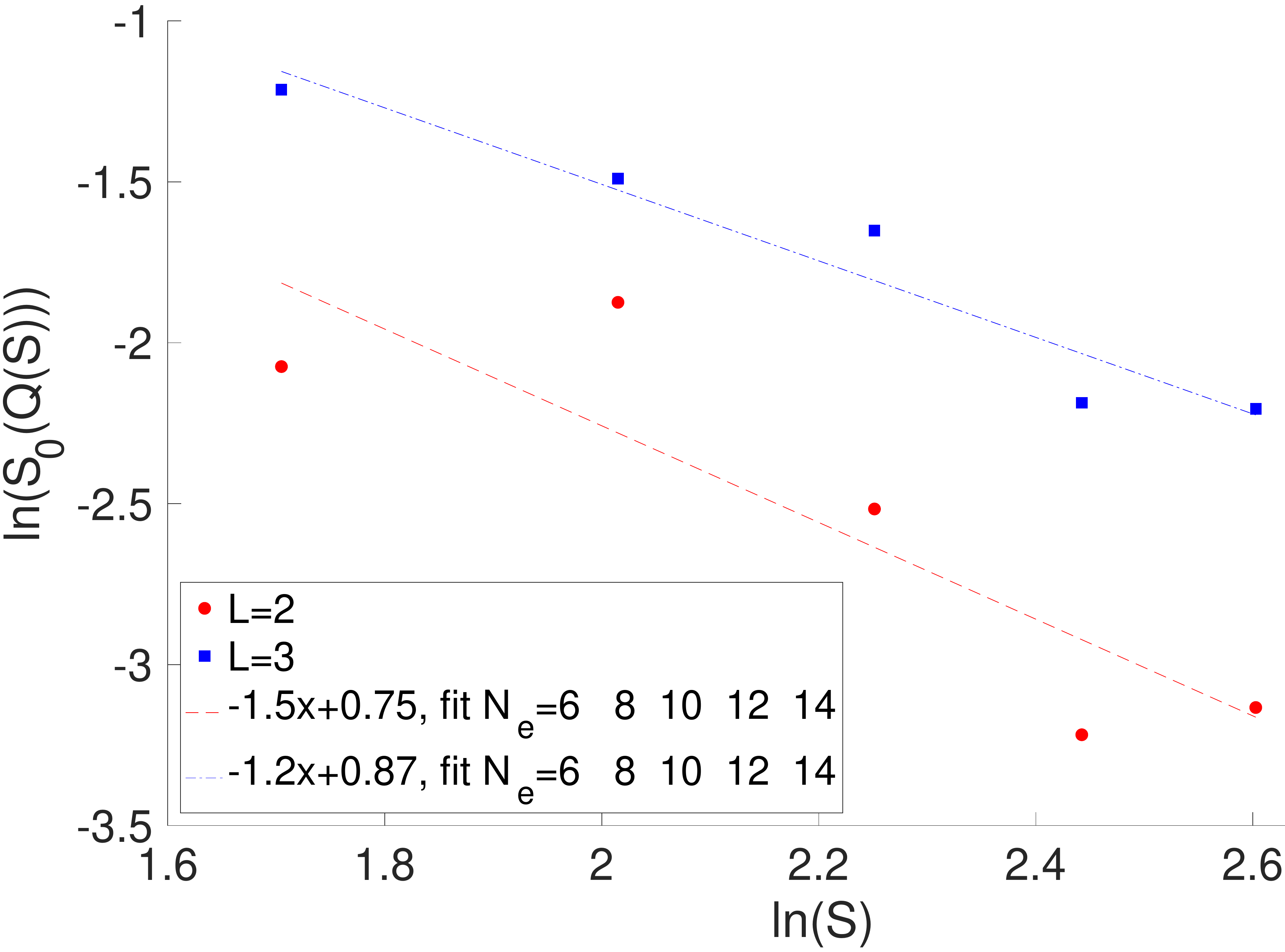}
    \caption{\label{fig:strFactExtr}
     Structure factor extrapolation over the ground states of the CV7 (Top Panel) and MV3 (Middle Panel) learned Hamiltonians. Several possible linear fits including subsets of the available system sizes are displayed. Data for the PH-Pfaffian model wavefunction is shown in the Bottom Panel.
}
\end{figure}

Large finite-size effects in the CV7 and MV3 data for PH-Pfaffian case and the special behaviour of the closed-shell system sizes lead to several possible ways to linearly fit the available data some of which are shown \footnote{Fitting using all the system sizes (not shown) would give $\alpha\approx3.6$ for $L=2$ and $\alpha\approx 2.6$ for $L=3$} in Fig.~\ref{fig:strFactExtr}. The extracted $\alpha$ and the qualitative result would depend on which data is used for the fit. In particular, a fit for the closed-shell sizes $N_e=6,12,20$ gives $\alpha<=3.6$ suggesting gapless state while a fit using the four largest available system sizes $N_e=14,16,18,20$ leads to $\alpha>=4.6$ which would be consistent with a gapped state. Given this uncertainty we are not able to draw a solid conclusion and leave it (along with all the raw data available in SM) to the reader.

Analogous data for the PH-Pfaffian wavefunction for $N_p\le 14$ is shown in the Bottom Panel of Fig. \ref{fig:strFactExtr} but is likely to not be representative of the thermodynamic limit behaviour as the data for anti-Pfaffian (dashed lines in Fig. \ref{fig:strFactPhPfVSAPF}) suggests.

A related issue are the oscillations of the structure factor at large $Q$ and the two-peak structure noted earlier \cite{rezayi2021energetics} for the model PH-Pfaffian at some $N_e$. With the approximate wavefunctions we have access to larger system sizes and observe in Fig. \ref{fig:strFactsLearnedH} that the oscillations decrease and the two-peak structure becomes less pronounced for larger system sizes. It is thus possible that these two artefacts would be gone completely in larger systems. This assumption is substantiated by the comparison to the anti-Pfaffian data (bottom left panel of Fig. \ref{fig:strFactsLearnedH}) where the 8-hole structure factor resembles the double-peak structure and the 12-14-hole data exhibits remainder oscillations similar to the 20-electron PH-Pfaffian data.

\section{Conclusions and Outlook}  

We presented two 2-body Hamiltonians that reasonably well approximate an implementation \cite{rezayi2021energetics} of the PH-Pfaffian wavefunction on a sphere for all the system sizes where it is available. One of the Hamiltonians is a deformation of the second Landau level Coulomb interaction, the other - of a non-interacting model. Both Hamiltonians belong to a finite region of the four-dimensional Hamiltonian variational space where each point approximately generates PH-Pfaffian. Access to these microscopical models of PH-Pfaffian will enable multiple future studies of its relevance for the 5/2 fractional quantum Hall effect. Diagonalizing the Hamiltonians for up to 20 electrons we find that the finite-size effects improve; remain however present consistent with a gap much smaller than the one for anti-Pfaffian. The available data neither asserts nor excludes the possibility that the ground state of the approximate PH-Pfaffian-generating Hamiltonian is gapped. Larger system sizes are more consistent with a valid FQHE state by some measures. Access to several higher sizes of the model wavefunction and exact learned Hamiltonian eigenstates would be needed to gain certainty while we do not expect more than one additional system size to become accessible in the near future due to the computational complexity of the problem. Approximate methods might therefore be worth considering.

There are several interesting directions for the future investigation. Since there is no preferred way of constructing a PH-Pfaffian it would be reasonable to study other implementations than used in this work \cite{Feldman2016} and all prior literature. The presented Hamiltonian is a simplest 4-parameter model while adjusting further pseudopotentials would improve the approximation precision. An important open question is whether the required deformation of the Coulomb interaction may be obtained within some effective realistic model? For example, the pseudopotentials that perturbatively account for the Landau level mixing and finite width \cite{Simon-Rezayi-2013,Peterson-Nayak-2013,Sodemann-2013} to lowest order do not to our knowledge produce the suitable two-body corrections. It has been argued that 3-body pseudopotentials may be required to stabilize PH-Pfaffian \cite{PhysRevB.98.115107,PhysRevB.100.195303,Milovanovic2020MPLB} and it would be interesting to include 3-body and higher-order pseudopotentials into the variational Hamiltonian ansatz. 
The 2-body Hamiltonians presented here will be a valuable starting point for such a study.

It is also possible to modify the search/optimisation criteria that could lead to a PH-Pfaffian approximation with smaller overlaps but larger gaps or include new terms that would enforce the expected topological properties of PH-Pfaffian.

In case one is able to find a 3- plus 2-body Hamiltonian (breaking PH-symmetry) of which the PH-symmetric model wavefunction is an exact eigenstate it may also be possible to further deform this Hamiltonian \cite{pakrouski2020GroupInvariantScars,pakrouski2021group} such that the model wavefunction would not thermalize with the rest of the Hilbert space and become an example of a many-body scar state, the phenomenon also known as weak ergodicity breaking \cite{Serbyn:2020wys,Moudgalya:2021xlu}.

\emph{Acknowledgements.} I gratefully acknowledge many useful discussions and the collaboration with E. Rezayi and F. D. M. Haldane during which the wavefunction used in this work was constructed. I also thank A. Balram for valuable discussions and comments on a version of this manuscript.
This work was supported by DOE grant \protect{DE-SC0002140} and by the Swiss National Science Foundation through the Early Postdoc.Mobility grant P2EZP2$\_$172168.
The simulations presented in this article were performed on computational resources managed and supported by Princeton's Institute for Computational Science $\&$ Engineering and OIT Research Computing. At the initial stages this work was supported by a grant from the Swiss National Supercomputing Centre (CSCS) under project IDs s395 and s551.

\appendix

\renewcommand{\thefigure}{A\arabic{figure}}
\renewcommand{\thetable}{A\arabic{table}}

\section{Pseudopotentials}

Table \ref{tab:ppsUsed} lists the pseudopotentials defining the learned Hamiltonians (CV7 and MV3) along with the pseudopotential corresponding to the 2nd Landau level Coulomb interaction. Coulomb pseudopotentials in the lowest Landau level are given in Table \ref{tab:LLLpps} together with the Hamiltonian $H^{SLLtoCV7}(\alpha=0.7)$ - it is the closest to the SLL Coulomb Hamiltonian that reasonably approximates the PH Pfaffian by various measures.

\begin{table}
\caption{\label{tab:ppsUsed} Pseudopotentials corresponding to the reference Coulomb interaction in the 2nd Landau level for 20 electrons and for the two learned Hamiltonians CV7 and MV3.}
\begin{center}
\begin{tabular}{|c|c|c|c|c|c|}
\hline
	 	& 	SLL Coulomb 			& 	CV7						& 	MV3\\
\hline
$V_1$	&	1					&	1						&	1\\
\hline
$V_3$	&	0.773278825612203 	&	0.694456627311176			&	0.433617799341989\\
\hline
$V_5$	&	0.576859542105588		&	0.665960300533016			&	0.370884676389928\\
\hline
$V_7$	&	0.487302680505104		&	0.448785272954577			&	0\\
\hline
$V_9$	& 	0.433005097996708		&	0.52955224410569			&	0.164743667925535\\
\hline
$V_{11}$	&	0.395897298672305		&	0.395897298672305			&	0\\
\hline
$V_{13}$	&	0.368758810329097		&	0.368758810329097			&	0\\
\hline
$V_{15}$	& 	0.348037679627816		&	0.348037679627816			&	0\\
\hline
$V_{17}$	& 	0.331753451882209		&	0.331753451882209			&	0\\
\hline
$V_{19}$	& 	0.318704133349498		&	0.318704133349498			&	0\\
\hline
$V_{21}$	& 	0.308114220905112		&	0.308114220905112			&	0\\
\hline
$V_{23}$	& 	0.299460095621806		&	0.299460095621806			&	0\\
\hline
$V_{25}$	& 	0.292375954106269		&	0.292375954106269			&	0\\
\hline
$V_{27}$	& 	0.286599807671959		&	0.286599807671959			&	0\\
\hline
$V_{29}$	& 	0.281940903418088		&	0.281940903418088			&	0\\
\hline
$V_{31}$	& 	0.278259296382938		&	0.278259296382938			&	0\\
\hline
$V_{33}$	& 	0.275452674381172		&	0.275452674381172			&	0\\
\hline
$V_{35}$	& 	0.273447719307893		&	0.273447719307893			&	0\\
\hline
$V_{37}$	& 	0.272194443356685		&	0.272194443356685			&	0\\
\hline
$V_{39}$	& 	0.271662583897594		&	0.271662583897594			&	0\\
\hline
\end{tabular}
\end{center}
\end{table}

\begin{table}
\caption{\label{tab:LLLpps}  Pseudopotentials corresponding to the lowest Landau level Coulomb interaction and the closest to the SLL Coulomb Hamiltonian that still approximates PH-Pfaffian}
\begin{center}
\begin{tabular}{|c|c|c|c|c|c|}
\hline
	 	& 	LLL Coulomb 			& 	$H^{SLLtoCV7}(\alpha=0.7)$\\
\hline
$V_1$	&	1					&	    1 \\
\hline
$V_3$	&	0.634091267132277		&	0.71934784782728967 \\
\hline
$V_5$	&	0.506828037378007		&	0.63782321892435445\\
\hline
$V_7$	&	0.438070153616146		&	0.46094866481263819\\
\hline
$V_9$	& 	0.394056175977230		&	0.49906367165022203\\
\hline
$V_{11}$	&	0.363219697006843		&	0.39589729867230494\\
\hline
$V_{13}$	&	0.340394925412771		&	 0.36875881032909702\\
\hline
$V_{15}$	& 	0.322888152268689		&	0.34803767962781601\\
\hline
$V_{17}$	& 	0.309143471292678		&	0.33175345188220901\\
\hline
$V_{19}$	& 	0.298194956815023		&	0.31870413334949799\\
\hline
$V_{21}$	& 	0.289410570149645		&	0.30811422090511198\\
\hline
$V_{23}$	& 	0.282360535479264		&	0.29946009562180598\\
\hline
$V_{25}$	& 	0.276744644896987		&	0.29237595410626899\\
\hline
$V_{27}$	& 	0.272349828383384		&	0.28659980767195897\\
\hline
$V_{29}$	& 	0.269024340615875		&	0.28194090341808797\\
\hline
$V_{31}$	& 	0.266661617786382		&	0.27825929638293800\\
\hline
$V_{33}$	& 	0.265190083071634		&	0.27545267438117199\\
\hline
$V_{35}$	& 	0.264566839059207		&  	0.27344771930789302\\
\hline
\end{tabular}
\end{center}
\end{table}

\section{Similarities between MV3 and CV7}

\begin{figure}[t]
    \centering
   \includegraphics[width=\columnwidth]{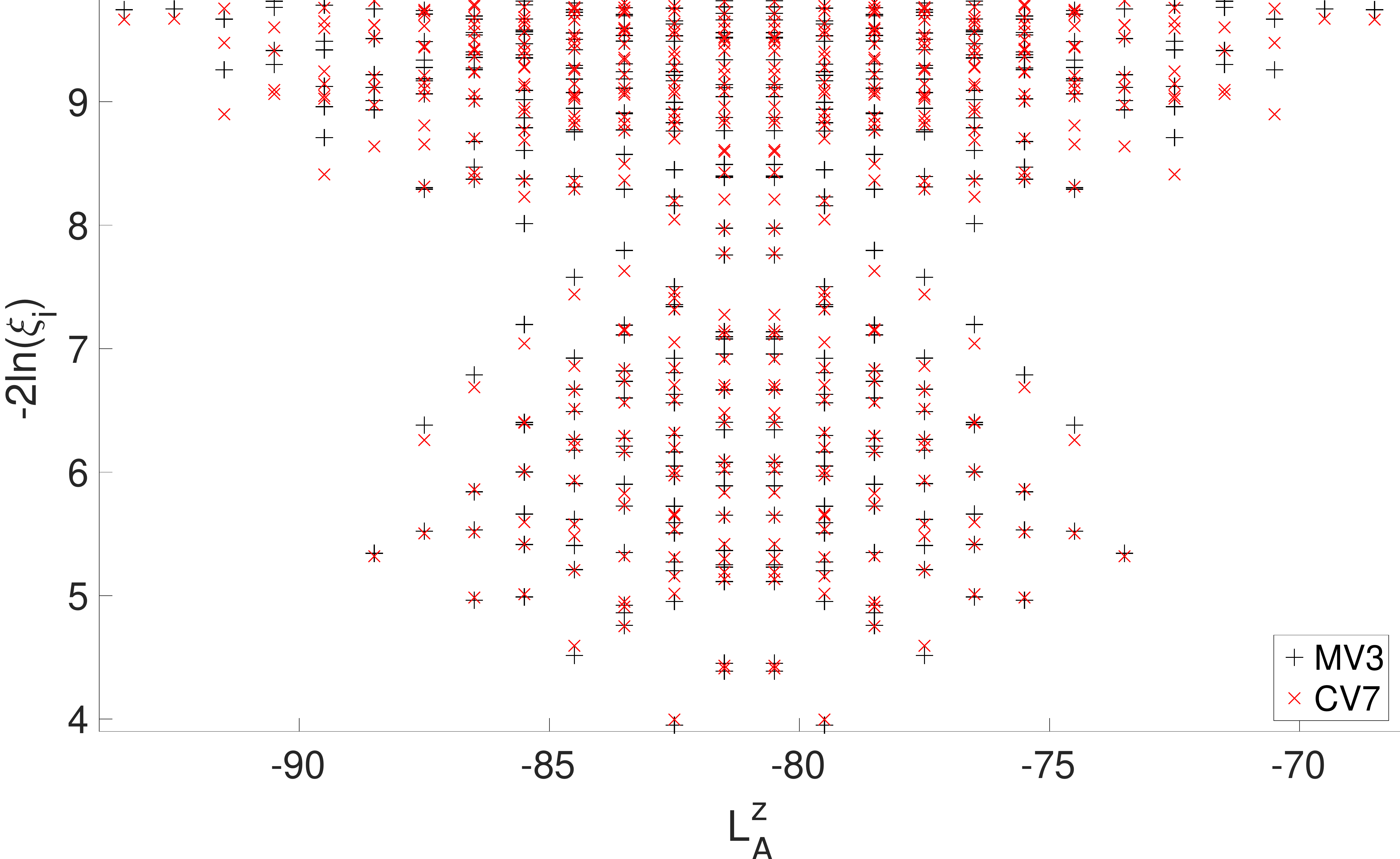}
    \includegraphics[width=\columnwidth]{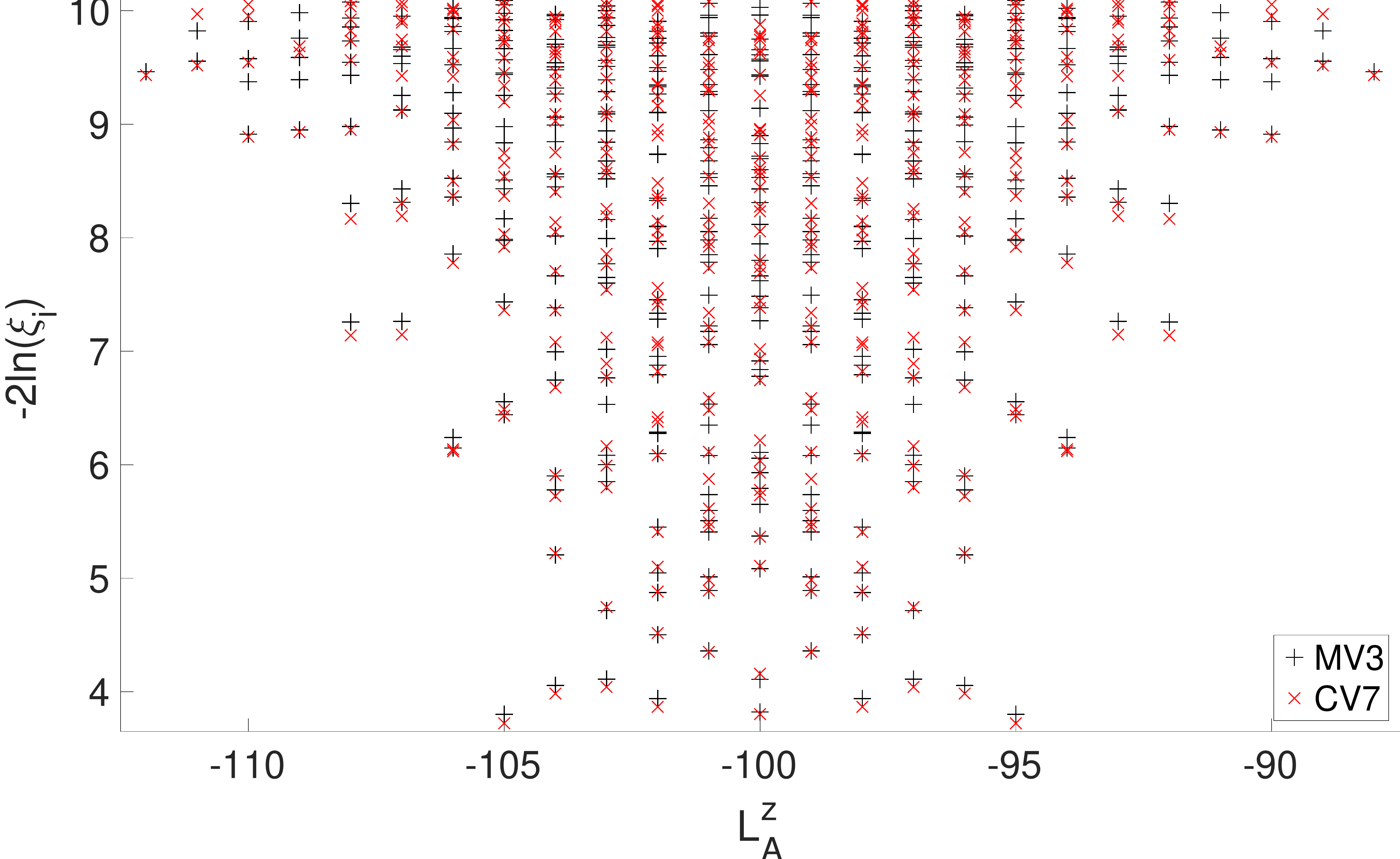}\\
    \caption{\label{fig:EntSpectrumCV7vsMV3}
    Entanglement spectrum for 18 (Top) and 20 (Bottom) electrons in the ground states of the two learned Hamiltonians.}
\end{figure}

\begin{figure}
	\centering
	\includegraphics[width=0.45\columnwidth]{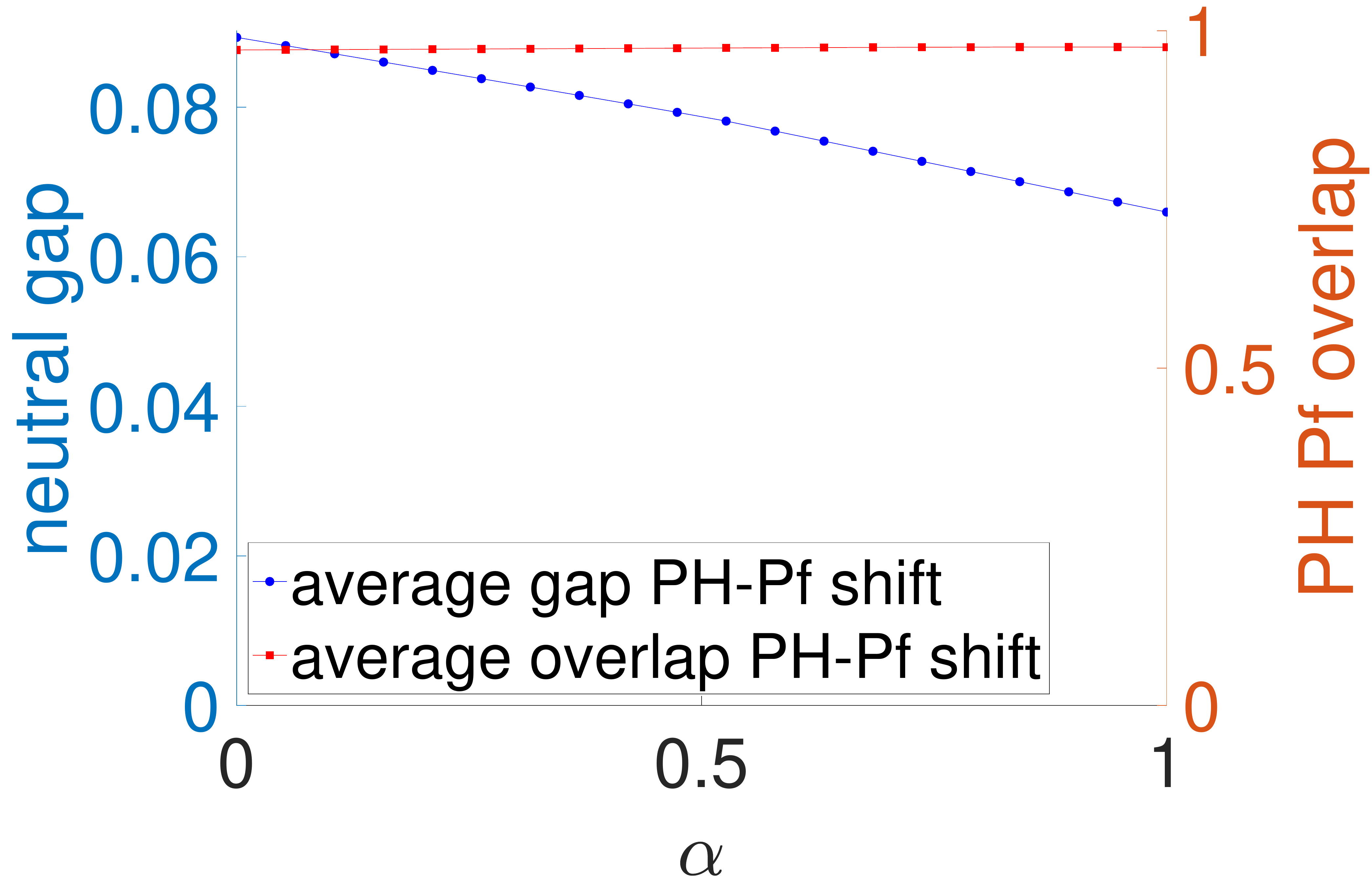}
	\includegraphics[width=0.45\columnwidth]{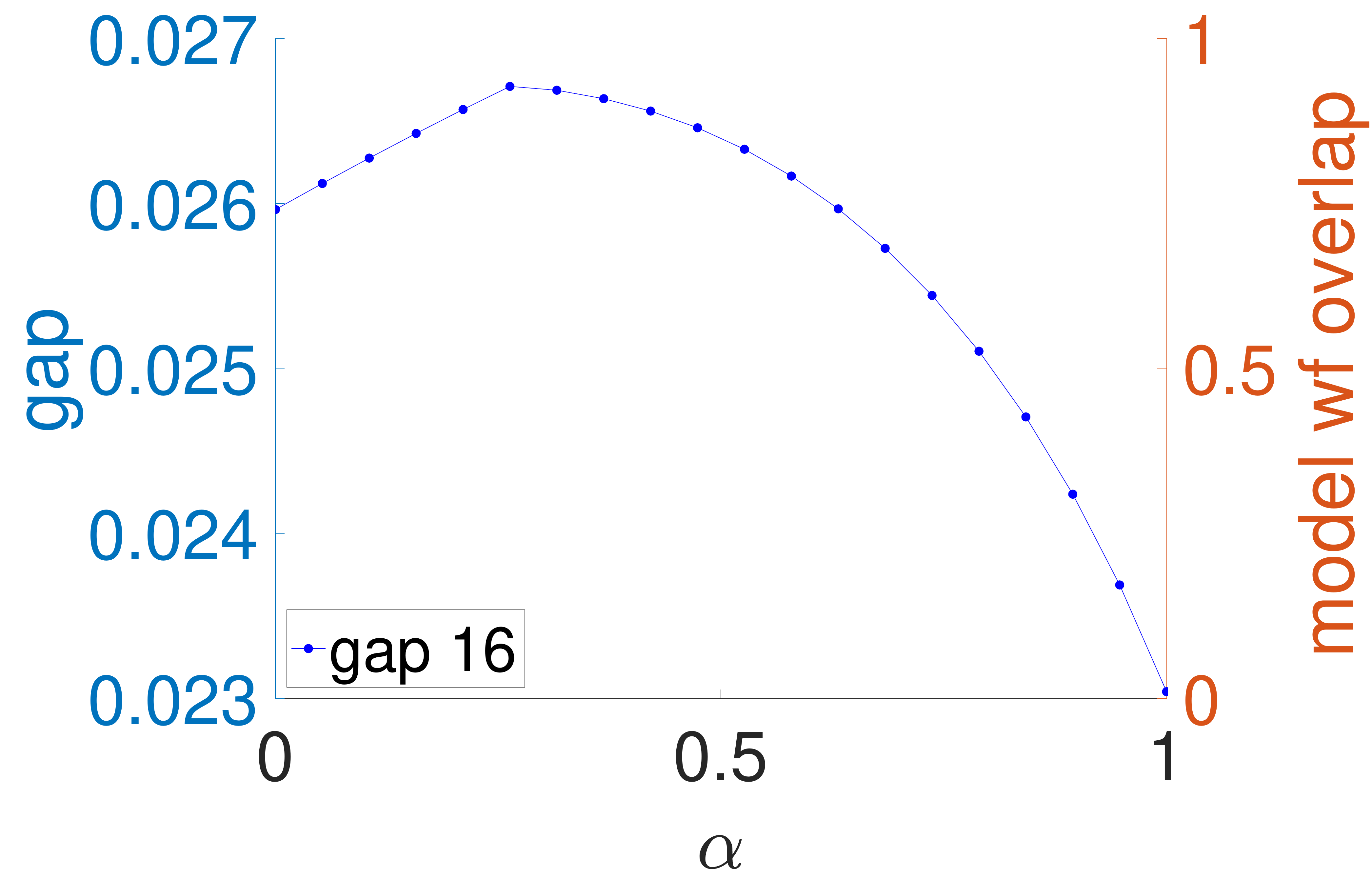}\\
	
	\includegraphics[width=0.45\columnwidth]{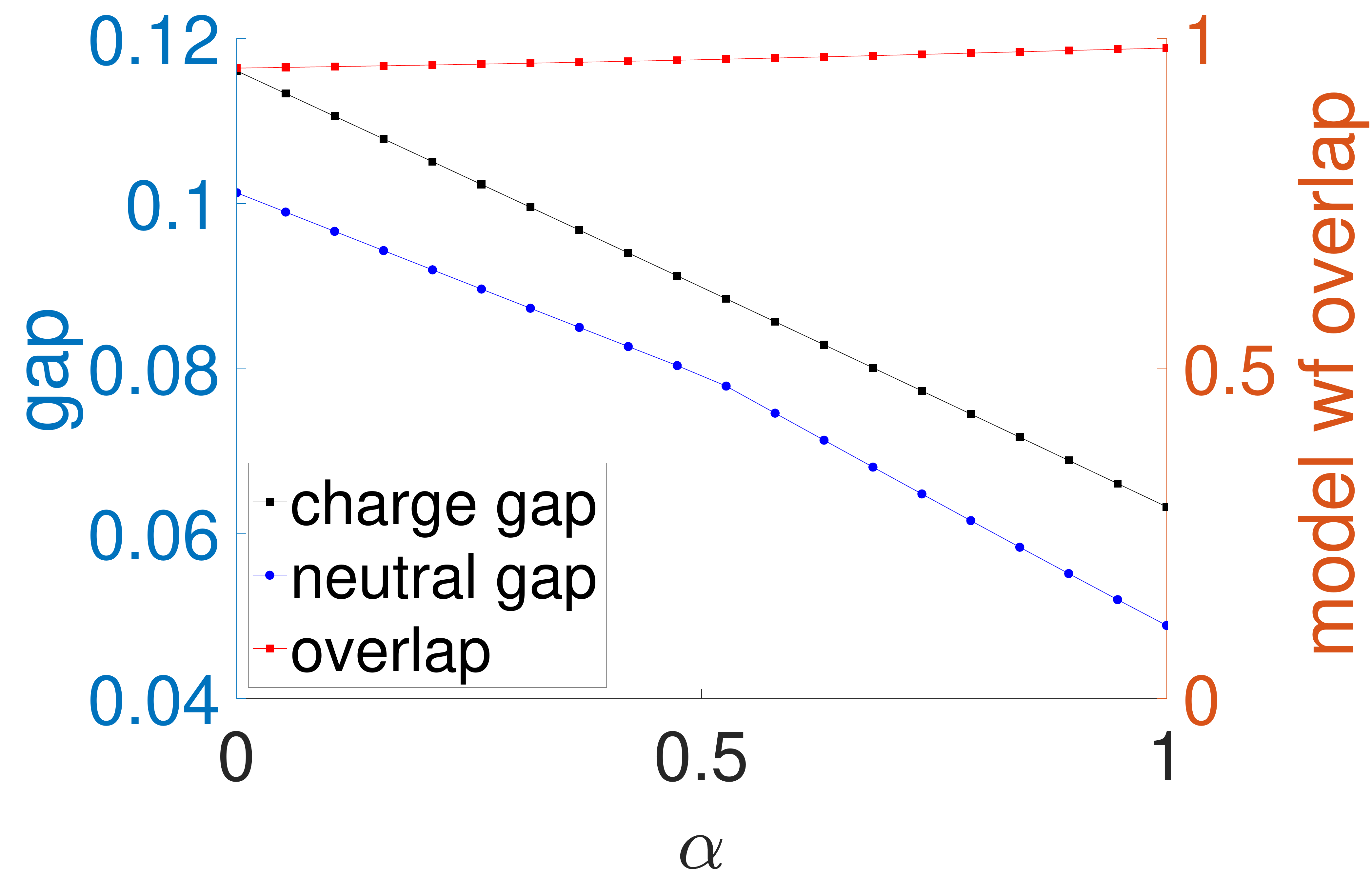}
	\includegraphics[width=0.45\columnwidth]{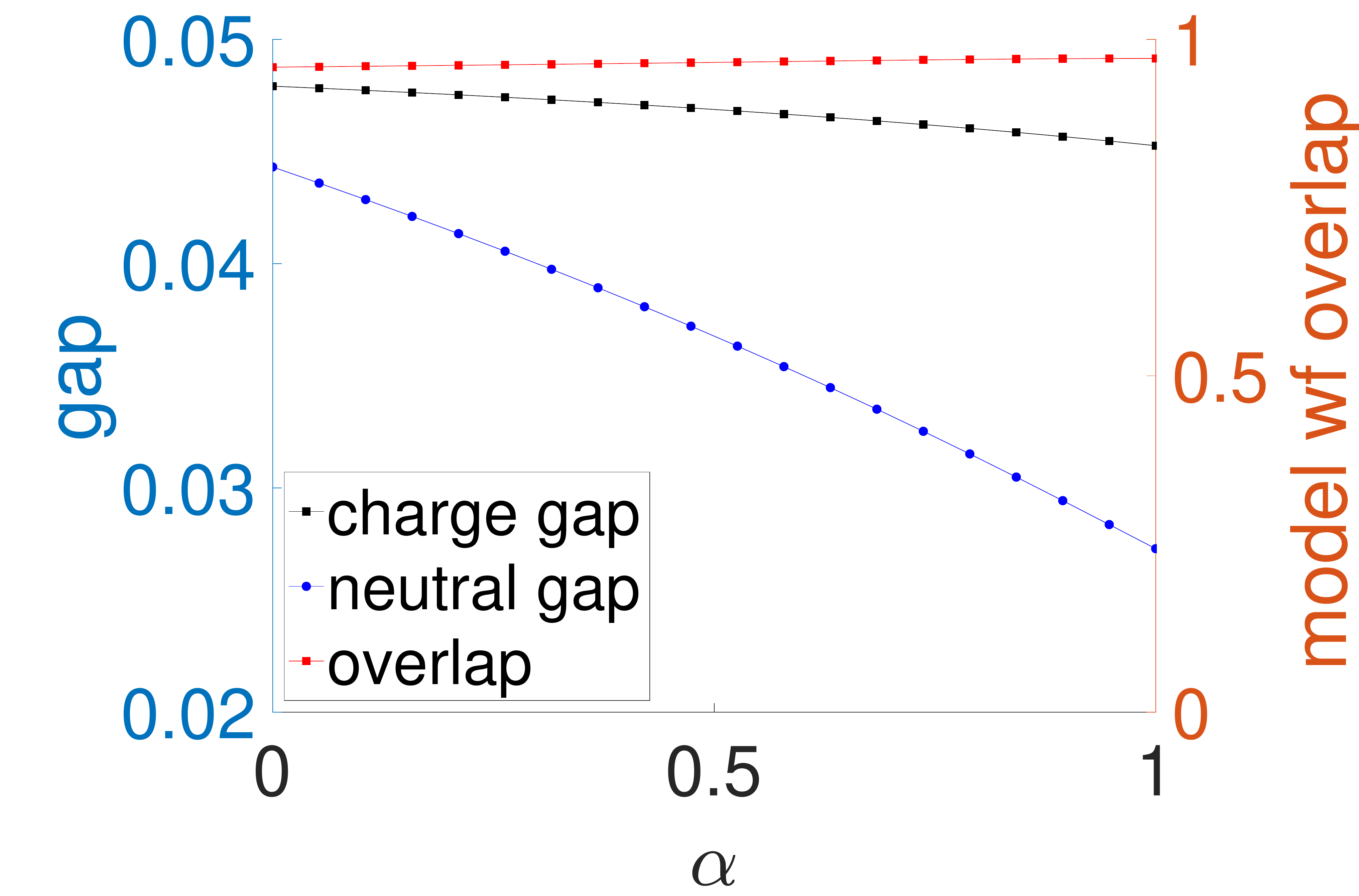}
	
	 \caption{\label{fig:interpolateMV3toCV7}%
    Interpolation between the two learned Hamiltonians MV3 ($\alpha=0$) and CV7 ($\alpha=1$). Shown are PH-Pfaffian overlap with the ground state and the neutral gap for $N_e$=12 (Bottom Left), $N_e$=14 (Bottom Right), $N_e$=16 (Top Right) and averaged for $N_e$ 6 to 16  (Top Left). The charge gap is estimated as $\Delta_c = 0.5\left[E_0(N^{PHPf}_{\phi}+1) + E_0(N^{PHPf}_{\phi}-1)\right] - E_0(N^{PHPf}_{\phi})$ without any background corrections. 
}
\end{figure}

\begin{figure}
	\centering
	\includegraphics[width=\columnwidth]{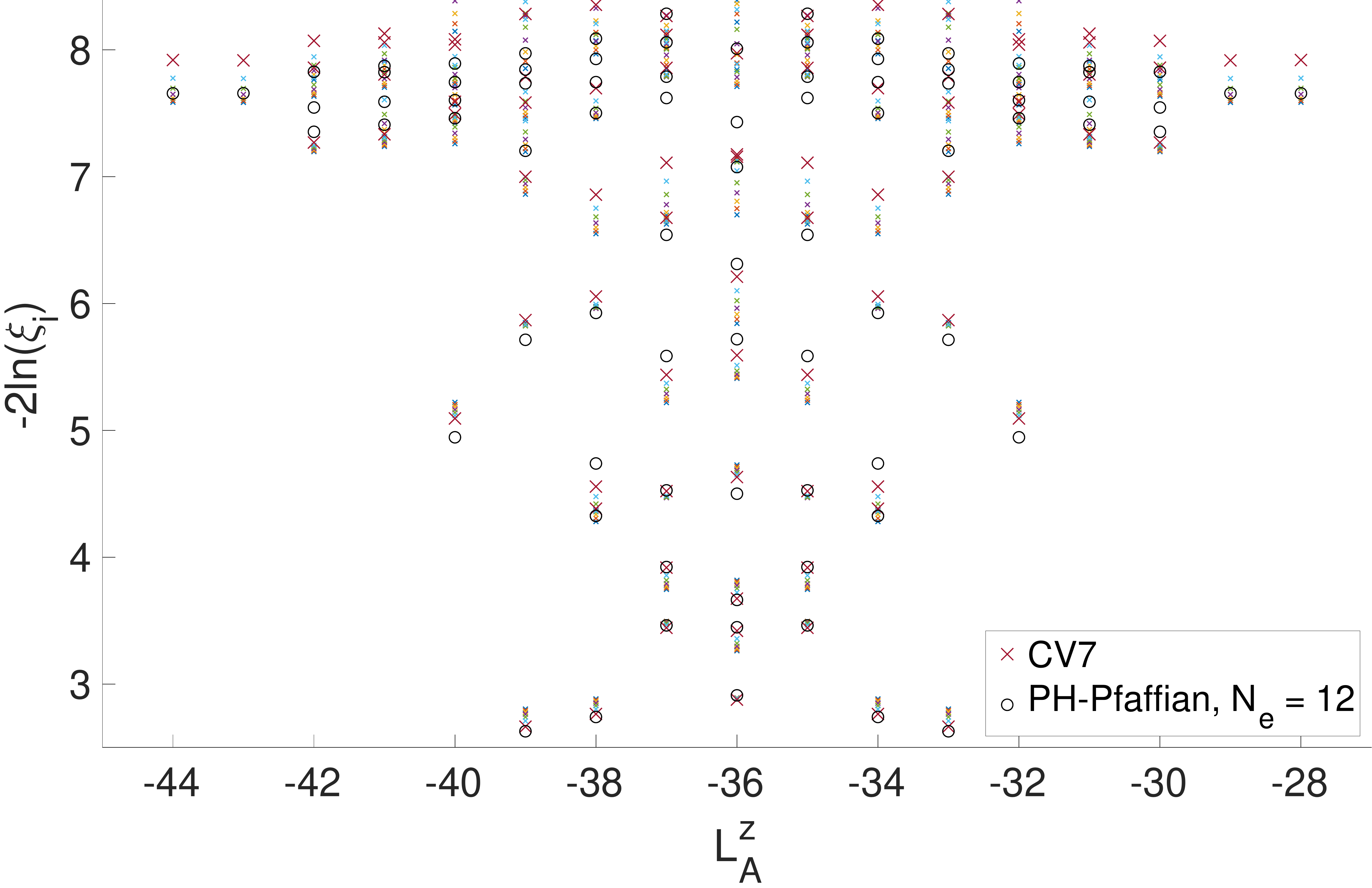}
	\includegraphics[width=\columnwidth]{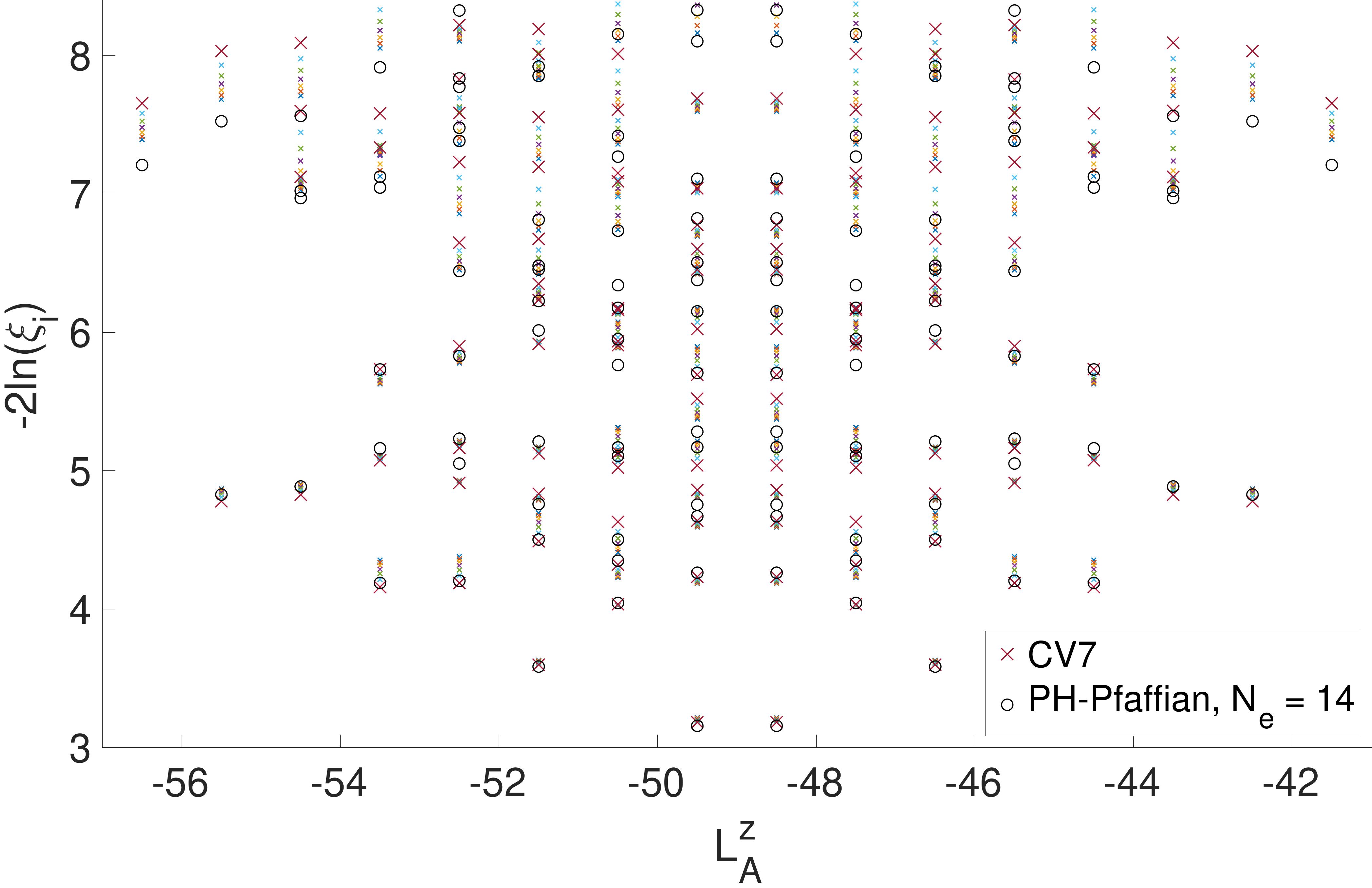}
	
	 \caption{\label{fig:entanglementMV3toCV7}
    Overlay of the lowest part of the entanglement spectra computed for every third step of MV3-to-CV7 interpolation plotted with small crosses. Large crosses indicate the data for CV7. Large circles indicate the reference data for the model PH-Pfaffian wavefunction. $N_e=12$ (Top Panel) and $N_e=14$ (Bottom Panel).
}
\end{figure}

The learned Hamiltonians MV3 and CV7 appear to belong to the same continuous region of Hamiltonians in the parameter space that approximate the PH Pfaffian. This is supported by the high overlaps between the corresponding ground states (Table \ref{tab:CV7vsMV3}) and similarities between their entanglement spectra (Fig. \ref{fig:EntSpectrumCV7vsMV3}). Furthermore the structure of the low-lying entanglement spectrum is preserved if we interpolate between MV3 and CV7 as shown in Fig. \ref{fig:entanglementMV3toCV7}. In the course of such interpolation the gap does not appear to close in the systems studied (Fig. \ref{fig:interpolateMV3toCV7}).

\begin{table}
\caption{\label{tab:CV7vsMV3} Overlaps between the ground states of the learned Hamiltonians CV7 and MV3.}
\begin{center}
\begin{tabular}{|c|c|c|c|c|c|}
\hline
	 					& 	CV7 and MV3 			\\
\hline
$\braket{\psi_{CV7}|\psi_{MV3}}(6)$	&	0.99998591277202543				\\
\hline
$\braket{\psi_{CV7}|\psi_{MV3}}(8)$	&	0.99510909086629048	\\
\hline
$\braket{\psi_{CV7}|\psi_{MV3}}(10)$	&	0.99786862142518831			\\
\hline
$\braket{\psi_{CV7}|\psi_{MV3}}(12)$	&	0.98280787588118113		\\
\hline
$\braket{\psi_{CV7}|\psi_{MV3}}(14)$	&	0.98474351735983609			\\
\hline
$\braket{\psi_{CV7}|\psi_{MV3}}(16)$	&	0.98701364088021704			\\
\hline
$\braket{\psi_{CV7}|\psi_{MV3}}(18)$	&	0.99005884851780076		\\
\hline
$\braket{\psi_{CV7}|\psi_{MV3}}(20)$	&	0.98352526730226702		\\
\hline

\end{tabular}
\end{center}
\end{table}

\section{Stability and structure factors}

Scaling of the structure factor and its form bear information on the stability of the underlying state.

\begin{figure}[t]
    \centering
    
    \includegraphics[width=\columnwidth]{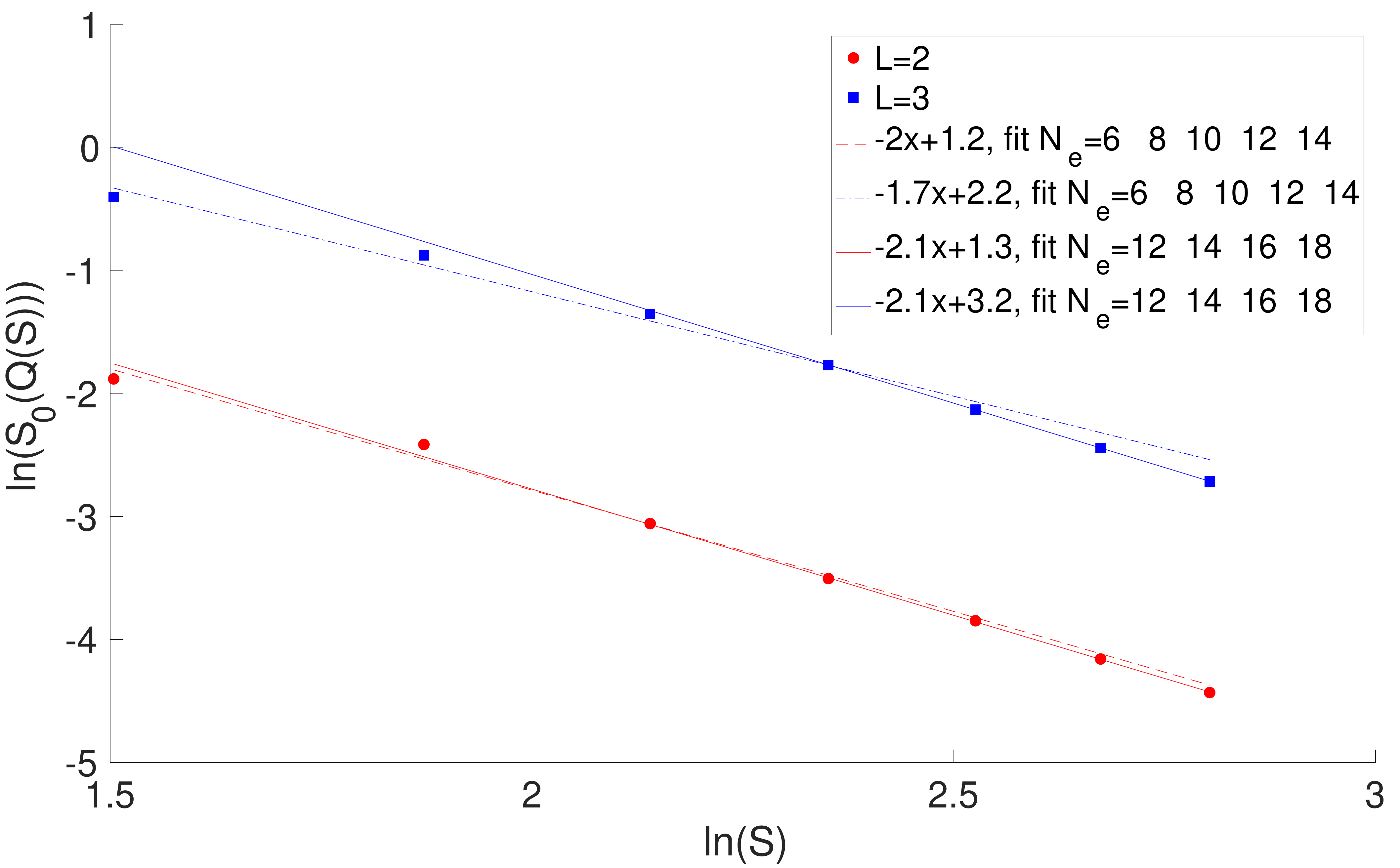}
    \caption{\label{fig:strFactPhPfVSAPF}
    Structure factor scaling analysis. The wavefunctions are model anti-Pfaffian states. Dashed lines fit the smaller and solid lines - the larger system sizes.
}
\end{figure}

Scaling analysis for the anti-Pfaffian model wavefunction is presented in Fig. \ref{fig:strFactPhPfVSAPF}. The state would appear gapless if only systems with less than 16 particles (as available for PH-Pfaffian) were considered.

\begin{figure}[t]
    \centering
    \includegraphics[width=0.49\columnwidth]{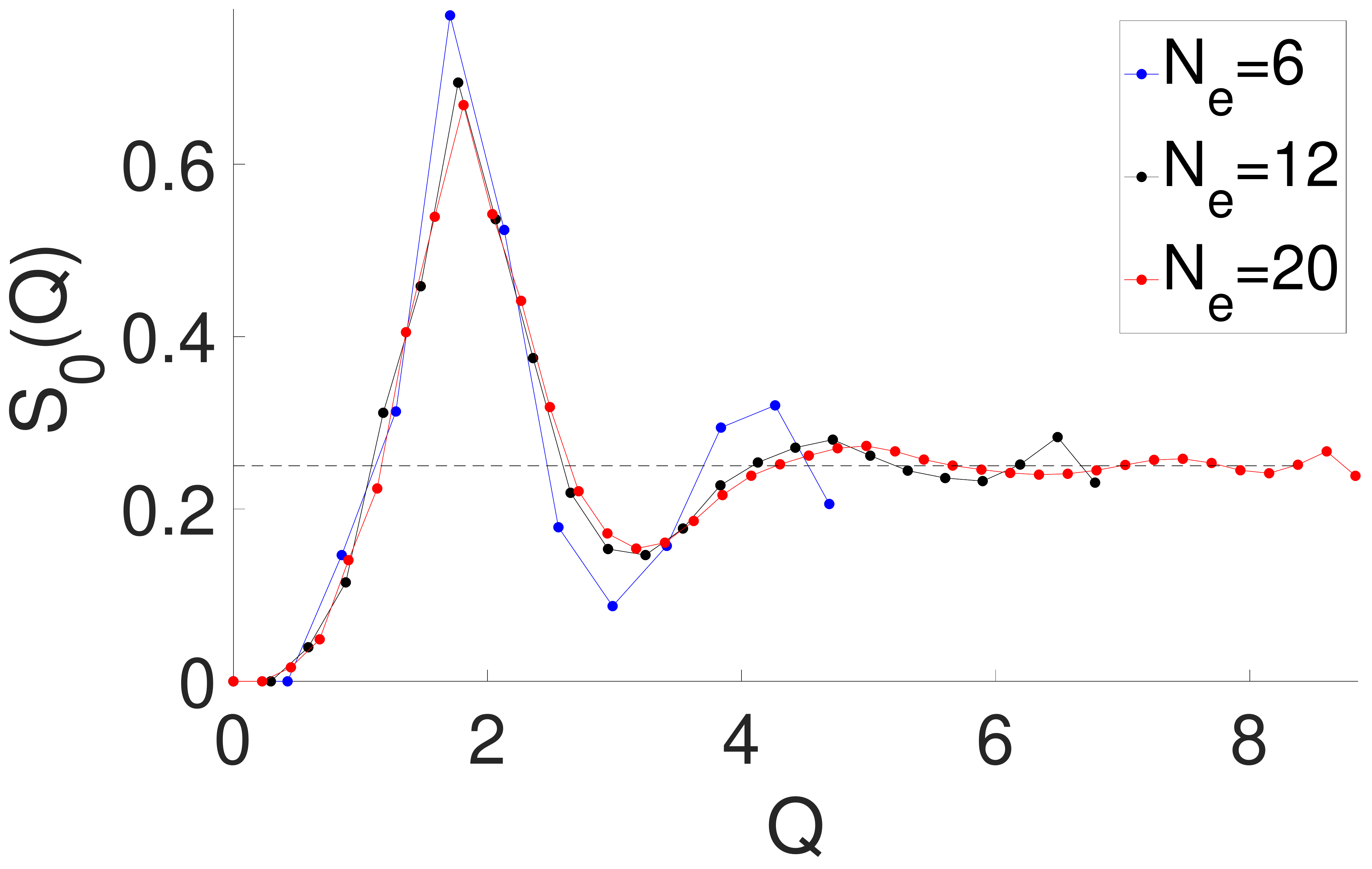}
    \includegraphics[width=0.49\columnwidth]{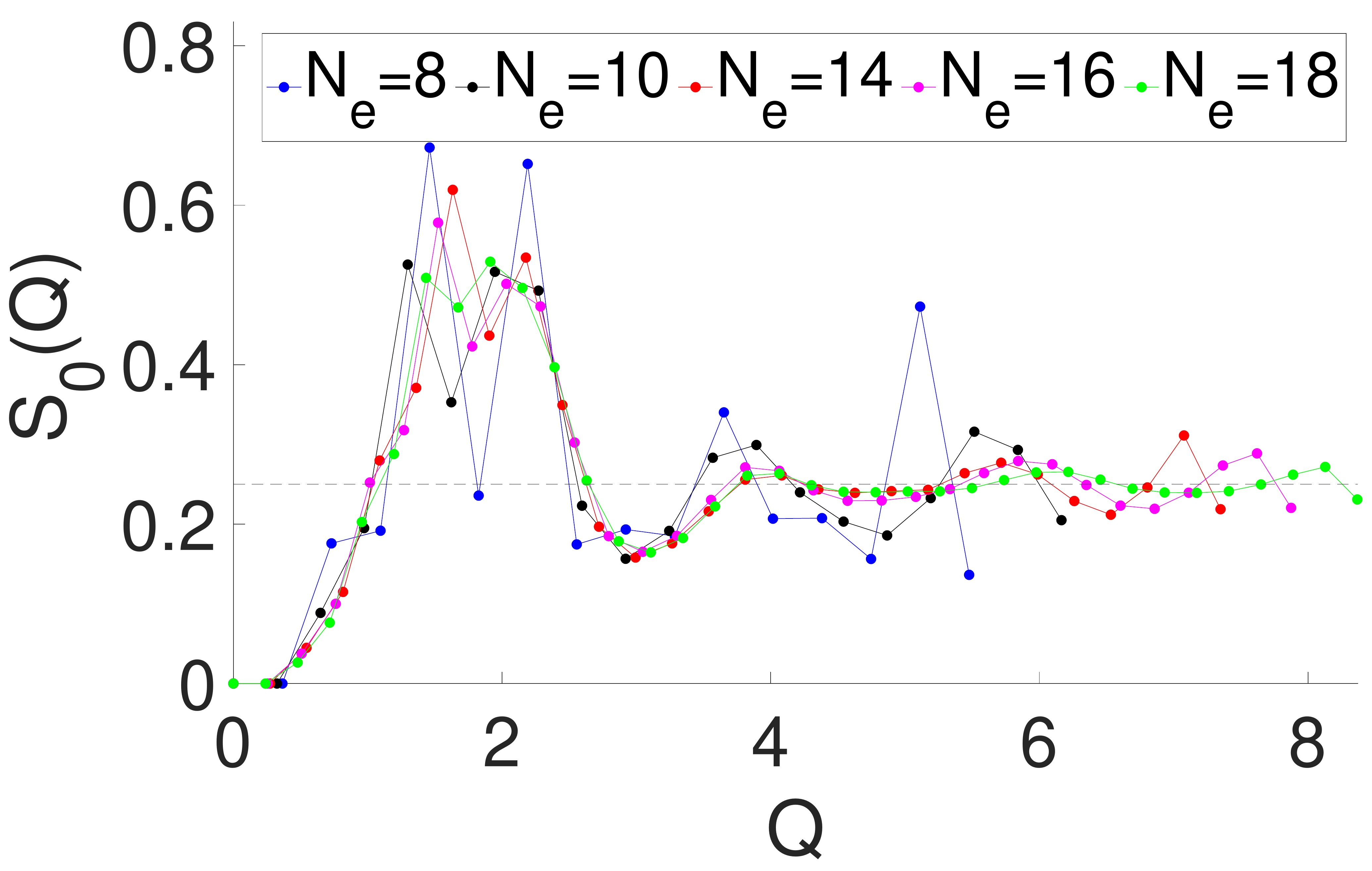}\\
        \includegraphics[width=0.49\columnwidth]{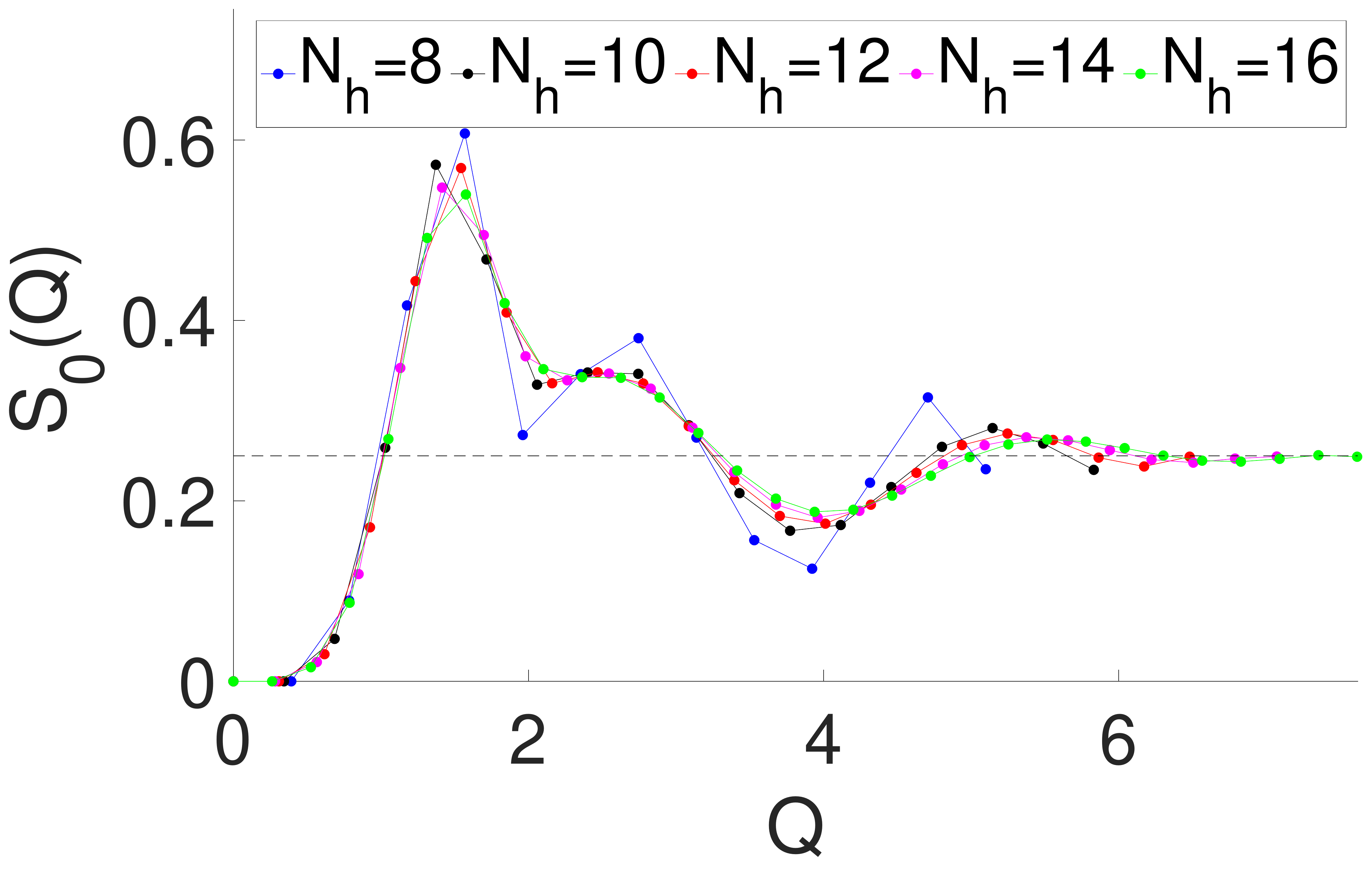}
        \includegraphics[width=0.49\columnwidth]{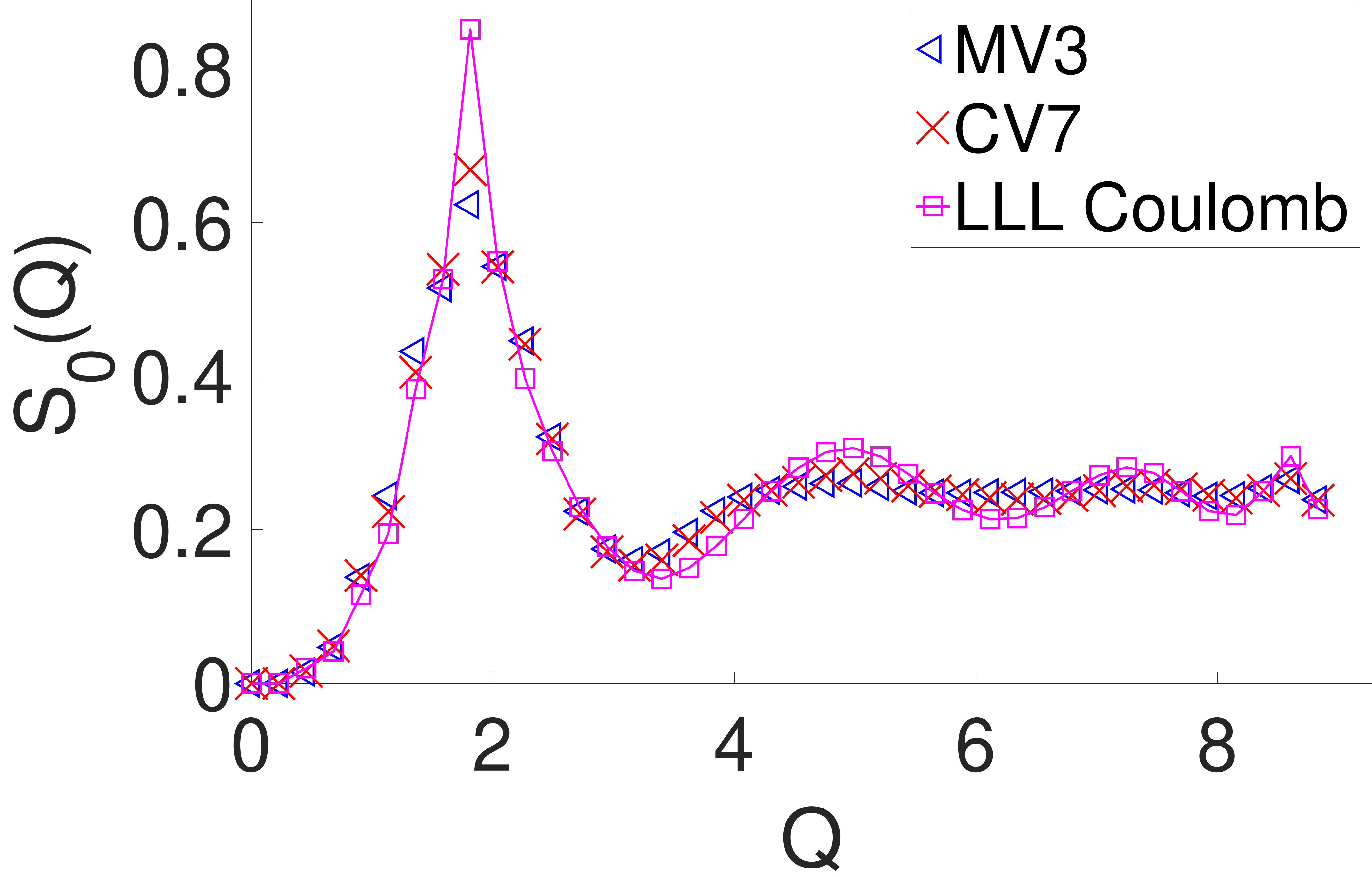}
    \caption{\label{fig:strFactsLearnedH}
    Top: Structure factor calculated for all available system sizes in the ground state of the learned Hamiltonian CV7. Left Panel: $N_e=6,12,20$ 
Right Panel: $N_e=8,10,14,16,18$. Bottom Left: Structure factor calculated for anti-Pfaffian model wavefunction. Bottom Right: Structure factor calculated for $N_e=20$ in the ground states of learned Hamiltonians and of the LLL Coulomb interaction.
}
\end{figure}

 In Fig. \ref{fig:strFactsLearnedH} we plot the structure factors for all available ground states of the learned Hamiltonian CV7 and for the model anti-Pfaffian wavefunction. 

Different topological states correspond to different "shifts" on the sphere that determines the number of fluxes corresponding to a certain $N_e$ in finite systems. Therefore one can extract information about the stability of the PH Pfaffian state (for which $2N_e=N_{\phi}$+1) under certain interaction by comparing it's ground state energy to the lowest energy at the nearby shifts where one flux quantum is added or subtracted. The necessary data is given in Table \ref{tab:sweepFlux} and confirms the stability of the PH-Pfaffian state for the Hamiltonians CV7 and MV3.

\begin{table}
\caption{\label{tab:sweepFlux} Learned Hamiltonians diagonalized at nearby shifts. Following system sizes are aliased: 6,12 with $\nu=2/5$; 12,24 with $\nu=3/7$; 16,31 with $\nu=4/9$; 12,22 with $\nu=4/9$; 18,38 with $\nu=3/7$; 20,40 with $\nu=4/9$. Extra stability of 16,31 for MV3 is likely because the state there corresponds to the LLL $\nu=4/9$ state. For all other sizes in case of MV3 and for all available sizes for CV7 the PH-Pfaffian shift corresponds to the lowest ground state energy w.r.t. the nearby shifts at $\pm$2 flux quanta.}
\begin{center}
\begin{tabular}{|c|c|c|c|c|c|c|}
\hline
	 $N_e$,$N_{\phi}$+1		& 	$E$, CV7  			&	L 			& 	$E$, MV3	&	L\\
\hline
12,22					&	37.75779031942652		&	0			&	17.56284812507863	&	0\\
\hline
12,23					&	37.0547351356873		&	2			&	16.65270506190967	&	2\\
\hline
12,24					&	36.29631579008422		&	0			&	15.62965756265555 &	0\\
\hline
12,25					&	35.66441319316203		&	2			&	14.83880981496597 &	2\\
\hline
12,26					&	35.0112559096271		&	2			&	14.02912633787132	&	2\\
\hline
\hline
14,26					&	49.16430588912498		&	2			&	20.88725047201897&	2\\
\hline
14,27					&	48.33780059455478		&	1			&	19.83464758433945&	1\\
\hline
14,28					&	47.51196100367208		&	0			&	18.8376266029838	&	0\\
\hline
14,29					&	46.77664603481952		&	3			&	17.93643762732407	&	3\\
\hline
14,30					&	46.07713591920228		&	4			&	17.12495015897343	&	4\\
\hline
\hline
16,30					&	61.65843181189761		&	4			&	24.11892235185186	&	4\\
\hline
16,31					&	60.71255531205486		&	0			&	22.96860300250802	&	0\\
\hline
16,32					&	59.88278337534194		&	0			&	22.03566015418639	&	0\\
\hline
16,33					&	59.0611893949843		&	4			&	21.07776538314662	&	4\\
\hline
16,34					&	58.28352775461818		&	0			&	20.19318847939453	&	0\\
\hline
\hline
18,34					&	75.22172551641899		&	0			&	27.29104448189041	&	0	\\
\hline
18,35					&	74.26146141886335		&	1			&	26.22629910173576	&	1\\
\hline
18,36					&	73.35589800221878		&	0			&	25.19975494811957	&	0	\\
\hline
18,37					&	72.48241065038177		&	3			&	24.25618648940761	&	3	\\
\hline
18,38					&	71.5903121524222		&	0			&	23.25195072313593	&	0\\
\hline
\hline

\end{tabular}
\end{center}
\end{table}

\section{Interpolation between Coulomb and learned Hamiltonians}

\begin{figure}
	\centering
	\includegraphics[width=0.49\columnwidth]{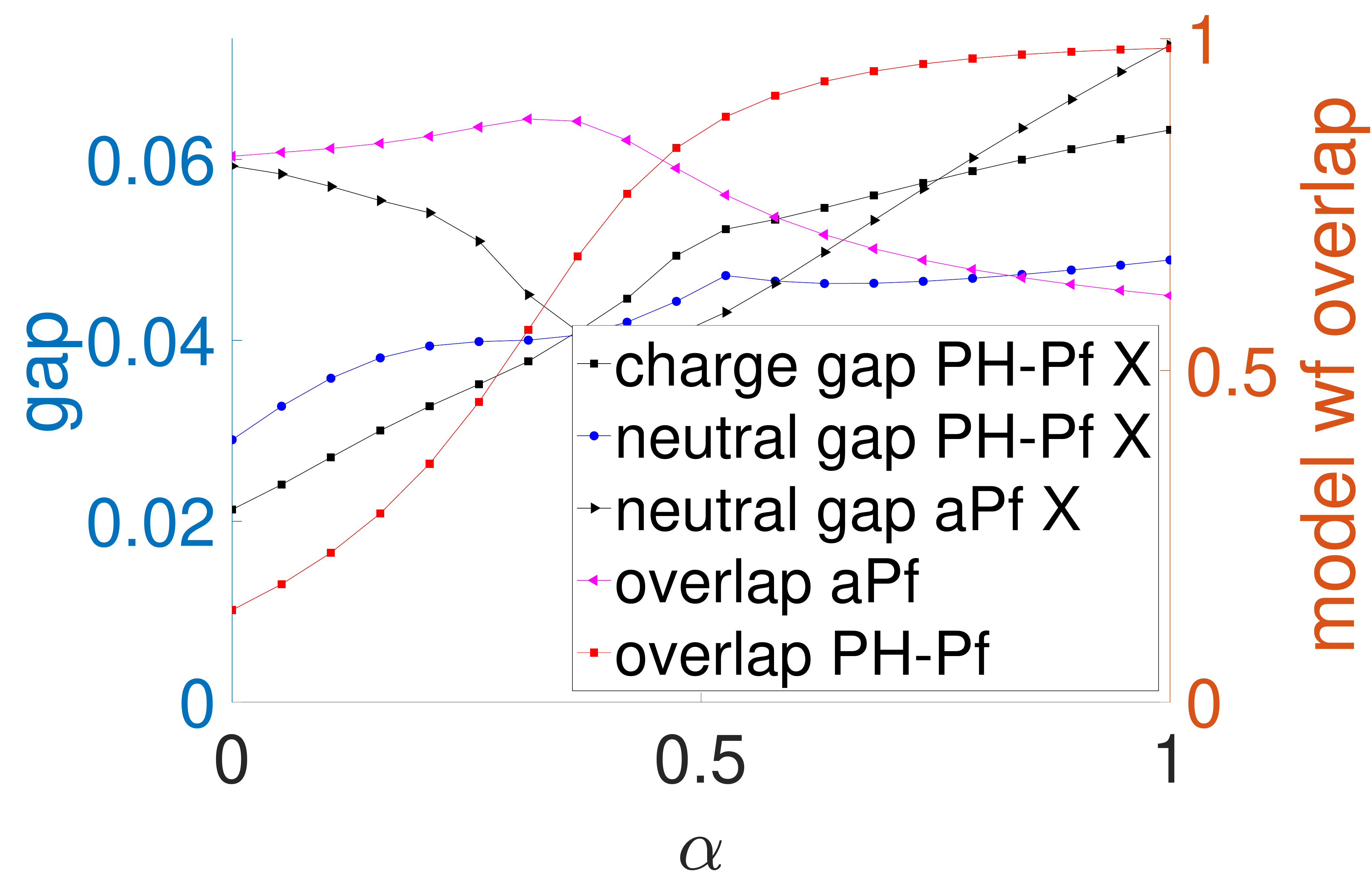}
	\includegraphics[width=0.49\columnwidth]{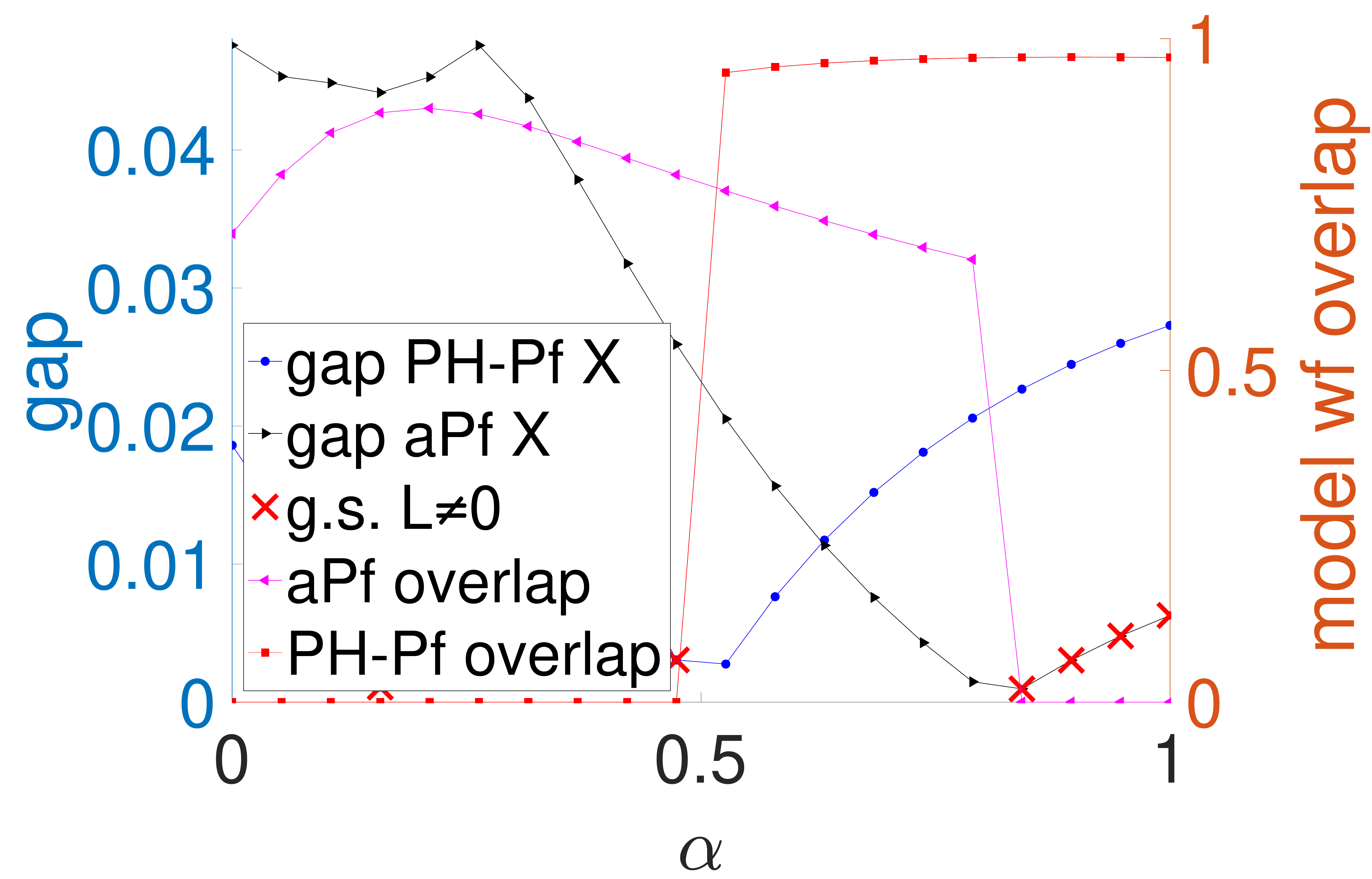}\\
	
	\includegraphics[width=0.49\columnwidth]{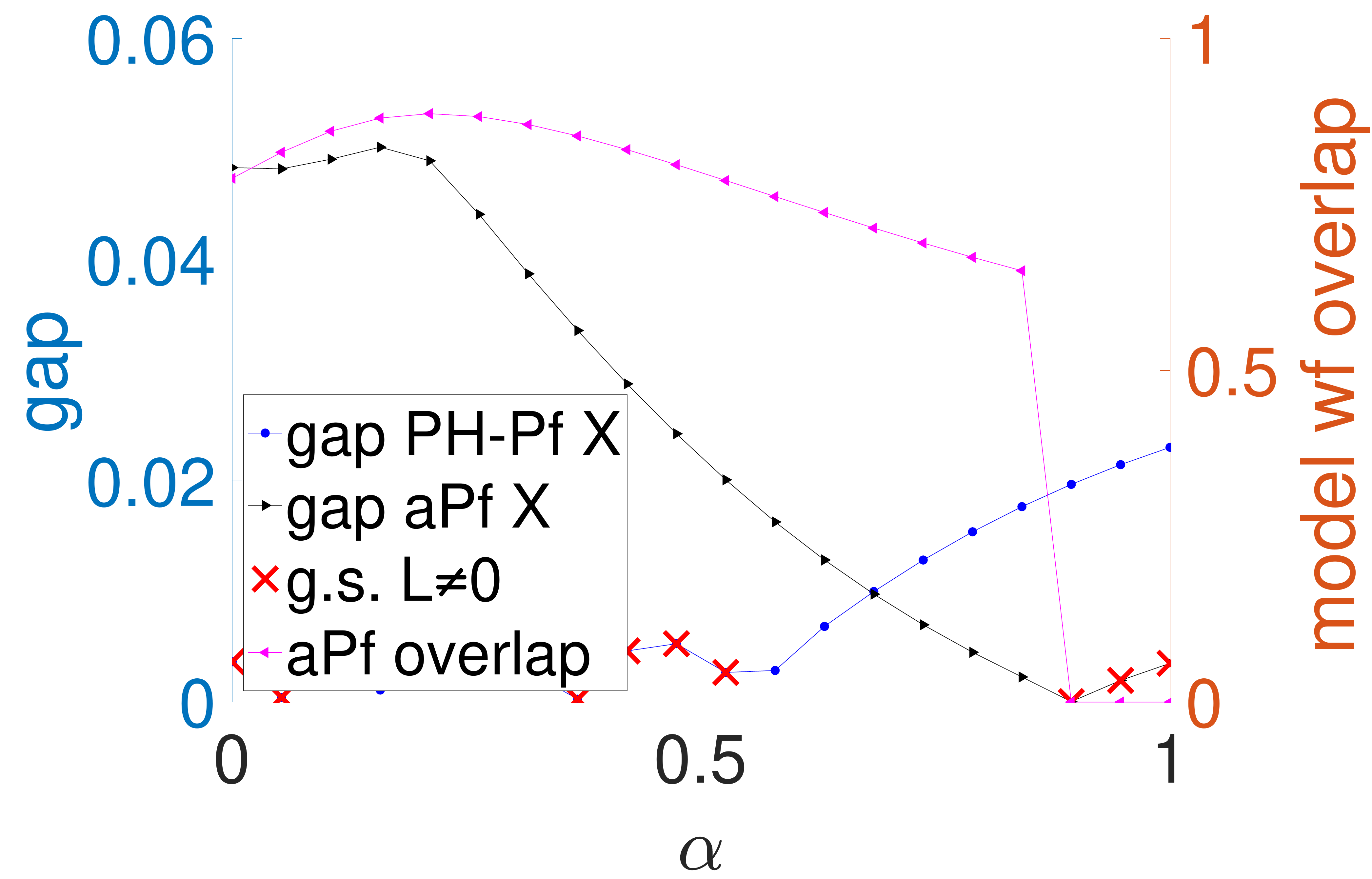}
	\includegraphics[width=0.49\columnwidth]{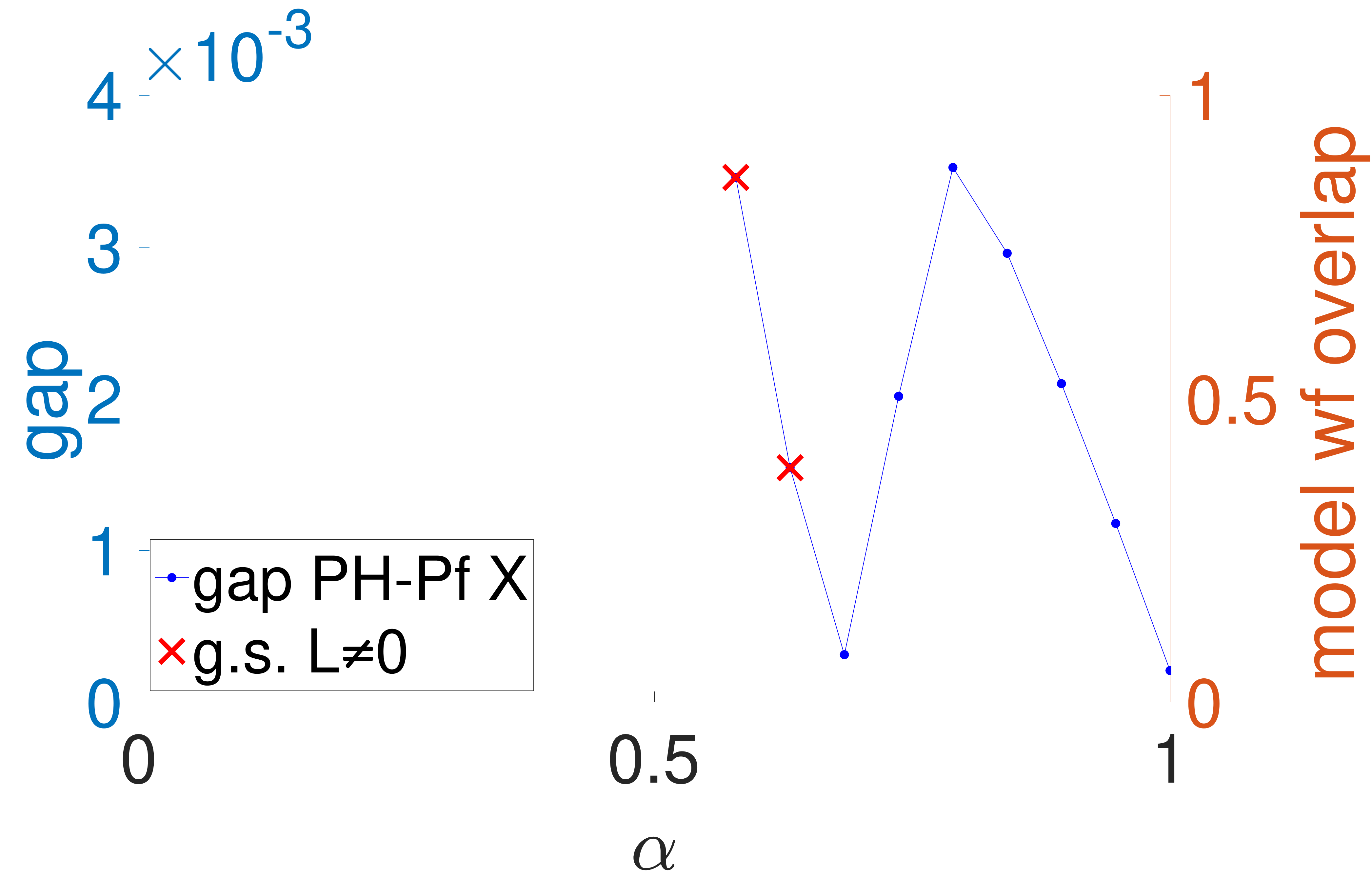}
	
	\caption{\label{fig:interpolateToCV7SM}
	Interpolation between the Coulomb interaction in second Landau level ($\alpha=0$) and the learned Hamiltonian CV7 ($\alpha=1$). Shown are model wavefunction overlap with the ground state and the neutral gap for $N_e$=12 (Top Left), $N_e$=14 (Top Right), $N_e$=16 (Bottom Left) and $N_e$=18 (Bottom Right). Charge gap is shown additionally for 12 electrons. Red crosses indicate the datapoints where the total angular momentum $L\ne 0$ for the lowest energy eigenstate. For 12 electrons neither neutral nor charge gap appear to close.
}
\end{figure}

\begin{figure}
	\centering
	\includegraphics[width=0.49\columnwidth]{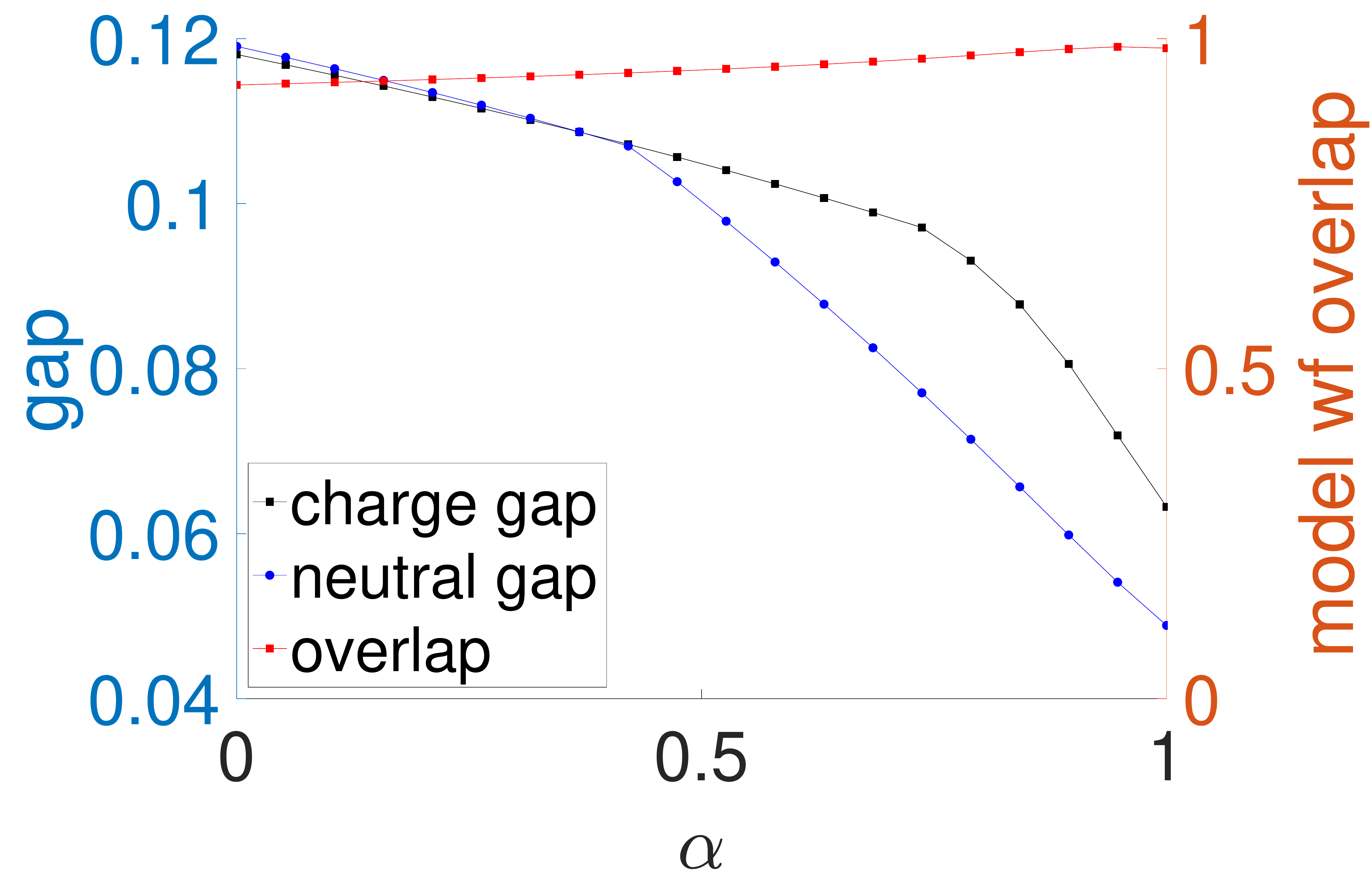}
	\includegraphics[width=0.49\columnwidth]{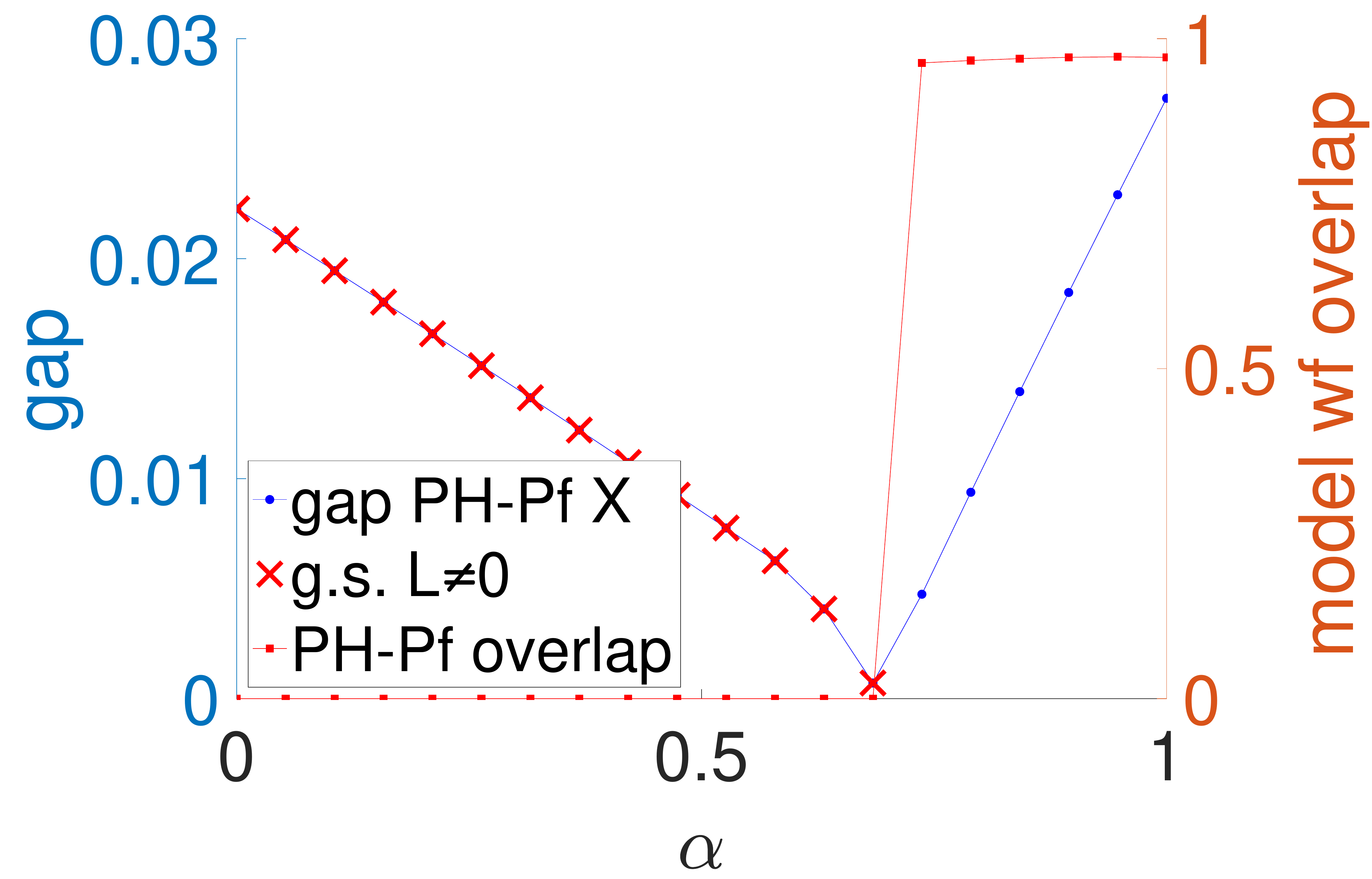}
	
	 \caption{\label{fig:interpolateLLLtoCV7}%
    Interpolation between the Coulomb interaction in the lowest Landau level ($\alpha=0$) and the learned Hamiltonian CV7 ($\alpha=1$). Shown are model wavefunction overlap with the ground state and the neutral gap for $N_e$=12 (Left Panel), $N_e$=14 (Right Panel). Red crosses indicate the datapoints where the total angular momentum $L\ne 0$ for the lowest energy eigenstate. For 12 electrons the lowest excitation has $L=2$ for the learned Hamiltonian and $L=6$ for LLL Coulomb. Change of the lowest excitation coincides with the point where the neutral gap changes the slope in the left panel. }
\end{figure}

In Figs. \ref{fig:interpolateToCV7SM} and \ref{fig:interpolateLLLtoCV7} we show the model wavefunction overlap and gap as the Hamiltonian interpolates between Coulomb interaction in the 2nd and lowest Landau levels and the learned CV7 interaction. Besides the neutral gap (difference between the lowest eigenvalues), for some systems we also show the charge gap estimated as $\Delta_c = 0.5\left[E_0(N^{PHPf}_{\phi}+1) + E_0(N^{PHPf}_{\phi}-1)\right] - E_0(N^{PHPf}_{\phi})$ without any background corrections.

\section{Optimization parameters}

The following optimisation parameters were used to obtain CV7.
Step scheme: Hestenes; overlap weight: 1.5; (neutral) gap weight: 0.725; weight for the deviation of the variational parameters from the reference point (SLL Coulomb pseudopotentials): 0.0001; weight for the term enforcing uniform $L=0$ ground state: 50000; weight for energy variance: 4.   

\section{Spectra and composite fermions interpretation \label{sec:AppendixCFInterpret}}

\begin{figure}[t]
    \centering
    \includegraphics[width=0.49\columnwidth]{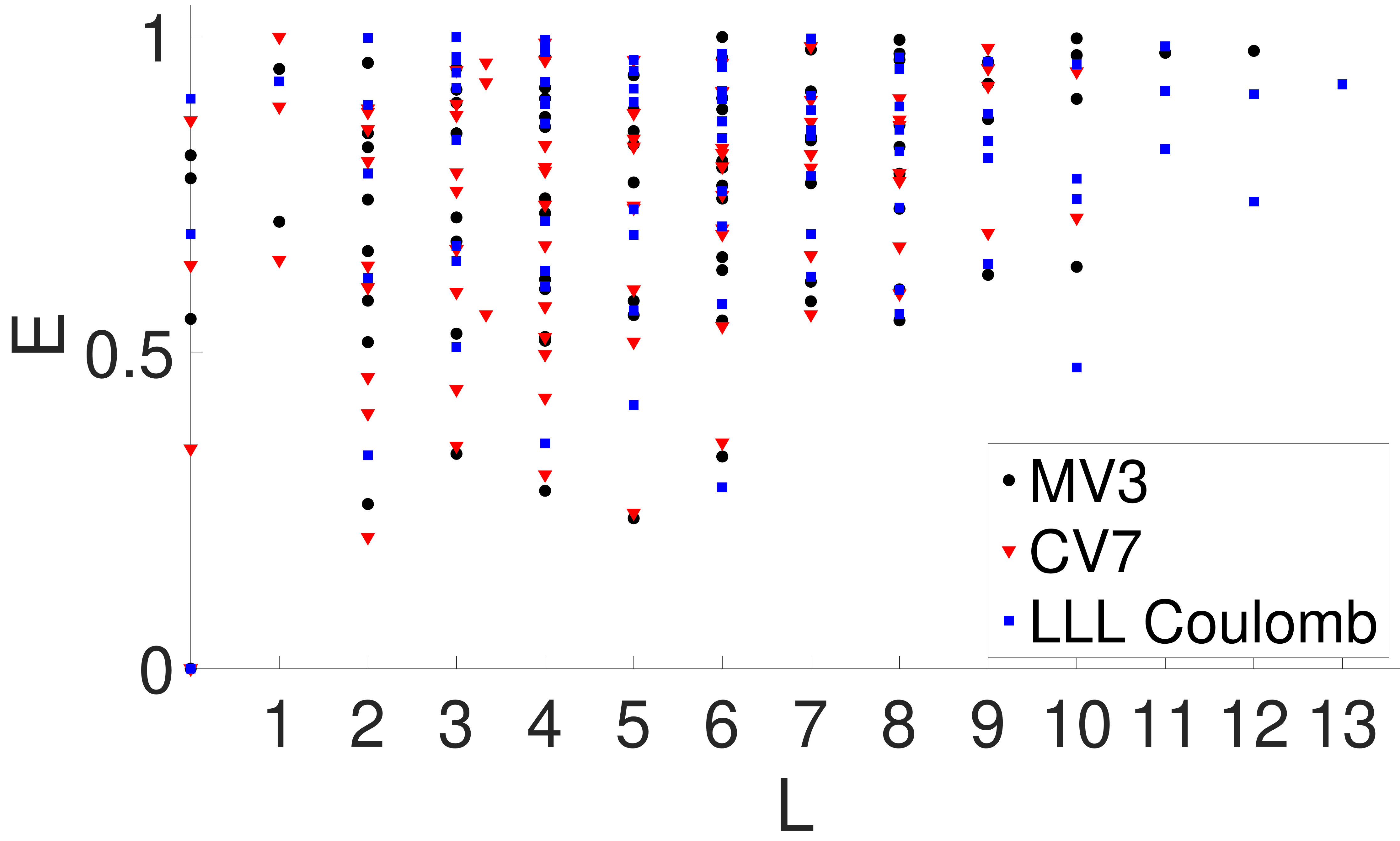}
    \includegraphics[width=0.49\columnwidth]{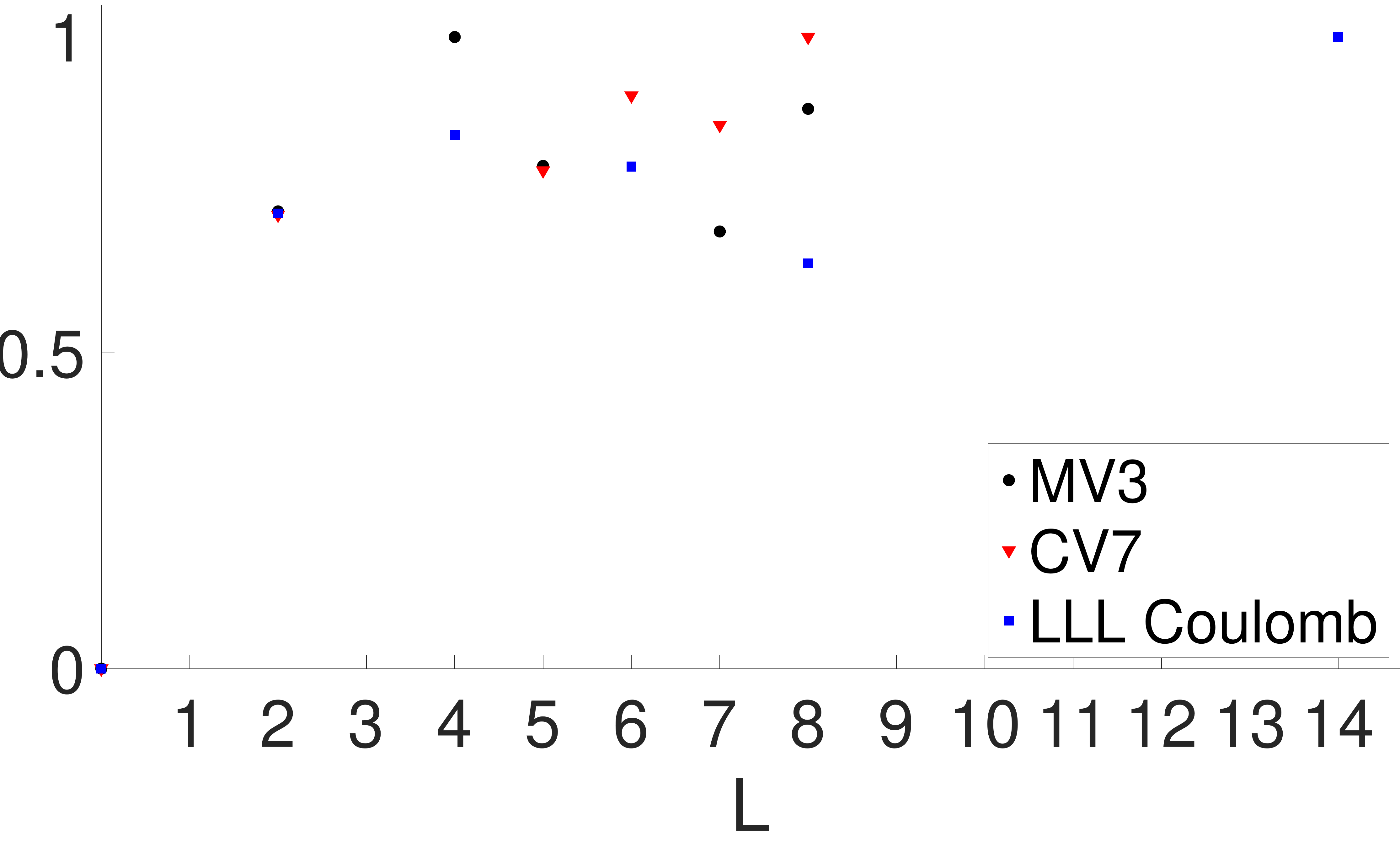}\\
    
     \includegraphics[width=0.49\columnwidth]{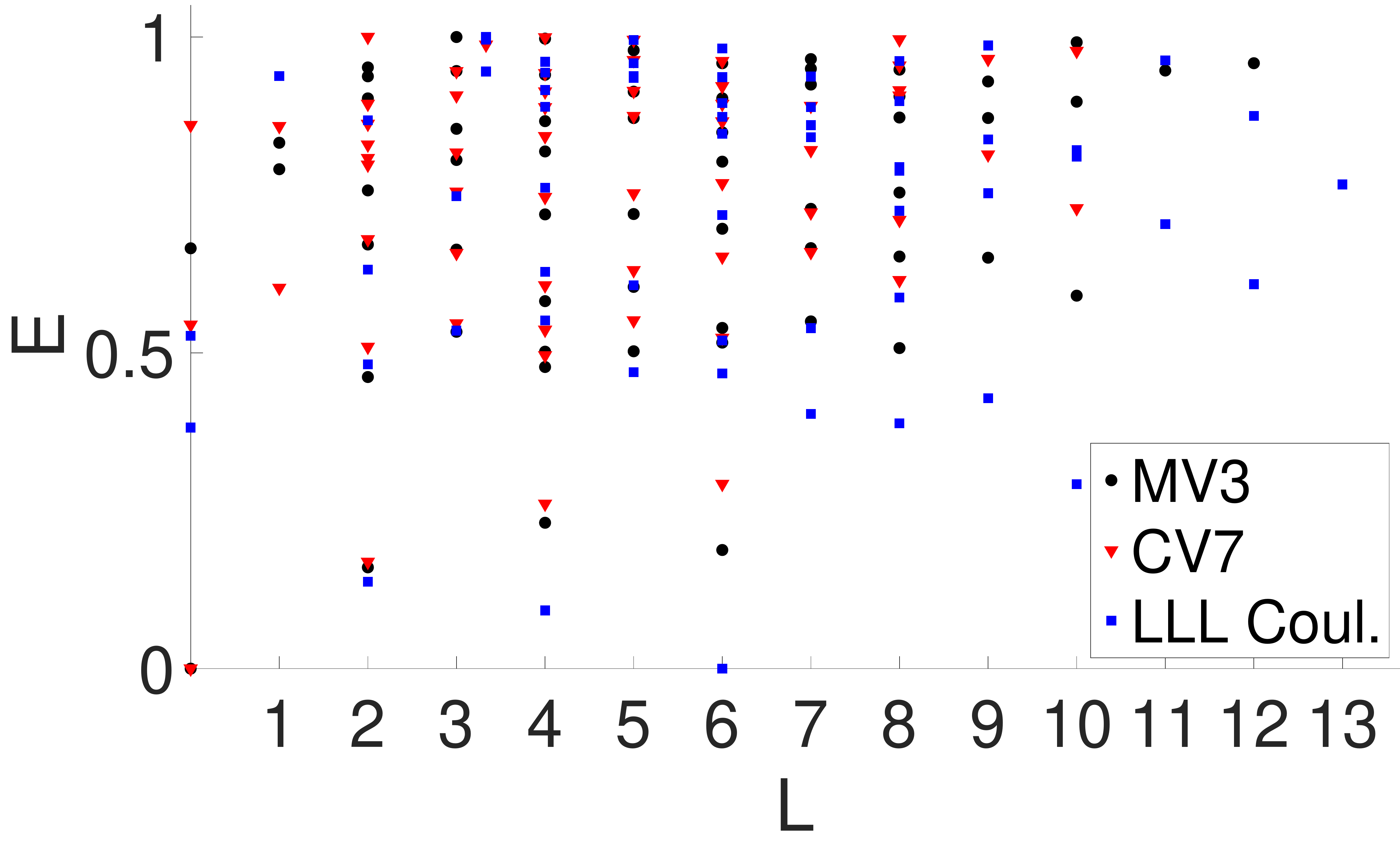}
   \includegraphics[width=0.49\columnwidth]{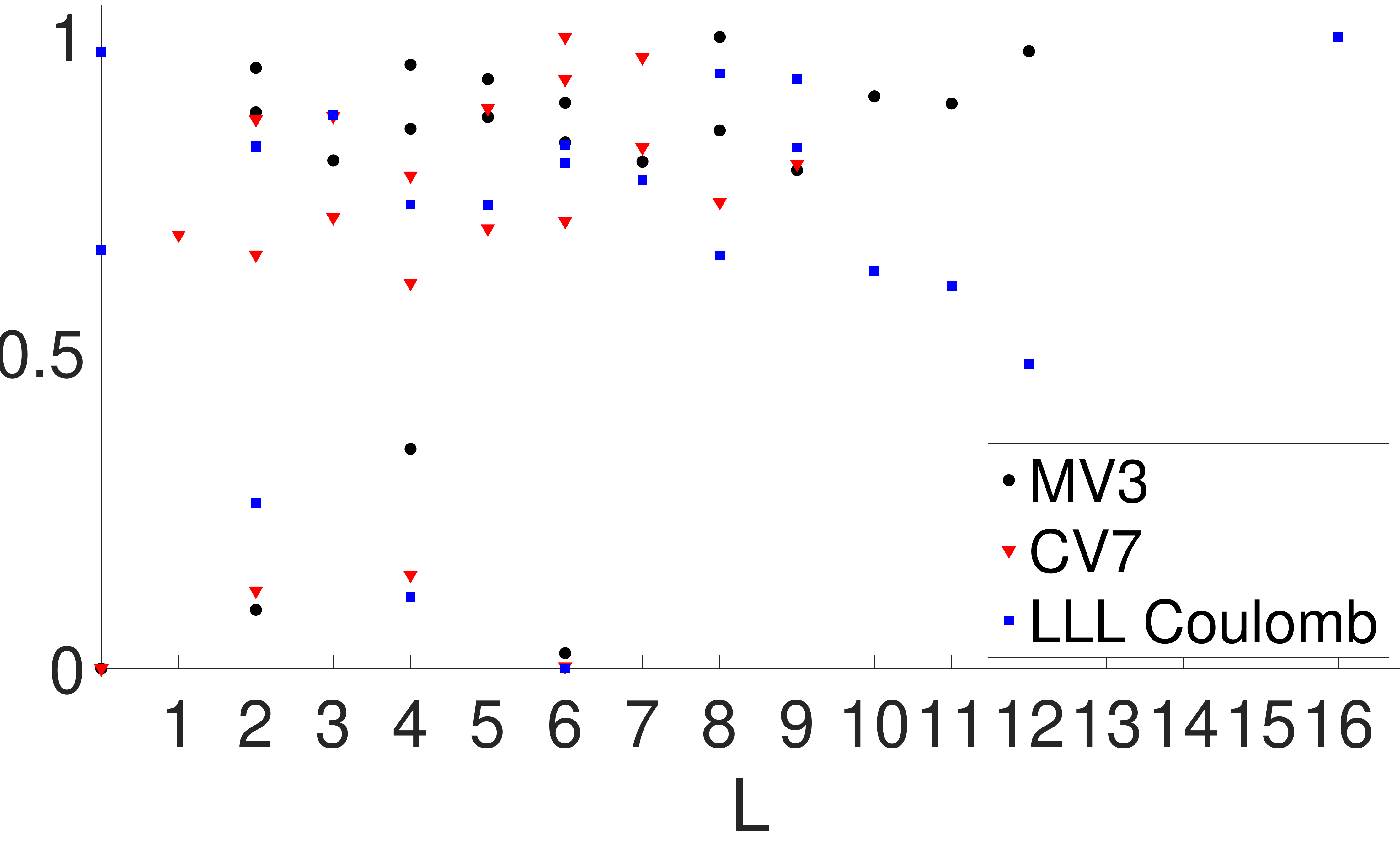}
   
    \caption{\label{fig:learnedSpect1220}
Learned Hamiltonian spectrum for 12 (Top Left), 20 (Top Right), 14 (Bottom Left) and 18 (Bottom Right) electrons. For each individual system size the spectra for different interactions are shifted by a constant to make the ground state be at $E=0$ and rescaled such that their "bandwidth" is equal.
}
\end{figure}

Pairing of composite fermions and existence of closed shell could also restrict the allowed total momentum $L$ for the low-energy excited many-body states for not-closed-shell $N_e$. The problem reduces to finding possible fermionic many-body states that can be formed from the particles in the "valent" $\Lambda$-level. The resulting numbers appear consistent with $L$ of the excited states that we obtain by diagonalising the learned Hamiltonians (see Fig.~\ref{fig:learnedSpect1220}). 

When diagonalizing systems with odd electron number for both learned Hamiltonians we get the following angular momentum of the ground state: (($N_e;L$)): (7;2.5), (9;2.5), (11;2.5), (13;3.5), (15;3.5), (17;3.5). This is consistent with any even number of electrons pairing to form an $L=0$ state and the single remaining electron being in the state with highest $\tilde{L}$ available to it: $\tilde{L}=2.5$ for $\tilde{n}=2$ and $\tilde{L}=3.5$ for $\tilde{n}=3$.

In the composite fermions picture the systems with 14/18 electrons have two electrons/holes in the 8-orbital $\tilde{n}=3$ $\Lambda$ level with angular momentum projection from -7/2 to 7/2. It is natural to expect that the low energy spectrum of the two systems is similar as they may be approximately related by the particle-hole transformation within the $\tilde{n}=3$ $\Lambda$ level. Comparison of the data (Bottom Panel of Fig. \ref{fig:learnedSpect1220}) for 14 and 18 electrons for the same interaction, say CV7, confirm this. The low-energy states are separated from the bulk of the spectrum by a visible gap and only have even $L=2,4,6$ (also true for the not shown data with $N_e$=16). The states with odd $L$ correspond to the symmetric ("bosonic") two-body states and are absent in the low-energy spectrum.

The systems with 12 and 20 electrons correspond to filled $\Lambda$-levels. Here (see Fig. \ref{fig:learnedSpect1220}), the low lying excitations are one-electron states in an excited $\Lambda$-level. They have both odd and even angular momenta and form a distinct dispersion curve separated from higher excitations where the second electron is placed in the excited $\Lambda$-level.

Although the 20 electron ground states are similar between the learned Hamiltonian CV7 and LLL Coulomb (overlap between them is 0.8557) we observe the difference in the spectrum (Fig. \ref{fig:learnedSpect1220}) : both odd and even L low-energy states occur for the learned Hamiltonians while for LLL Coulomb only even L states are present. Another striking difference is in the structure factor (bottom panel of Fig. \ref{fig:strFactsLearnedH}): the oscillations at larger $Q$ are much more pronounced in case of LLL Coulomb.

\begin{figure}
	\centering
	\includegraphics[width=\columnwidth]{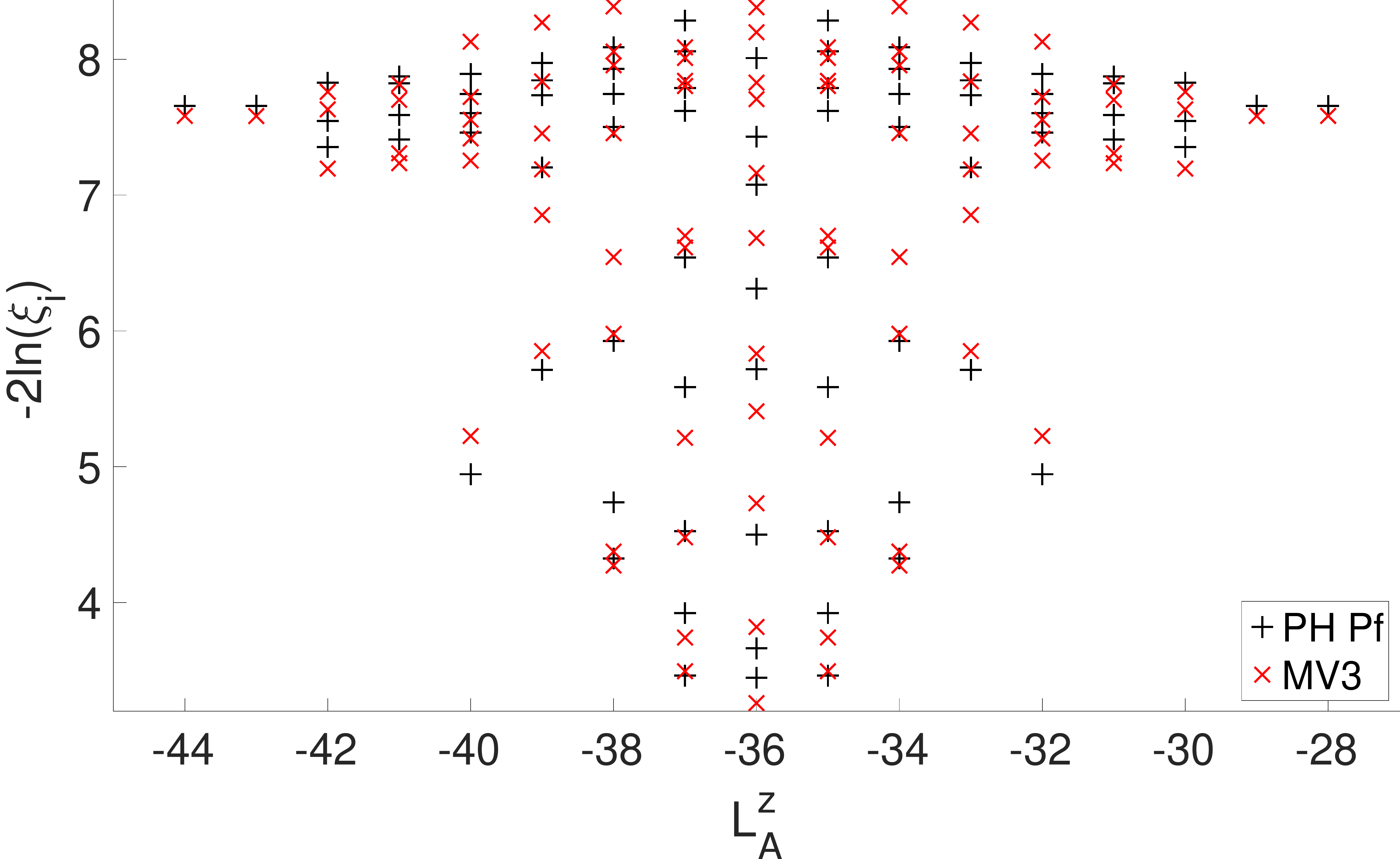}
	\includegraphics[width=\columnwidth]{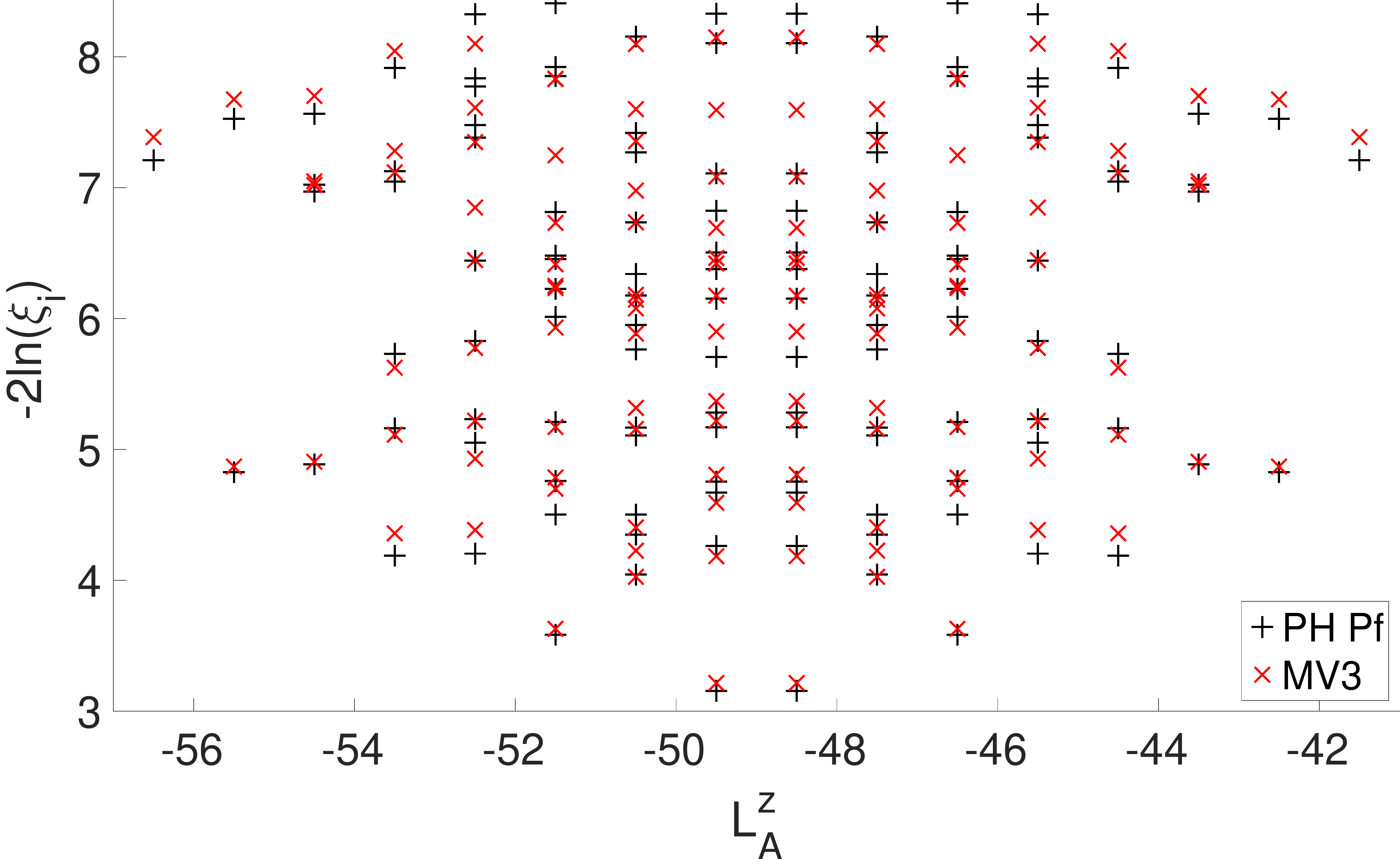}
	\caption{Entanglement spectrum in the ground state of MV3. Top: 12el; Bottom: 14el. In both plots the data for the model PH Pf wavefunction is shown with black $+$ symbols.}
		\label{fig:entr12CV7vsLLL}
\end{figure}

\begin{table}
\caption{\label{tab:SLLPLusAPF} Overlaps and energy variances for the second Landau level Coulomb interaction and anti-Pfaffian wavefunction. The indicated system sizes from 8 to 14 is the number of holes in the system.}
\begin{center}
\begin{tabular}{|c|c|c|c|c|c|}
\hline
	 					& 	SLL Coulomb 			\\
\hline
$\braket{\psi_{o}|\psi_{r}}(8)$	&	0.936407935222018				\\
\hline
$\braket{\psi_{o}|\psi_{r}}(10)$	&	0.8910370857075266			\\
\hline
$\braket{\psi_{o}|\psi_{r}}(12)$	&	0.8228169487809225		\\
\hline
$\braket{\psi_{o}|\psi_{r}}(14)$	&	0.7060756836847729			\\
\hline
$(\sigma^{rel}_E)^2(8)$		& 	1.724761519337407e-05			\\
\hline
$(\sigma^{rel}_E)^2(10)$		&	1.0722752038401e-05		\\
\hline
$(\sigma^{rel}_E)^2(12)$		&	1.35787594338842e-05		\\
\hline
$(\sigma^{rel}_E)^2(14)$		& 	7.215644779918406e-06		\\
\hline
\end{tabular}
\end{center}
\end{table}

\bibliography{phPfGeneratingH}

\begin{thebibliography}{74}%
\makeatletter
\providecommand \@ifxundefined [1]{%
 \@ifx{#1\undefined}
}%
\providecommand \@ifnum [1]{%
 \ifnum #1\expandafter \@firstoftwo
 \else \expandafter \@secondoftwo
 \fi
}%
\providecommand \@ifx [1]{%
 \ifx #1\expandafter \@firstoftwo
 \else \expandafter \@secondoftwo
 \fi
}%
\providecommand \natexlab [1]{#1}%
\providecommand \enquote  [1]{``#1''}%
\providecommand \bibnamefont  [1]{#1}%
\providecommand \bibfnamefont [1]{#1}%
\providecommand \citenamefont [1]{#1}%
\providecommand \href@noop [0]{\@secondoftwo}%
\providecommand \href [0]{\begingroup \@sanitize@url \@href}%
\providecommand \@href[1]{\@@startlink{#1}\@@href}%
\providecommand \@@href[1]{\endgroup#1\@@endlink}%
\providecommand \@sanitize@url [0]{\catcode `\\12\catcode `\$12\catcode
  `\&12\catcode `\#12\catcode `\^12\catcode `\_12\catcode `\%12\relax}%
\providecommand \@@startlink[1]{}%
\providecommand \@@endlink[0]{}%
\providecommand \url  [0]{\begingroup\@sanitize@url \@url }%
\providecommand \@url [1]{\endgroup\@href {#1}{\urlprefix }}%
\providecommand \urlprefix  [0]{URL }%
\providecommand \Eprint [0]{\href }%
\providecommand \doibase [0]{https://doi.org/}%
\providecommand \selectlanguage [0]{\@gobble}%
\providecommand \bibinfo  [0]{\@secondoftwo}%
\providecommand \bibfield  [0]{\@secondoftwo}%
\providecommand \translation [1]{[#1]}%
\providecommand \BibitemOpen [0]{}%
\providecommand \bibitemStop [0]{}%
\providecommand \bibitemNoStop [0]{.\EOS\space}%
\providecommand \EOS [0]{\spacefactor3000\relax}%
\providecommand \BibitemShut  [1]{\csname bibitem#1\endcsname}%
\let\auto@bib@innerbib\@empty
\bibitem [{\citenamefont {Willett}\ \emph {et~al.}(1987)\citenamefont
  {Willett}, \citenamefont {Eisenstein}, \citenamefont {St\"ormer},
  \citenamefont {Tsui}, \citenamefont {Gossard},\ and\ \citenamefont
  {English}}]{Willett1987}%
  \BibitemOpen
  \bibfield  {author} {\bibinfo {author} {\bibfnamefont {R.}~\bibnamefont
  {Willett}}, \bibinfo {author} {\bibfnamefont {J.~P.}\ \bibnamefont
  {Eisenstein}}, \bibinfo {author} {\bibfnamefont {H.~L.}\ \bibnamefont
  {St\"ormer}}, \bibinfo {author} {\bibfnamefont {D.~C.}\ \bibnamefont {Tsui}},
  \bibinfo {author} {\bibfnamefont {A.~C.}\ \bibnamefont {Gossard}},\ and\
  \bibinfo {author} {\bibfnamefont {J.~H.}\ \bibnamefont {English}},\
  }\bibfield  {title} {\bibinfo {title} {Observation of an even-denominator
  quantum number in the fractional quantum hall effect},\ }\href
  {https://doi.org/10.1103/PhysRevLett.59.1776} {\bibfield  {journal} {\bibinfo
   {journal} {Phys. Rev. Lett.}\ }\textbf {\bibinfo {volume} {59}},\ \bibinfo
  {pages} {1776} (\bibinfo {year} {1987})}\BibitemShut {NoStop}%
\bibitem [{\citenamefont {Wen}(1990)}]{Wen1990}%
  \BibitemOpen
  \bibfield  {author} {\bibinfo {author} {\bibfnamefont {X.~G.}\ \bibnamefont
  {Wen}},\ }\bibfield  {title} {\bibinfo {title} {Topological orders in rigid
  states},\ }\href {https://doi.org/10.1142/S0217979290000139} {\bibfield
  {journal} {\bibinfo  {journal} {International Journal of Modern Physics B}\
  }\textbf {\bibinfo {volume} {04}},\ \bibinfo {pages} {239} (\bibinfo {year}
  {1990})}\BibitemShut {NoStop}%
\bibitem [{\citenamefont {Wen}(1992)}]{Wen1992}%
  \BibitemOpen
  \bibfield  {author} {\bibinfo {author} {\bibfnamefont {X.-G.}\ \bibnamefont
  {Wen}},\ }\bibfield  {title} {\bibinfo {title} {Theory of the edge states in
  fractional quantum hall effects},\ }\href
  {https://doi.org/10.1142/S0217979292000840} {\bibfield  {journal} {\bibinfo
  {journal} {International Journal of Modern Physics B}\ }\textbf {\bibinfo
  {volume} {06}},\ \bibinfo {pages} {1711} (\bibinfo {year}
  {1992})}\BibitemShut {NoStop}%
\bibitem [{\citenamefont {Nayak}\ \emph {et~al.}(2008)\citenamefont {Nayak},
  \citenamefont {Simon}, \citenamefont {Stern}, \citenamefont {Freedman},\ and\
  \citenamefont {Das~Sarma}}]{RevModPhys-Nayak-Simon-2008}%
  \BibitemOpen
  \bibfield  {author} {\bibinfo {author} {\bibfnamefont {C.}~\bibnamefont
  {Nayak}}, \bibinfo {author} {\bibfnamefont {S.~H.}\ \bibnamefont {Simon}},
  \bibinfo {author} {\bibfnamefont {A.}~\bibnamefont {Stern}}, \bibinfo
  {author} {\bibfnamefont {M.}~\bibnamefont {Freedman}},\ and\ \bibinfo
  {author} {\bibfnamefont {S.}~\bibnamefont {Das~Sarma}},\ }\bibfield  {title}
  {\bibinfo {title} {Non-abelian anyons and topological quantum computation},\
  }\href {https://doi.org/10.1103/RevModPhys.80.1083} {\bibfield  {journal}
  {\bibinfo  {journal} {Rev. Mod. Phys.}\ }\textbf {\bibinfo {volume} {80}},\
  \bibinfo {pages} {1083} (\bibinfo {year} {2008})}\BibitemShut {NoStop}%
\bibitem [{\citenamefont {Hansson}\ \emph {et~al.}(2017)\citenamefont
  {Hansson}, \citenamefont {Hermanns}, \citenamefont {Simon},\ and\
  \citenamefont {Viefers}}]{RevModPhys-Hansson-Simon-2017}%
  \BibitemOpen
  \bibfield  {author} {\bibinfo {author} {\bibfnamefont {T.~H.}\ \bibnamefont
  {Hansson}}, \bibinfo {author} {\bibfnamefont {M.}~\bibnamefont {Hermanns}},
  \bibinfo {author} {\bibfnamefont {S.~H.}\ \bibnamefont {Simon}},\ and\
  \bibinfo {author} {\bibfnamefont {S.~F.}\ \bibnamefont {Viefers}},\
  }\bibfield  {title} {\bibinfo {title} {Quantum hall physics: Hierarchies and
  conformal field theory techniques},\ }\href
  {https://doi.org/10.1103/RevModPhys.89.025005} {\bibfield  {journal}
  {\bibinfo  {journal} {Rev. Mod. Phys.}\ }\textbf {\bibinfo {volume} {89}},\
  \bibinfo {pages} {025005} (\bibinfo {year} {2017})}\BibitemShut {NoStop}%
\bibitem [{\citenamefont {Moore}\ and\ \citenamefont
  {Read}(1991)}]{MR1991nonabelions}%
  \BibitemOpen
  \bibfield  {author} {\bibinfo {author} {\bibfnamefont {G.}~\bibnamefont
  {Moore}}\ and\ \bibinfo {author} {\bibfnamefont {N.}~\bibnamefont {Read}},\
  }\bibfield  {title} {\bibinfo {title} {Nonabelions in the fractional quantum
  hall effect},\ }\href@noop {} {\bibfield  {journal} {\bibinfo  {journal}
  {Nuclear Physics B}\ }\textbf {\bibinfo {volume} {360}},\ \bibinfo {pages}
  {362} (\bibinfo {year} {1991})}\BibitemShut {NoStop}%
\bibitem [{\citenamefont {Levin}\ \emph {et~al.}(2007)\citenamefont {Levin},
  \citenamefont {Halperin},\ and\ \citenamefont {Rosenow}}]{Levin-Apf}%
  \BibitemOpen
  \bibfield  {author} {\bibinfo {author} {\bibfnamefont {M.}~\bibnamefont
  {Levin}}, \bibinfo {author} {\bibfnamefont {B.~I.}\ \bibnamefont
  {Halperin}},\ and\ \bibinfo {author} {\bibfnamefont {B.}~\bibnamefont
  {Rosenow}},\ }\bibfield  {title} {\bibinfo {title} {Particle-hole symmetry
  and the pfaffian state},\ }\href
  {https://doi.org/10.1103/PhysRevLett.99.236806} {\bibfield  {journal}
  {\bibinfo  {journal} {Phys. Rev. Lett.}\ }\textbf {\bibinfo {volume} {99}},\
  \bibinfo {pages} {236806} (\bibinfo {year} {2007})}\BibitemShut {NoStop}%
\bibitem [{\citenamefont {Lee}\ \emph {et~al.}(2007)\citenamefont {Lee},
  \citenamefont {Ryu}, \citenamefont {Nayak},\ and\ \citenamefont
  {Fisher}}]{Nayak-Apf}%
  \BibitemOpen
  \bibfield  {author} {\bibinfo {author} {\bibfnamefont {S.-S.}\ \bibnamefont
  {Lee}}, \bibinfo {author} {\bibfnamefont {S.}~\bibnamefont {Ryu}}, \bibinfo
  {author} {\bibfnamefont {C.}~\bibnamefont {Nayak}},\ and\ \bibinfo {author}
  {\bibfnamefont {M.~P.~A.}\ \bibnamefont {Fisher}},\ }\bibfield  {title}
  {\bibinfo {title} {Particle-hole symmetry and the
  $\ensuremath{\nu}=\frac{5}{2}$ quantum hall state},\ }\href
  {https://doi.org/10.1103/PhysRevLett.99.236807} {\bibfield  {journal}
  {\bibinfo  {journal} {Phys. Rev. Lett.}\ }\textbf {\bibinfo {volume} {99}},\
  \bibinfo {pages} {236807} (\bibinfo {year} {2007})}\BibitemShut {NoStop}%
\bibitem [{Note1()}]{Note1}%
  \BibitemOpen
  \bibinfo {note} {That may be on the order of the intra-Landau-level Coulomb
  energy scale}\BibitemShut {NoStop}%
\bibitem [{\citenamefont {Pakrouski}\ \emph {et~al.}(2015)\citenamefont
  {Pakrouski}, \citenamefont {Peterson}, \citenamefont {Jolicoeur},
  \citenamefont {Scarola}, \citenamefont {Nayak},\ and\ \citenamefont
  {Troyer}}]{Pakrouski-etal-2016}%
  \BibitemOpen
  \bibfield  {author} {\bibinfo {author} {\bibfnamefont {K.}~\bibnamefont
  {Pakrouski}}, \bibinfo {author} {\bibfnamefont {M.~R.}\ \bibnamefont
  {Peterson}}, \bibinfo {author} {\bibfnamefont {T.}~\bibnamefont {Jolicoeur}},
  \bibinfo {author} {\bibfnamefont {V.~W.}\ \bibnamefont {Scarola}}, \bibinfo
  {author} {\bibfnamefont {C.}~\bibnamefont {Nayak}},\ and\ \bibinfo {author}
  {\bibfnamefont {M.}~\bibnamefont {Troyer}},\ }\bibfield  {title} {\bibinfo
  {title} {Phase diagram of the $\ensuremath{\nu}=5/2$ fractional quantum hall
  effect: Effects of landau-level mixing and nonzero width},\ }\href
  {https://doi.org/10.1103/PhysRevX.5.021004} {\bibfield  {journal} {\bibinfo
  {journal} {Phys. Rev. X}\ }\textbf {\bibinfo {volume} {5}},\ \bibinfo {pages}
  {021004} (\bibinfo {year} {2015})}\BibitemShut {NoStop}%
\bibitem [{\citenamefont {Peterson}\ and\ \citenamefont
  {Nayak}(2013)}]{Peterson-Nayak-2013}%
  \BibitemOpen
  \bibfield  {author} {\bibinfo {author} {\bibfnamefont {M.~R.}\ \bibnamefont
  {Peterson}}\ and\ \bibinfo {author} {\bibfnamefont {C.}~\bibnamefont
  {Nayak}},\ }\bibfield  {title} {\bibinfo {title} {More realistic hamiltonians
  for the fractional quantum hall regime in gaas and graphene},\ }\href
  {https://doi.org/10.1103/PhysRevB.87.245129} {\bibfield  {journal} {\bibinfo
  {journal} {Phys. Rev. B}\ }\textbf {\bibinfo {volume} {87}},\ \bibinfo
  {pages} {245129} (\bibinfo {year} {2013})}\BibitemShut {NoStop}%
\bibitem [{\citenamefont {W\'ojs}\ \emph {et~al.}(2010)\citenamefont {W\'ojs},
  \citenamefont {Toke},\ and\ \citenamefont {Jain}}]{Jain-Toke-Wojs-2010}%
  \BibitemOpen
  \bibfield  {author} {\bibinfo {author} {\bibfnamefont {A.}~\bibnamefont
  {W\'ojs}}, \bibinfo {author} {\bibfnamefont {C.}~\bibnamefont {Toke}},\ and\
  \bibinfo {author} {\bibfnamefont {J.~K.}\ \bibnamefont {Jain}},\ }\bibfield
  {title} {\bibinfo {title} {Landau-level mixing and the emergence of pfaffian
  excitations for the $5/2$ fractional quantum hall effect},\ }\href
  {https://doi.org/10.1103/PhysRevLett.105.096802} {\bibfield  {journal}
  {\bibinfo  {journal} {Phys. Rev. Lett.}\ }\textbf {\bibinfo {volume} {105}},\
  \bibinfo {pages} {096802} (\bibinfo {year} {2010})}\BibitemShut {NoStop}%
\bibitem [{\citenamefont {Rezayi}(2017)}]{Rezayi-2017}%
  \BibitemOpen
  \bibfield  {author} {\bibinfo {author} {\bibfnamefont {E.~H.}\ \bibnamefont
  {Rezayi}},\ }\bibfield  {title} {\bibinfo {title} {Landau level mixing and
  the ground state of the $\ensuremath{\nu}=5/2$ quantum hall effect},\ }\href
  {https://doi.org/10.1103/PhysRevLett.119.026801} {\bibfield  {journal}
  {\bibinfo  {journal} {Phys. Rev. Lett.}\ }\textbf {\bibinfo {volume} {119}},\
  \bibinfo {pages} {026801} (\bibinfo {year} {2017})}\BibitemShut {NoStop}%
\bibitem [{\citenamefont {Simon}\ and\ \citenamefont
  {Rezayi}(2013)}]{Simon-Rezayi-2013}%
  \BibitemOpen
  \bibfield  {author} {\bibinfo {author} {\bibfnamefont {S.~H.}\ \bibnamefont
  {Simon}}\ and\ \bibinfo {author} {\bibfnamefont {E.~H.}\ \bibnamefont
  {Rezayi}},\ }\bibfield  {title} {\bibinfo {title} {Landau level mixing in the
  perturbative limit},\ }\href {https://doi.org/10.1103/PhysRevB.87.155426}
  {\bibfield  {journal} {\bibinfo  {journal} {Phys. Rev. B}\ }\textbf {\bibinfo
  {volume} {87}},\ \bibinfo {pages} {155426} (\bibinfo {year}
  {2013})}\BibitemShut {NoStop}%
\bibitem [{\citenamefont {Rezayi}\ and\ \citenamefont
  {Simon}(2011)}]{Rezayi-simon-2011}%
  \BibitemOpen
  \bibfield  {author} {\bibinfo {author} {\bibfnamefont {E.~H.}\ \bibnamefont
  {Rezayi}}\ and\ \bibinfo {author} {\bibfnamefont {S.~H.}\ \bibnamefont
  {Simon}},\ }\bibfield  {title} {\bibinfo {title} {Breaking of particle-hole
  symmetry by landau level mixing in the $\ensuremath{\nu}=5/2$ quantized hall
  state},\ }\href {https://doi.org/10.1103/PhysRevLett.106.116801} {\bibfield
  {journal} {\bibinfo  {journal} {Phys. Rev. Lett.}\ }\textbf {\bibinfo
  {volume} {106}},\ \bibinfo {pages} {116801} (\bibinfo {year}
  {2011})}\BibitemShut {NoStop}%
\bibitem [{\citenamefont {Zaletel}\ \emph {et~al.}(2015)\citenamefont
  {Zaletel}, \citenamefont {Mong}, \citenamefont {Pollmann},\ and\
  \citenamefont {Rezayi}}]{Zaletel-etal-2015}%
  \BibitemOpen
  \bibfield  {author} {\bibinfo {author} {\bibfnamefont {M.~P.}\ \bibnamefont
  {Zaletel}}, \bibinfo {author} {\bibfnamefont {R.~S.~K.}\ \bibnamefont
  {Mong}}, \bibinfo {author} {\bibfnamefont {F.}~\bibnamefont {Pollmann}},\
  and\ \bibinfo {author} {\bibfnamefont {E.~H.}\ \bibnamefont {Rezayi}},\
  }\bibfield  {title} {\bibinfo {title} {Infinite density matrix
  renormalization group for multicomponent quantum hall systems},\ }\href
  {https://doi.org/10.1103/PhysRevB.91.045115} {\bibfield  {journal} {\bibinfo
  {journal} {Phys. Rev. B}\ }\textbf {\bibinfo {volume} {91}},\ \bibinfo
  {pages} {045115} (\bibinfo {year} {2015})}\BibitemShut {NoStop}%
\bibitem [{\citenamefont {Son}(2015)}]{SonPRX.5.031027}%
  \BibitemOpen
  \bibfield  {author} {\bibinfo {author} {\bibfnamefont {D.~T.}\ \bibnamefont
  {Son}},\ }\bibfield  {title} {\bibinfo {title} {Is the composite fermion a
  dirac particle?},\ }\href {https://doi.org/10.1103/PhysRevX.5.031027}
  {\bibfield  {journal} {\bibinfo  {journal} {Phys. Rev. X}\ }\textbf {\bibinfo
  {volume} {5}},\ \bibinfo {pages} {031027} (\bibinfo {year}
  {2015})}\BibitemShut {NoStop}%
\bibitem [{\citenamefont {Jolicoeur}(2007)}]{Jolicoeur2007}%
  \BibitemOpen
  \bibfield  {author} {\bibinfo {author} {\bibfnamefont {T.}~\bibnamefont
  {Jolicoeur}},\ }\bibfield  {title} {\bibinfo {title} {Non-abelian states with
  negative flux: A new series of quantum hall states},\ }\href
  {https://doi.org/10.1103/PhysRevLett.99.036805} {\bibfield  {journal}
  {\bibinfo  {journal} {Phys. Rev. Lett.}\ }\textbf {\bibinfo {volume} {99}},\
  \bibinfo {pages} {036805} (\bibinfo {year} {2007})}\BibitemShut {NoStop}%
\bibitem [{\citenamefont {Banerjee}\ \emph {et~al.}(2018)\citenamefont
  {Banerjee}, \citenamefont {Heiblum}, \citenamefont {Umansky}, \citenamefont
  {Feldman}, \citenamefont {Oreg},\ and\ \citenamefont
  {Stern}}]{Banerjee-etal-2018}%
  \BibitemOpen
  \bibfield  {author} {\bibinfo {author} {\bibfnamefont {M.}~\bibnamefont
  {Banerjee}}, \bibinfo {author} {\bibfnamefont {M.}~\bibnamefont {Heiblum}},
  \bibinfo {author} {\bibfnamefont {V.}~\bibnamefont {Umansky}}, \bibinfo
  {author} {\bibfnamefont {D.~E.}\ \bibnamefont {Feldman}}, \bibinfo {author}
  {\bibfnamefont {Y.}~\bibnamefont {Oreg}},\ and\ \bibinfo {author}
  {\bibfnamefont {A.}~\bibnamefont {Stern}},\ }\bibfield  {title} {\bibinfo
  {title} {Observation of half-integer thermal hall conductance},\ }\href
  {https://doi.org/10.1038/s41586-018-0184-1} {\bibfield  {journal} {\bibinfo
  {journal} {Nature}\ }\textbf {\bibinfo {volume} {559}},\ \bibinfo {pages}
  {205} (\bibinfo {year} {2018})}\BibitemShut {NoStop}%
\bibitem [{\citenamefont {Dutta}\ \emph {et~al.}(2021)\citenamefont {Dutta},
  \citenamefont {Yang}, \citenamefont {Melcer}, \citenamefont {Kundu},
  \citenamefont {Heiblum}, \citenamefont {Umansky}, \citenamefont {Oreg},
  \citenamefont {Stern},\ and\ \citenamefont
  {Mross}}]{dutta2021NewPHPfThermal}%
  \BibitemOpen
  \bibfield  {author} {\bibinfo {author} {\bibfnamefont {B.}~\bibnamefont
  {Dutta}}, \bibinfo {author} {\bibfnamefont {W.}~\bibnamefont {Yang}},
  \bibinfo {author} {\bibfnamefont {R.~A.}\ \bibnamefont {Melcer}}, \bibinfo
  {author} {\bibfnamefont {H.~K.}\ \bibnamefont {Kundu}}, \bibinfo {author}
  {\bibfnamefont {M.}~\bibnamefont {Heiblum}}, \bibinfo {author} {\bibfnamefont
  {V.}~\bibnamefont {Umansky}}, \bibinfo {author} {\bibfnamefont
  {Y.}~\bibnamefont {Oreg}}, \bibinfo {author} {\bibfnamefont {A.}~\bibnamefont
  {Stern}},\ and\ \bibinfo {author} {\bibfnamefont {D.}~\bibnamefont {Mross}},\
  }\href@noop {} {\bibinfo {title} {Novel method distinguishing between
  competing topological orders}} (\bibinfo {year} {2021}),\ \Eprint
  {https://arxiv.org/abs/2101.01419} {arXiv:2101.01419 [cond-mat.mes-hall]}
  \BibitemShut {NoStop}%
\bibitem [{\citenamefont {Mishmash}\ \emph {et~al.}(2018)\citenamefont
  {Mishmash}, \citenamefont {Mross}, \citenamefont {Alicea},\ and\
  \citenamefont {Motrunich}}]{Mishmash-etal-2018}%
  \BibitemOpen
  \bibfield  {author} {\bibinfo {author} {\bibfnamefont {R.~V.}\ \bibnamefont
  {Mishmash}}, \bibinfo {author} {\bibfnamefont {D.~F.}\ \bibnamefont {Mross}},
  \bibinfo {author} {\bibfnamefont {J.}~\bibnamefont {Alicea}},\ and\ \bibinfo
  {author} {\bibfnamefont {O.~I.}\ \bibnamefont {Motrunich}},\ }\bibfield
  {title} {\bibinfo {title} {Numerical exploration of trial wave functions for
  the particle-hole-symmetric pfaffian},\ }\href
  {https://doi.org/10.1103/PhysRevB.98.081107} {\bibfield  {journal} {\bibinfo
  {journal} {Phys. Rev. B}\ }\textbf {\bibinfo {volume} {98}},\ \bibinfo
  {pages} {081107} (\bibinfo {year} {2018})}\BibitemShut {NoStop}%
\bibitem [{\citenamefont {Balram}\ \emph {et~al.}(2018)\citenamefont {Balram},
  \citenamefont {Barkeshli},\ and\ \citenamefont {Rudner}}]{Balram-etal-2018}%
  \BibitemOpen
  \bibfield  {author} {\bibinfo {author} {\bibfnamefont {A.~C.}\ \bibnamefont
  {Balram}}, \bibinfo {author} {\bibfnamefont {M.}~\bibnamefont {Barkeshli}},\
  and\ \bibinfo {author} {\bibfnamefont {M.~S.}\ \bibnamefont {Rudner}},\
  }\bibfield  {title} {\bibinfo {title} {Parton construction of a wave function
  in the anti-pfaffian phase},\ }\href
  {https://doi.org/10.1103/PhysRevB.98.035127} {\bibfield  {journal} {\bibinfo
  {journal} {Phys. Rev. B}\ }\textbf {\bibinfo {volume} {98}},\ \bibinfo
  {pages} {035127} (\bibinfo {year} {2018})}\BibitemShut {NoStop}%
\bibitem [{\citenamefont {Yutushui}\ and\ \citenamefont
  {Mross}(2020)}]{Mross2020QMC56PHPf}%
  \BibitemOpen
  \bibfield  {author} {\bibinfo {author} {\bibfnamefont {M.}~\bibnamefont
  {Yutushui}}\ and\ \bibinfo {author} {\bibfnamefont {D.~F.}\ \bibnamefont
  {Mross}},\ }\bibfield  {title} {\bibinfo {title} {Large-scale simulations of
  particle-hole-symmetric pfaffian trial wave functions},\ }\href
  {https://doi.org/10.1103/PhysRevB.102.195153} {\bibfield  {journal} {\bibinfo
   {journal} {Phys. Rev. B}\ }\textbf {\bibinfo {volume} {102}},\ \bibinfo
  {pages} {195153} (\bibinfo {year} {2020})}\BibitemShut {NoStop}%
\bibitem [{\citenamefont {Morf}(1998)}]{Morf1998-spin}%
  \BibitemOpen
  \bibfield  {author} {\bibinfo {author} {\bibfnamefont {R.~H.}\ \bibnamefont
  {Morf}},\ }\bibfield  {title} {\bibinfo {title} {Transition from quantum hall
  to compressible states in the second landau level: New light on the $\nu=5/2$
  enigma},\ }\href {https://doi.org/10.1103/PhysRevLett.80.1505} {\bibfield
  {journal} {\bibinfo  {journal} {Phys. Rev. Lett.}\ }\textbf {\bibinfo
  {volume} {80}},\ \bibinfo {pages} {1505} (\bibinfo {year}
  {1998})}\BibitemShut {NoStop}%
\bibitem [{\citenamefont {Rezayi}\ and\ \citenamefont
  {Haldane}(2000)}]{Rezayi-Haldane2000}%
  \BibitemOpen
  \bibfield  {author} {\bibinfo {author} {\bibfnamefont {E.~H.}\ \bibnamefont
  {Rezayi}}\ and\ \bibinfo {author} {\bibfnamefont {F.~D.~M.}\ \bibnamefont
  {Haldane}},\ }\bibfield  {title} {\bibinfo {title} {Incompressible paired
  hall state, stripe order, and the composite fermion liquid phase in
  half-filled landau levels},\ }\href
  {https://doi.org/10.1103/PhysRevLett.84.4685} {\bibfield  {journal} {\bibinfo
   {journal} {Phys. Rev. Lett.}\ }\textbf {\bibinfo {volume} {84}},\ \bibinfo
  {pages} {4685} (\bibinfo {year} {2000})}\BibitemShut {NoStop}%
\bibitem [{\citenamefont {Feiguin}\ \emph {et~al.}(2008)\citenamefont
  {Feiguin}, \citenamefont {Rezayi}, \citenamefont {Nayak},\ and\ \citenamefont
  {Das~Sarma}}]{feiguin-DMRG-2008}%
  \BibitemOpen
  \bibfield  {author} {\bibinfo {author} {\bibfnamefont {A.~E.}\ \bibnamefont
  {Feiguin}}, \bibinfo {author} {\bibfnamefont {E.}~\bibnamefont {Rezayi}},
  \bibinfo {author} {\bibfnamefont {C.}~\bibnamefont {Nayak}},\ and\ \bibinfo
  {author} {\bibfnamefont {S.}~\bibnamefont {Das~Sarma}},\ }\bibfield  {title}
  {\bibinfo {title} {Density matrix renormalization group study of
  incompressible fractional quantum hall states},\ }\href
  {https://doi.org/10.1103/PhysRevLett.100.166803} {\bibfield  {journal}
  {\bibinfo  {journal} {Phys. Rev. Lett.}\ }\textbf {\bibinfo {volume} {100}},\
  \bibinfo {pages} {166803} (\bibinfo {year} {2008})}\BibitemShut {NoStop}%
\bibitem [{\citenamefont {Peterson}\ \emph {et~al.}(2008)\citenamefont
  {Peterson}, \citenamefont {Jolicoeur},\ and\ \citenamefont
  {Das~Sarma}}]{Peterson-Jolicoeur-2008}%
  \BibitemOpen
  \bibfield  {author} {\bibinfo {author} {\bibfnamefont {M.~R.}\ \bibnamefont
  {Peterson}}, \bibinfo {author} {\bibfnamefont {T.}~\bibnamefont
  {Jolicoeur}},\ and\ \bibinfo {author} {\bibfnamefont {S.}~\bibnamefont
  {Das~Sarma}},\ }\bibfield  {title} {\bibinfo {title} {Finite-layer thickness
  stabilizes the pfaffian state for the 5/2 fractional quantum hall effect:
  Wave function overlap and topological degeneracy},\ }\href
  {https://doi.org/10.1103/PhysRevLett.101.016807} {\bibfield  {journal}
  {\bibinfo  {journal} {Phys. Rev. Lett.}\ }\textbf {\bibinfo {volume} {101}},\
  \bibinfo {pages} {016807} (\bibinfo {year} {2008})}\BibitemShut {NoStop}%
\bibitem [{\citenamefont {Wan}\ \emph {et~al.}(2008)\citenamefont {Wan},
  \citenamefont {Hu}, \citenamefont {Rezayi},\ and\ \citenamefont
  {Yang}}]{XinWan-etal-disk-2008}%
  \BibitemOpen
  \bibfield  {author} {\bibinfo {author} {\bibfnamefont {X.}~\bibnamefont
  {Wan}}, \bibinfo {author} {\bibfnamefont {Z.-X.}\ \bibnamefont {Hu}},
  \bibinfo {author} {\bibfnamefont {E.~H.}\ \bibnamefont {Rezayi}},\ and\
  \bibinfo {author} {\bibfnamefont {K.}~\bibnamefont {Yang}},\ }\bibfield
  {title} {\bibinfo {title} {Fractional quantum hall effect at
  $\ensuremath{\nu}=5/2$: Ground states, non-abelian quasiholes, and edge modes
  in a microscopic model},\ }\href {https://doi.org/10.1103/PhysRevB.77.165316}
  {\bibfield  {journal} {\bibinfo  {journal} {Phys. Rev. B}\ }\textbf {\bibinfo
  {volume} {77}},\ \bibinfo {pages} {165316} (\bibinfo {year}
  {2008})}\BibitemShut {NoStop}%
\bibitem [{\citenamefont {Zhao}\ \emph {et~al.}(2011)\citenamefont {Zhao},
  \citenamefont {Sheng},\ and\ \citenamefont {Haldane}}]{Sheng-Haldane-2011}%
  \BibitemOpen
  \bibfield  {author} {\bibinfo {author} {\bibfnamefont {J.}~\bibnamefont
  {Zhao}}, \bibinfo {author} {\bibfnamefont {D.~N.}\ \bibnamefont {Sheng}},\
  and\ \bibinfo {author} {\bibfnamefont {F.~D.~M.}\ \bibnamefont {Haldane}},\
  }\bibfield  {title} {\bibinfo {title} {Fractional quantum hall states at
  $\frac{1}{3}$ and $\frac{5}{2}$ filling: Density-matrix renormalization group
  calculations},\ }\href {https://doi.org/10.1103/PhysRevB.83.195135}
  {\bibfield  {journal} {\bibinfo  {journal} {Phys. Rev. B}\ }\textbf {\bibinfo
  {volume} {83}},\ \bibinfo {pages} {195135} (\bibinfo {year}
  {2011})}\BibitemShut {NoStop}%
\bibitem [{\citenamefont {Zucker}\ and\ \citenamefont
  {Feldman}(2016)}]{Feldman2016}%
  \BibitemOpen
  \bibfield  {author} {\bibinfo {author} {\bibfnamefont {P.~T.}\ \bibnamefont
  {Zucker}}\ and\ \bibinfo {author} {\bibfnamefont {D.~E.}\ \bibnamefont
  {Feldman}},\ }\bibfield  {title} {\bibinfo {title} {Stabilization of the
  particle-hole pfaffian order by landau-level mixing and impurities that break
  particle-hole symmetry},\ }\href
  {https://doi.org/10.1103/PhysRevLett.117.096802} {\bibfield  {journal}
  {\bibinfo  {journal} {Phys. Rev. Lett.}\ }\textbf {\bibinfo {volume} {117}},\
  \bibinfo {pages} {096802} (\bibinfo {year} {2016})}\BibitemShut {NoStop}%
\bibitem [{\citenamefont {Simon}(2018{\natexlab{a}})}]{Simon-aPfGS-2018}%
  \BibitemOpen
  \bibfield  {author} {\bibinfo {author} {\bibfnamefont {S.~H.}\ \bibnamefont
  {Simon}},\ }\bibfield  {title} {\bibinfo {title} {Interpretation of thermal
  conductance of the $\mathbf{\ensuremath{\nu}}=\mathbf{5}/\mathbf{2}$ edge},\
  }\href {https://doi.org/10.1103/PhysRevB.97.121406} {\bibfield  {journal}
  {\bibinfo  {journal} {Phys. Rev. B}\ }\textbf {\bibinfo {volume} {97}},\
  \bibinfo {pages} {121406} (\bibinfo {year} {2018}{\natexlab{a}})}\BibitemShut
  {NoStop}%
\bibitem [{\citenamefont {Wang}\ \emph {et~al.}(2018)\citenamefont {Wang},
  \citenamefont {Vishwanath},\ and\ \citenamefont {Halperin}}]{Wang-etal-2018}%
  \BibitemOpen
  \bibfield  {author} {\bibinfo {author} {\bibfnamefont {C.}~\bibnamefont
  {Wang}}, \bibinfo {author} {\bibfnamefont {A.}~\bibnamefont {Vishwanath}},\
  and\ \bibinfo {author} {\bibfnamefont {B.~I.}\ \bibnamefont {Halperin}},\
  }\bibfield  {title} {\bibinfo {title} {Topological order from disorder and
  the quantized hall thermal metal: Possible applications to the
  $\ensuremath{\nu}=5/2$ state},\ }\href
  {https://doi.org/10.1103/PhysRevB.98.045112} {\bibfield  {journal} {\bibinfo
  {journal} {Phys. Rev. B}\ }\textbf {\bibinfo {volume} {98}},\ \bibinfo
  {pages} {045112} (\bibinfo {year} {2018})}\BibitemShut {NoStop}%
\bibitem [{\citenamefont {Mross}\ \emph {et~al.}(2018)\citenamefont {Mross},
  \citenamefont {Oreg}, \citenamefont {Stern}, \citenamefont {Margalit},\ and\
  \citenamefont {Heiblum}}]{Mross-etal-2018}%
  \BibitemOpen
  \bibfield  {author} {\bibinfo {author} {\bibfnamefont {D.~F.}\ \bibnamefont
  {Mross}}, \bibinfo {author} {\bibfnamefont {Y.}~\bibnamefont {Oreg}},
  \bibinfo {author} {\bibfnamefont {A.}~\bibnamefont {Stern}}, \bibinfo
  {author} {\bibfnamefont {G.}~\bibnamefont {Margalit}},\ and\ \bibinfo
  {author} {\bibfnamefont {M.}~\bibnamefont {Heiblum}},\ }\bibfield  {title}
  {\bibinfo {title} {Theory of disorder-induced half-integer thermal hall
  conductance},\ }\href {https://doi.org/10.1103/PhysRevLett.121.026801}
  {\bibfield  {journal} {\bibinfo  {journal} {Phys. Rev. Lett.}\ }\textbf
  {\bibinfo {volume} {121}},\ \bibinfo {pages} {026801} (\bibinfo {year}
  {2018})}\BibitemShut {NoStop}%
\bibitem [{\citenamefont {Lian}\ and\ \citenamefont
  {Wang}(2018)}]{Lian-Wang-2018}%
  \BibitemOpen
  \bibfield  {author} {\bibinfo {author} {\bibfnamefont {B.}~\bibnamefont
  {Lian}}\ and\ \bibinfo {author} {\bibfnamefont {J.}~\bibnamefont {Wang}},\
  }\bibfield  {title} {\bibinfo {title} {Theory of the disordered
  $\ensuremath{\nu}=\frac{5}{2}$ quantum thermal hall state: Emergent symmetry
  and phase diagram},\ }\href {https://doi.org/10.1103/PhysRevB.97.165124}
  {\bibfield  {journal} {\bibinfo  {journal} {Phys. Rev. B}\ }\textbf {\bibinfo
  {volume} {97}},\ \bibinfo {pages} {165124} (\bibinfo {year}
  {2018})}\BibitemShut {NoStop}%
\bibitem [{\citenamefont {Simon}\ \emph {et~al.}(2020)\citenamefont {Simon},
  \citenamefont {Ippoliti}, \citenamefont {Zaletel},\ and\ \citenamefont
  {Rezayi}}]{Simon-etal-pf-apf-2020}%
  \BibitemOpen
  \bibfield  {author} {\bibinfo {author} {\bibfnamefont {S.~H.}\ \bibnamefont
  {Simon}}, \bibinfo {author} {\bibfnamefont {M.}~\bibnamefont {Ippoliti}},
  \bibinfo {author} {\bibfnamefont {M.~P.}\ \bibnamefont {Zaletel}},\ and\
  \bibinfo {author} {\bibfnamefont {E.~H.}\ \bibnamefont {Rezayi}},\ }\bibfield
   {title} {\bibinfo {title} {Energetics of pfaffian--anti-pfaffian domains},\
  }\href {https://doi.org/10.1103/PhysRevB.101.041302} {\bibfield  {journal}
  {\bibinfo  {journal} {Phys. Rev. B}\ }\textbf {\bibinfo {volume} {101}},\
  \bibinfo {pages} {041302} (\bibinfo {year} {2020})}\BibitemShut {NoStop}%
\bibitem [{\citenamefont {Feldman}(2018)}]{PhysRevB.98.167401}%
  \BibitemOpen
  \bibfield  {author} {\bibinfo {author} {\bibfnamefont {D.~E.}\ \bibnamefont
  {Feldman}},\ }\bibfield  {title} {\bibinfo {title} {Comment on
  ``interpretation of thermal conductance of the $\ensuremath{\nu}=5/2$
  edge''},\ }\href {https://doi.org/10.1103/PhysRevB.98.167401} {\bibfield
  {journal} {\bibinfo  {journal} {Phys. Rev. B}\ }\textbf {\bibinfo {volume}
  {98}},\ \bibinfo {pages} {167401} (\bibinfo {year} {2018})}\BibitemShut
  {NoStop}%
\bibitem [{\citenamefont {Simon}(2018{\natexlab{b}})}]{PhysRevB.98.167402}%
  \BibitemOpen
  \bibfield  {author} {\bibinfo {author} {\bibfnamefont {S.~H.}\ \bibnamefont
  {Simon}},\ }\bibfield  {title} {\bibinfo {title} {Reply to ``comment on
  `interpretation of thermal conductance of the $\ensuremath{\nu}=5/2$ edge'
  ''},\ }\href {https://doi.org/10.1103/PhysRevB.98.167402} {\bibfield
  {journal} {\bibinfo  {journal} {Phys. Rev. B}\ }\textbf {\bibinfo {volume}
  {98}},\ \bibinfo {pages} {167402} (\bibinfo {year}
  {2018}{\natexlab{b}})}\BibitemShut {NoStop}%
\bibitem [{\citenamefont {Ma}\ and\ \citenamefont
  {Feldman}(2019)}]{PhysRevB.99.085309}%
  \BibitemOpen
  \bibfield  {author} {\bibinfo {author} {\bibfnamefont {K.~K.~W.}\
  \bibnamefont {Ma}}\ and\ \bibinfo {author} {\bibfnamefont {D.~E.}\
  \bibnamefont {Feldman}},\ }\bibfield  {title} {\bibinfo {title} {Partial
  equilibration of integer and fractional edge channels in the thermal quantum
  hall effect},\ }\href {https://doi.org/10.1103/PhysRevB.99.085309} {\bibfield
   {journal} {\bibinfo  {journal} {Phys. Rev. B}\ }\textbf {\bibinfo {volume}
  {99}},\ \bibinfo {pages} {085309} (\bibinfo {year} {2019})}\BibitemShut
  {NoStop}%
\bibitem [{\citenamefont {Park}\ \emph {et~al.}(2020)\citenamefont {Park},
  \citenamefont {Sp\aa{}nsl\"att}, \citenamefont {Gefen},\ and\ \citenamefont
  {Mirlin}}]{PhysRevLett.125.157702}%
  \BibitemOpen
  \bibfield  {author} {\bibinfo {author} {\bibfnamefont {J.}~\bibnamefont
  {Park}}, \bibinfo {author} {\bibfnamefont {C.}~\bibnamefont
  {Sp\aa{}nsl\"att}}, \bibinfo {author} {\bibfnamefont {Y.}~\bibnamefont
  {Gefen}},\ and\ \bibinfo {author} {\bibfnamefont {A.~D.}\ \bibnamefont
  {Mirlin}},\ }\bibfield  {title} {\bibinfo {title} {Noise on the non-abelian
  $\ensuremath{\nu}=5/2$ fractional quantum hall edge},\ }\href
  {https://doi.org/10.1103/PhysRevLett.125.157702} {\bibfield  {journal}
  {\bibinfo  {journal} {Phys. Rev. Lett.}\ }\textbf {\bibinfo {volume} {125}},\
  \bibinfo {pages} {157702} (\bibinfo {year} {2020})}\BibitemShut {NoStop}%
\bibitem [{\citenamefont {Fulga}\ \emph {et~al.}(2020)\citenamefont {Fulga},
  \citenamefont {Oreg}, \citenamefont {Mirlin}, \citenamefont {Stern},\ and\
  \citenamefont {Mross}}]{PhysRevLett.125.236802}%
  \BibitemOpen
  \bibfield  {author} {\bibinfo {author} {\bibfnamefont {I.~C.}\ \bibnamefont
  {Fulga}}, \bibinfo {author} {\bibfnamefont {Y.}~\bibnamefont {Oreg}},
  \bibinfo {author} {\bibfnamefont {A.~D.}\ \bibnamefont {Mirlin}}, \bibinfo
  {author} {\bibfnamefont {A.}~\bibnamefont {Stern}},\ and\ \bibinfo {author}
  {\bibfnamefont {D.~F.}\ \bibnamefont {Mross}},\ }\bibfield  {title} {\bibinfo
  {title} {Temperature enhancement of thermal hall conductance quantization},\
  }\href {https://doi.org/10.1103/PhysRevLett.125.236802} {\bibfield  {journal}
  {\bibinfo  {journal} {Phys. Rev. Lett.}\ }\textbf {\bibinfo {volume} {125}},\
  \bibinfo {pages} {236802} (\bibinfo {year} {2020})}\BibitemShut {NoStop}%
\bibitem [{\citenamefont {Simon}\ and\ \citenamefont
  {Rosenow}(2020)}]{PhysRevLett.124.126801}%
  \BibitemOpen
  \bibfield  {author} {\bibinfo {author} {\bibfnamefont {S.~H.}\ \bibnamefont
  {Simon}}\ and\ \bibinfo {author} {\bibfnamefont {B.}~\bibnamefont
  {Rosenow}},\ }\bibfield  {title} {\bibinfo {title} {Partial equilibration of
  the anti-pfaffian edge due to majorana disorder},\ }\href
  {https://doi.org/10.1103/PhysRevLett.124.126801} {\bibfield  {journal}
  {\bibinfo  {journal} {Phys. Rev. Lett.}\ }\textbf {\bibinfo {volume} {124}},\
  \bibinfo {pages} {126801} (\bibinfo {year} {2020})}\BibitemShut {NoStop}%
\bibitem [{\citenamefont {Asasi}\ and\ \citenamefont
  {Mulligan}(2020)}]{PhysRevB.102.205104}%
  \BibitemOpen
  \bibfield  {author} {\bibinfo {author} {\bibfnamefont {H.}~\bibnamefont
  {Asasi}}\ and\ \bibinfo {author} {\bibfnamefont {M.}~\bibnamefont
  {Mulligan}},\ }\bibfield  {title} {\bibinfo {title} {Partial equilibration of
  anti-pfaffian edge modes at $\ensuremath{\nu}=5/2$},\ }\href
  {https://doi.org/10.1103/PhysRevB.102.205104} {\bibfield  {journal} {\bibinfo
   {journal} {Phys. Rev. B}\ }\textbf {\bibinfo {volume} {102}},\ \bibinfo
  {pages} {205104} (\bibinfo {year} {2020})}\BibitemShut {NoStop}%
\bibitem [{\citenamefont {Ma}\ and\ \citenamefont
  {Feldman}(2020)}]{PhysRevLett.125.016801}%
  \BibitemOpen
  \bibfield  {author} {\bibinfo {author} {\bibfnamefont {K.~K.~W.}\
  \bibnamefont {Ma}}\ and\ \bibinfo {author} {\bibfnamefont {D.~E.}\
  \bibnamefont {Feldman}},\ }\bibfield  {title} {\bibinfo {title} {Thermal
  equilibration on the edges of topological liquids},\ }\href
  {https://doi.org/10.1103/PhysRevLett.125.016801} {\bibfield  {journal}
  {\bibinfo  {journal} {Phys. Rev. Lett.}\ }\textbf {\bibinfo {volume} {125}},\
  \bibinfo {pages} {016801} (\bibinfo {year} {2020})}\BibitemShut {NoStop}%
\bibitem [{\citenamefont {Rezayi}\ \emph {et~al.}(2021)\citenamefont {Rezayi},
  \citenamefont {Pakrouski},\ and\ \citenamefont
  {Haldane}}]{rezayi2021energetics}%
  \BibitemOpen
  \bibfield  {author} {\bibinfo {author} {\bibfnamefont {E.~H.}\ \bibnamefont
  {Rezayi}}, \bibinfo {author} {\bibfnamefont {K.}~\bibnamefont {Pakrouski}},\
  and\ \bibinfo {author} {\bibfnamefont {F.~D.~M.}\ \bibnamefont {Haldane}},\
  }\bibfield  {title} {\bibinfo {title} {Stability of the particle-hole
  pfaffian state and the $\frac{5}{2}$-fractional quantum hall effect},\ }\href
  {https://doi.org/10.1103/PhysRevB.104.L081407} {\bibfield  {journal}
  {\bibinfo  {journal} {Phys. Rev. B}\ }\textbf {\bibinfo {volume} {104}},\
  \bibinfo {pages} {L081407} (\bibinfo {year} {2021})}\BibitemShut {NoStop}%
\bibitem [{Note2()}]{Note2}%
  \BibitemOpen
  \bibinfo {note} {Although it has been argued \cite {Sun2020PRB} that in
  principle the PH-Pfaffian topological order can exist in a translationally
  and rotationally invariant system}\BibitemShut {NoStop}%
\bibitem [{\citenamefont {Fano}\ \emph {et~al.}(1986)\citenamefont {Fano},
  \citenamefont {Ortolani},\ and\ \citenamefont {Colombo}}]{Fano-etal-1986}%
  \BibitemOpen
  \bibfield  {author} {\bibinfo {author} {\bibfnamefont {G.}~\bibnamefont
  {Fano}}, \bibinfo {author} {\bibfnamefont {F.}~\bibnamefont {Ortolani}},\
  and\ \bibinfo {author} {\bibfnamefont {E.}~\bibnamefont {Colombo}},\
  }\bibfield  {title} {\bibinfo {title} {Configuration-interaction calculations
  on the fractional quantum hall effect},\ }\href
  {https://doi.org/10.1103/PhysRevB.34.2670} {\bibfield  {journal} {\bibinfo
  {journal} {Phys. Rev. B}\ }\textbf {\bibinfo {volume} {34}},\ \bibinfo
  {pages} {2670} (\bibinfo {year} {1986})}\BibitemShut {NoStop}%
\bibitem [{\citenamefont {Haldane}(1983)}]{Haldane-1983}%
  \BibitemOpen
  \bibfield  {author} {\bibinfo {author} {\bibfnamefont {F.~D.~M.}\
  \bibnamefont {Haldane}},\ }\bibfield  {title} {\bibinfo {title} {Fractional
  quantization of the hall effect: A hierarchy of incompressible quantum fluid
  states},\ }\href {https://doi.org/10.1103/PhysRevLett.51.605} {\bibfield
  {journal} {\bibinfo  {journal} {Phys. Rev. Lett.}\ }\textbf {\bibinfo
  {volume} {51}},\ \bibinfo {pages} {605} (\bibinfo {year} {1983})}\BibitemShut
  {NoStop}%
\bibitem [{\citenamefont {Wen}\ and\ \citenamefont {Niu}(1990)}]{Wen1990shift}%
  \BibitemOpen
  \bibfield  {author} {\bibinfo {author} {\bibfnamefont {X.~G.}\ \bibnamefont
  {Wen}}\ and\ \bibinfo {author} {\bibfnamefont {Q.}~\bibnamefont {Niu}},\
  }\bibfield  {title} {\bibinfo {title} {Ground-state degeneracy of the
  fractional quantum hall states in the presence of a random potential and on
  high-genus riemann surfaces},\ }\href
  {https://doi.org/10.1103/PhysRevB.41.9377} {\bibfield  {journal} {\bibinfo
  {journal} {Phys. Rev. B}\ }\textbf {\bibinfo {volume} {41}},\ \bibinfo
  {pages} {9377} (\bibinfo {year} {1990})}\BibitemShut {NoStop}%
\bibitem [{\citenamefont {Pakrouski}(2020)}]{pakrouski2019automatic}%
  \BibitemOpen
  \bibfield  {author} {\bibinfo {author} {\bibfnamefont {K.}~\bibnamefont
  {Pakrouski}},\ }\bibfield  {title} {\bibinfo {title} {Automatic design of
  {H}amiltonians},\ }\href {https://doi.org/10.22331/q-2020-09-02-315}
  {\bibfield  {journal} {\bibinfo  {journal} {{Quantum}}\ }\textbf {\bibinfo
  {volume} {4}},\ \bibinfo {pages} {315} (\bibinfo {year} {2020})}\BibitemShut
  {NoStop}%
\bibitem [{\citenamefont {Schuld}\ and\ \citenamefont
  {Killoran}(2019)}]{Killoran2019QMLInFeatureHilbSp}%
  \BibitemOpen
  \bibfield  {author} {\bibinfo {author} {\bibfnamefont {M.}~\bibnamefont
  {Schuld}}\ and\ \bibinfo {author} {\bibfnamefont {N.}~\bibnamefont
  {Killoran}},\ }\bibfield  {title} {\bibinfo {title} {Quantum machine learning
  in feature hilbert spaces},\ }\href
  {https://doi.org/10.1103/PhysRevLett.122.040504} {\bibfield  {journal}
  {\bibinfo  {journal} {Phys. Rev. Lett.}\ }\textbf {\bibinfo {volume} {122}},\
  \bibinfo {pages} {040504} (\bibinfo {year} {2019})}\BibitemShut {NoStop}%
\bibitem [{\citenamefont {Chertkov}\ and\ \citenamefont
  {Clark}(2018)}]{Chertkov2018InvMethod}%
  \BibitemOpen
  \bibfield  {author} {\bibinfo {author} {\bibfnamefont {E.}~\bibnamefont
  {Chertkov}}\ and\ \bibinfo {author} {\bibfnamefont {B.~K.}\ \bibnamefont
  {Clark}},\ }\bibfield  {title} {\bibinfo {title} {Computational inverse
  method for constructing spaces of quantum models from wave functions},\
  }\href {https://doi.org/10.1103/PhysRevX.8.031029} {\bibfield  {journal}
  {\bibinfo  {journal} {Phys. Rev. X}\ }\textbf {\bibinfo {volume} {8}},\
  \bibinfo {pages} {031029} (\bibinfo {year} {2018})}\BibitemShut {NoStop}%
\bibitem [{\citenamefont {Greiter}\ \emph {et~al.}(2018)\citenamefont
  {Greiter}, \citenamefont {Schnells},\ and\ \citenamefont
  {Thomale}}]{Thomale2018HMethod}%
  \BibitemOpen
  \bibfield  {author} {\bibinfo {author} {\bibfnamefont {M.}~\bibnamefont
  {Greiter}}, \bibinfo {author} {\bibfnamefont {V.}~\bibnamefont {Schnells}},\
  and\ \bibinfo {author} {\bibfnamefont {R.}~\bibnamefont {Thomale}},\
  }\bibfield  {title} {\bibinfo {title} {Method to identify parent hamiltonians
  for trial states},\ }\href {https://doi.org/10.1103/PhysRevB.98.081113}
  {\bibfield  {journal} {\bibinfo  {journal} {Phys. Rev. B}\ }\textbf {\bibinfo
  {volume} {98}},\ \bibinfo {pages} {081113} (\bibinfo {year}
  {2018})}\BibitemShut {NoStop}%
\bibitem [{\citenamefont {Qi}\ and\ \citenamefont
  {Ranard}(2019)}]{Qi2019HMethod}%
  \BibitemOpen
  \bibfield  {author} {\bibinfo {author} {\bibfnamefont {X.-L.}\ \bibnamefont
  {Qi}}\ and\ \bibinfo {author} {\bibfnamefont {D.}~\bibnamefont {Ranard}},\
  }\bibfield  {title} {\bibinfo {title} {Determining a local {H}amiltonian from
  a single eigenstate},\ }\href {https://doi.org/10.22331/q-2019-07-08-159}
  {\bibfield  {journal} {\bibinfo  {journal} {{Quantum}}\ }\textbf {\bibinfo
  {volume} {3}},\ \bibinfo {pages} {159} (\bibinfo {year} {2019})}\BibitemShut
  {NoStop}%
\bibitem [{\citenamefont {T~\H{o}ke}\ and\ \citenamefont
  {Jain}(2006)}]{TokeJain2006NLLCoulPPs}%
  \BibitemOpen
  \bibfield  {author} {\bibinfo {author} {\bibfnamefont {C.}~\bibnamefont
  {T~\H{o}ke}}\ and\ \bibinfo {author} {\bibfnamefont {J.~K.}\ \bibnamefont
  {Jain}},\ }\bibfield  {title} {\bibinfo {title} {Understanding the
  $\frac{5}{2}$ fractional quantum hall effect without the pfaffian wave
  function},\ }\href {https://doi.org/10.1103/PhysRevLett.96.246805} {\bibfield
   {journal} {\bibinfo  {journal} {Phys. Rev. Lett.}\ }\textbf {\bibinfo
  {volume} {96}},\ \bibinfo {pages} {246805} (\bibinfo {year}
  {2006})}\BibitemShut {NoStop}%
\bibitem [{\citenamefont {Li}\ and\ \citenamefont
  {Haldane}(2008)}]{LiHaldane2008EntSpec}%
  \BibitemOpen
  \bibfield  {author} {\bibinfo {author} {\bibfnamefont {H.}~\bibnamefont
  {Li}}\ and\ \bibinfo {author} {\bibfnamefont {F.~D.~M.}\ \bibnamefont
  {Haldane}},\ }\bibfield  {title} {\bibinfo {title} {Entanglement spectrum as
  a generalization of entanglement entropy: Identification of topological order
  in non-abelian fractional quantum hall effect states},\ }\href
  {https://doi.org/10.1103/PhysRevLett.101.010504} {\bibfield  {journal}
  {\bibinfo  {journal} {Phys. Rev. Lett.}\ }\textbf {\bibinfo {volume} {101}},\
  \bibinfo {pages} {010504} (\bibinfo {year} {2008})}\BibitemShut {NoStop}%
\bibitem [{\citenamefont {Girvin}\ \emph {et~al.}(1985)\citenamefont {Girvin},
  \citenamefont {MacDonald},\ and\ \citenamefont
  {Platzman}}]{GMPQtoThe4thPRL1985}%
  \BibitemOpen
  \bibfield  {author} {\bibinfo {author} {\bibfnamefont {S.~M.}\ \bibnamefont
  {Girvin}}, \bibinfo {author} {\bibfnamefont {A.~H.}\ \bibnamefont
  {MacDonald}},\ and\ \bibinfo {author} {\bibfnamefont {P.~M.}\ \bibnamefont
  {Platzman}},\ }\bibfield  {title} {\bibinfo {title} {Collective-excitation
  gap in the fractional quantum hall effect},\ }\href
  {https://doi.org/10.1103/PhysRevLett.54.581} {\bibfield  {journal} {\bibinfo
  {journal} {Phys. Rev. Lett.}\ }\textbf {\bibinfo {volume} {54}},\ \bibinfo
  {pages} {581} (\bibinfo {year} {1985})}\BibitemShut {NoStop}%
\bibitem [{\citenamefont {Girvin}\ \emph {et~al.}(1986)\citenamefont {Girvin},
  \citenamefont {MacDonald},\ and\ \citenamefont {Platzman}}]{GMPQtoThe4th}%
  \BibitemOpen
  \bibfield  {author} {\bibinfo {author} {\bibfnamefont {S.~M.}\ \bibnamefont
  {Girvin}}, \bibinfo {author} {\bibfnamefont {A.~H.}\ \bibnamefont
  {MacDonald}},\ and\ \bibinfo {author} {\bibfnamefont {P.~M.}\ \bibnamefont
  {Platzman}},\ }\bibfield  {title} {\bibinfo {title} {Magneto-roton theory of
  collective excitations in the fractional quantum hall effect},\ }\href
  {https://doi.org/10.1103/PhysRevB.33.2481} {\bibfield  {journal} {\bibinfo
  {journal} {Phys. Rev. B}\ }\textbf {\bibinfo {volume} {33}},\ \bibinfo
  {pages} {2481} (\bibinfo {year} {1986})}\BibitemShut {NoStop}%
\bibitem [{\citenamefont {Haldane}(1990)}]{haldane_quantum_1990}%
  \BibitemOpen
  \bibfield  {author} {\bibinfo {author} {\bibfnamefont {F.~D.~M.}\
  \bibnamefont {Haldane}},\ }\bibinfo {title} {The quantum {Hall} effect}\
  (\bibinfo  {publisher} {Springer, New York},\ \bibinfo {year} {1990})\ Chap.\
  \bibinfo {chapter} {The Hierarchy of Fractional States and Numerical
  Studies}, p.\ \bibinfo {pages} {303}\BibitemShut {NoStop}%
\bibitem [{Note3()}]{Note3}%
  \BibitemOpen
  \bibinfo {note} {$S_0(0)=0$}\BibitemShut {NoStop}%
\bibitem [{\citenamefont {Haldane}()}]{HaldaneUnpublished}%
  \BibitemOpen
  \bibfield  {author} {\bibinfo {author} {\bibfnamefont {F.~D.~M.}\
  \bibnamefont {Haldane}},\ }\bibinfo {note} {unpublished}\BibitemShut
  {NoStop}%
\bibitem [{Note4()}]{Note4}%
  \BibitemOpen
  \bibinfo {note} {While the structure factor is defined to be 0 for $L=0$ the
  equation evaluates to $\protect \frac {N^2}{2Q+1}$ for $L=0$}\BibitemShut
  {NoStop}%
\bibitem [{Note5()}]{Note5}%
  \BibitemOpen
  \bibinfo {note} {Conditioned on the state at the two points MV3 and CV7 being
  gapped in the first place, as discussed later finite size effects prohibit a
  definitive answer}\BibitemShut {NoStop}%
\bibitem [{\citenamefont {Storni}\ \emph {et~al.}(2010)\citenamefont {Storni},
  \citenamefont {Morf},\ and\ \citenamefont
  {Das~Sarma}}]{Morf2010CoulSameUCAsMR}%
  \BibitemOpen
  \bibfield  {author} {\bibinfo {author} {\bibfnamefont {M.}~\bibnamefont
  {Storni}}, \bibinfo {author} {\bibfnamefont {R.~H.}\ \bibnamefont {Morf}},\
  and\ \bibinfo {author} {\bibfnamefont {S.}~\bibnamefont {Das~Sarma}},\
  }\bibfield  {title} {\bibinfo {title} {Fractional quantum hall state at
  $\ensuremath{\nu}=\frac{5}{2}$ and the moore-read pfaffian},\ }\href
  {https://doi.org/10.1103/PhysRevLett.104.076803} {\bibfield  {journal}
  {\bibinfo  {journal} {Phys. Rev. Lett.}\ }\textbf {\bibinfo {volume} {104}},\
  \bibinfo {pages} {076803} (\bibinfo {year} {2010})}\BibitemShut {NoStop}%
\bibitem [{\citenamefont {Jain}(1989)}]{Jain-CF-1989}%
  \BibitemOpen
  \bibfield  {author} {\bibinfo {author} {\bibfnamefont {J.~K.}\ \bibnamefont
  {Jain}},\ }\bibfield  {title} {\bibinfo {title} {Composite-fermion approach
  for the fractional quantum {Hall} effect},\ }\href
  {https://doi.org/10.1103/PhysRevLett.63.199} {\bibfield  {journal} {\bibinfo
  {journal} {Phys. Rev. Lett.}\ }\textbf {\bibinfo {volume} {63}},\ \bibinfo
  {pages} {199} (\bibinfo {year} {1989})}\BibitemShut {NoStop}%
\bibitem [{Note6()}]{Note6}%
  \BibitemOpen
  \bibinfo {note} {Fitting using all the system sizes (not shown) would give
  $\alpha \approx 3.6$ for $L=2$ and $\alpha \approx 2.6$ for
  $L=3$}\BibitemShut {NoStop}%
\bibitem [{\citenamefont {Sodemann}\ and\ \citenamefont
  {MacDonald}(2013)}]{Sodemann-2013}%
  \BibitemOpen
  \bibfield  {author} {\bibinfo {author} {\bibfnamefont {I.}~\bibnamefont
  {Sodemann}}\ and\ \bibinfo {author} {\bibfnamefont {A.~H.}\ \bibnamefont
  {MacDonald}},\ }\bibfield  {title} {\bibinfo {title} {Landau level mixing and
  the fractional quantum hall effect},\ }\href
  {https://doi.org/10.1103/PhysRevB.87.245425} {\bibfield  {journal} {\bibinfo
  {journal} {Phys. Rev. B}\ }\textbf {\bibinfo {volume} {87}},\ \bibinfo
  {pages} {245425} (\bibinfo {year} {2013})}\BibitemShut {NoStop}%
\bibitem [{\citenamefont {Antoni\'{c}}\ \emph {et~al.}(2018)\citenamefont
  {Antoni\'{c}}, \citenamefont {Vucicevi\'{c}},\ and\ \citenamefont
  {Milovanovi\'{c}}}]{PhysRevB.98.115107}%
  \BibitemOpen
  \bibfield  {author} {\bibinfo {author} {\bibfnamefont {L.}~\bibnamefont
  {Antoni\'{c}}}, \bibinfo {author} {\bibfnamefont {J.}~\bibnamefont
  {Vucicevi\'{c}}},\ and\ \bibinfo {author} {\bibfnamefont {M.~V.}\
  \bibnamefont {Milovanovi\'{c}}},\ }\bibfield  {title} {\bibinfo {title}
  {Paired states at 5/2: Particle-hole pfaffian and particle-hole symmetry
  breaking},\ }\href {https://doi.org/10.1103/PhysRevB.98.115107} {\bibfield
  {journal} {\bibinfo  {journal} {Phys. Rev. B}\ }\textbf {\bibinfo {volume}
  {98}},\ \bibinfo {pages} {115107} (\bibinfo {year} {2018})}\BibitemShut
  {NoStop}%
\bibitem [{\citenamefont {Djurdjevi\'{c}}\ and\ \citenamefont
  {Milovanovi\'{c}}(2019)}]{PhysRevB.100.195303}%
  \BibitemOpen
  \bibfield  {author} {\bibinfo {author} {\bibfnamefont {S.}~\bibnamefont
  {Djurdjevi\'{c}}}\ and\ \bibinfo {author} {\bibfnamefont {M.~V.}\
  \bibnamefont {Milovanovi\'{c}}},\ }\bibfield  {title} {\bibinfo {title}
  {Model interactions for pfaffian paired states based on chern-simons field
  theory description},\ }\href {https://doi.org/10.1103/PhysRevB.100.195303}
  {\bibfield  {journal} {\bibinfo  {journal} {Phys. Rev. B}\ }\textbf {\bibinfo
  {volume} {100}},\ \bibinfo {pages} {195303} (\bibinfo {year}
  {2019})}\BibitemShut {NoStop}%
\bibitem [{\citenamefont {Milovanovi\'{c}}\ \emph {et~al.}(2020)\citenamefont
  {Milovanovi\'{c}}, \citenamefont {Djurdjevi\'{c}}, \citenamefont
  {Vucicevi\'{c}},\ and\ \citenamefont {Antoni\'{c}}}]{Milovanovic2020MPLB}%
  \BibitemOpen
  \bibfield  {author} {\bibinfo {author} {\bibfnamefont {M.~V.}\ \bibnamefont
  {Milovanovi\'{c}}}, \bibinfo {author} {\bibfnamefont {S.}~\bibnamefont
  {Djurdjevi\'{c}}}, \bibinfo {author} {\bibfnamefont {J.}~\bibnamefont
  {Vucicevi\'{c}}},\ and\ \bibinfo {author} {\bibfnamefont {L.}~\bibnamefont
  {Antoni\'{c}}},\ }\bibfield  {title} {\bibinfo {title} {Pfaffian paired
  states for half-integer fractional quantum hall effect},\ }\href
  {https://doi.org/10.1142/S0217984920300045} {\bibfield  {journal} {\bibinfo
  {journal} {Modern Physics Letters B}\ }\textbf {\bibinfo {volume} {34}},\
  \bibinfo {pages} {2030004} (\bibinfo {year} {2020})},\ \Eprint
  {https://arxiv.org/abs/https://doi.org/10.1142/S0217984920300045}
  {https://doi.org/10.1142/S0217984920300045} \BibitemShut {NoStop}%
\bibitem [{\citenamefont {Pakrouski}\ \emph {et~al.}(2020)\citenamefont
  {Pakrouski}, \citenamefont {Pallegar}, \citenamefont {Popov},\ and\
  \citenamefont {Klebanov}}]{pakrouski2020GroupInvariantScars}%
  \BibitemOpen
  \bibfield  {author} {\bibinfo {author} {\bibfnamefont {K.}~\bibnamefont
  {Pakrouski}}, \bibinfo {author} {\bibfnamefont {P.~N.}\ \bibnamefont
  {Pallegar}}, \bibinfo {author} {\bibfnamefont {F.~K.}\ \bibnamefont
  {Popov}},\ and\ \bibinfo {author} {\bibfnamefont {I.~R.}\ \bibnamefont
  {Klebanov}},\ }\bibfield  {title} {\bibinfo {title} {Many-body scars as a
  group invariant sector of hilbert space},\ }\href
  {https://doi.org/10.1103/PhysRevLett.125.230602} {\bibfield  {journal}
  {\bibinfo  {journal} {Phys. Rev. Lett.}\ }\textbf {\bibinfo {volume} {125}},\
  \bibinfo {pages} {230602} (\bibinfo {year} {2020})}\BibitemShut {NoStop}%
\bibitem [{\citenamefont {Pakrouski}\ \emph {et~al.}(2021)\citenamefont
  {Pakrouski}, \citenamefont {Pallegar}, \citenamefont {Popov},\ and\
  \citenamefont {Klebanov}}]{pakrouski2021group}%
  \BibitemOpen
  \bibfield  {author} {\bibinfo {author} {\bibfnamefont {K.}~\bibnamefont
  {Pakrouski}}, \bibinfo {author} {\bibfnamefont {P.~N.}\ \bibnamefont
  {Pallegar}}, \bibinfo {author} {\bibfnamefont {F.~K.}\ \bibnamefont
  {Popov}},\ and\ \bibinfo {author} {\bibfnamefont {I.~R.}\ \bibnamefont
  {Klebanov}},\ }\bibfield  {title} {\bibinfo {title} {Group theoretic approach
  to many-body scar states in fermionic lattice models},\ }\href
  {https://doi.org/10.1103/PhysRevResearch.3.043156} {\bibfield  {journal}
  {\bibinfo  {journal} {Phys. Rev. Research}\ }\textbf {\bibinfo {volume}
  {3}},\ \bibinfo {pages} {043156} (\bibinfo {year} {2021})}\BibitemShut
  {NoStop}%
\bibitem [{\citenamefont {Serbyn}\ \emph {et~al.}(2021)\citenamefont {Serbyn},
  \citenamefont {Abanin},\ and\ \citenamefont {Papi{\'c}}}]{Serbyn:2020wys}%
  \BibitemOpen
  \bibfield  {author} {\bibinfo {author} {\bibfnamefont {M.}~\bibnamefont
  {Serbyn}}, \bibinfo {author} {\bibfnamefont {D.~A.}\ \bibnamefont {Abanin}},\
  and\ \bibinfo {author} {\bibfnamefont {Z.}~\bibnamefont {Papi{\'c}}},\
  }\bibfield  {title} {\bibinfo {title} {Quantum many-body scars and weak
  breaking of ergodicity},\ }\bibfield  {journal} {\bibinfo  {journal} {Nature
  Physics}\ }\href {https://doi.org/10.1038/s41567-021-01230-2}
  {10.1038/s41567-021-01230-2} (\bibinfo {year} {2021})\BibitemShut {NoStop}%
\bibitem [{\citenamefont {Moudgalya}\ \emph {et~al.}(2021)\citenamefont
  {Moudgalya}, \citenamefont {Bernevig},\ and\ \citenamefont
  {Regnault}}]{Moudgalya:2021xlu}%
  \BibitemOpen
  \bibfield  {author} {\bibinfo {author} {\bibfnamefont {S.}~\bibnamefont
  {Moudgalya}}, \bibinfo {author} {\bibfnamefont {B.~A.}\ \bibnamefont
  {Bernevig}},\ and\ \bibinfo {author} {\bibfnamefont {N.}~\bibnamefont
  {Regnault}},\ }\bibfield  {title} {\bibinfo {title} {{Quantum Many-Body Scars
  and Hilbert Space Fragmentation: A Review of Exact Results}},\ }\href@noop {}
  {\  (\bibinfo {year} {2021})},\ \Eprint {https://arxiv.org/abs/2109.00548}
  {arXiv:2109.00548 [cond-mat.str-el]} \BibitemShut {NoStop}%
\bibitem [{\citenamefont {Sun}\ \emph {et~al.}(2020)\citenamefont {Sun},
  \citenamefont {Ma},\ and\ \citenamefont {Feldman}}]{Sun2020PRB}%
  \BibitemOpen
  \bibfield  {author} {\bibinfo {author} {\bibfnamefont {C.}~\bibnamefont
  {Sun}}, \bibinfo {author} {\bibfnamefont {K.~K.~W.}\ \bibnamefont {Ma}},\
  and\ \bibinfo {author} {\bibfnamefont {D.~E.}\ \bibnamefont {Feldman}},\
  }\bibfield  {title} {\bibinfo {title} {Particle-hole pfaffian order in a
  translationally and rotationally invariant system},\ }\href
  {https://doi.org/10.1103/PhysRevB.102.121303} {\bibfield  {journal} {\bibinfo
   {journal} {Phys. Rev. B}\ }\textbf {\bibinfo {volume} {102}},\ \bibinfo
  {pages} {121303} (\bibinfo {year} {2020})}\BibitemShut {NoStop}%
\end{thebibliography}%

\end{document}